\documentclass[aps, prb, twocolumn, amsmath, amssymb, amsfonts,superscriptaddress,preprintnumbers]{revtex4-2}
\usepackage{bm, ascmac, mathtools, braket}

\usepackage{times}
\usepackage{booktabs}
\usepackage{multirow}
\usepackage{graphicx}
\usepackage{float, color, xcolor}
\usepackage{bbold}
\usepackage[whole]{bxcjkjatype} 
\usepackage{cancel}
\usepackage{subfigure} 
\usepackage{amscd} 
\usepackage{makecell}
\usepackage{pifont}
\usepackage{relsize}
\usepackage{dsfont}

\newcommand{\xmark}{\ding{55}}
\usepackage[version=4]{mhchem}
\usepackage{appendix}
\usepackage{bbm}
\usepackage[normalem]{ulem}

\usepackage{comment}

\usepackage[
pagebackref=false,
colorlinks=true,
linkcolor=blue,
urlcolor=blue,
filecolor=black,
citecolor=red,
pdfstartview=FitV,
pdftitle={},
pdfauthor={},
pdfsubject={},
pdfkeywords={},
pdfpagemode=None,
bookmarksopen=true
    linkcolor={blue!80!black},
    citecolor={blue!80!black},
    urlcolor={blue!80!black}
]{hyperref}

\newcommand*{\rom}[1]{\expandafter\@slowromancap\romannumeral #1@}

\newcommand{\bk}{\mathbf{k}}

\usepackage{cellspace}
\begin{document}

\title{Quantum Geometric Origin of the 
Intrinsic Nonlinear Hall Effect}

\author{Yannis Ulrich}
\email{y.ulrich@fkf.mpg.de}
\affiliation{
Max Planck Institute for Solid State Research, D-70569 Stuttgart, Germany
}
\affiliation{School of Natural Sciences, Technische Universit\"at M\"unchen, D-85748 Garching, Germany}

\author{Johannes Mitscherling}
\affiliation{
Max Planck Institute for the Physics of Complex Systems, N\"othnitzer Strasse 38, D-01187 Dresden, Germany
}

\author{Laura Classen}
\affiliation{
Max Planck Institute for Solid State Research, D-70569 Stuttgart, Germany
}
\affiliation{School of Natural Sciences, Technische Universit\"at M\"unchen, D-85748 Garching, Germany}

\author{Andreas P. Schnyder}
\affiliation{
Max Planck Institute for Solid State Research, D-70569 Stuttgart, Germany
}

\date{\today}
\begin{abstract}
We decompose the intrinsic second-order nonlinear Hall effect (NLHE) of a generic multiband system into its quantum-geometric contributions within a fully quantum-mechanical, projector-based formalism. By expanding the nonlinear conductivity in powers of the quasiparticle lifetime $\tau$, we recover the established Berry curvature dipole at order $\tau$ and clarify discrepancies in previous literature concerning the (interband) quantum metric dipole (or Berry curvature polarizability) contribution at order $\tau^0$. Crucially, our method reveals an additional contribution at order $\tau^0$, determined by the {\it intraband} quantum metric dipole (intraQMD), arising from additional virtual interband transitions captured within the fully quantum-mechanical treatment. The intraQMD contribution is generically nonzero in systems with broken time-reversal symmetry and can be distinguished from other geometric contributions by symmetry. Analytical results for low-energy models of topological band crossings, which are hotspots of quantum geometry, demonstrate how band topology influences each contribution. In particular, the intraQMD contribution is especially large in gapped Dirac cones in antiferromagnets. Through a comprehensive symmetry classification of all magnetic space groups, we identify several candidate materials that are expected to exhibit large intrinsic NLHE, including the topological antiferromagnets Yb$_3$Pt$_4$, CuMnAs, and CoNb$_3$S$_6$, as well as the nodal-plane material MnNb$_3$S$_6$.
\end{abstract}

\maketitle

{\it Introduction.}---
Quantum geometry, such as the Berry curvature and the quantum metric, has become an essential tool for understanding wave-function properties of materials and relating them to physical responses~\cite{torma_PRL_essay_23, yu_queiroz_torma_arXiv_25, jiang_holder_review_arxiv_25,verma2025quantumgeometryrevisitingelectronic}. While the Berry curvature, whose integral is the Chern number, describes intrinsic linear Hall effects, the quantum metric is intimately connected to nonlinear responses, such as second-harmonic generation, bulk photovoltaic effects, and the nonlinear Hall effect~\cite{holder_consequences_2020, avdoshkin_multi-state_2024, ahn_low-frequency_2020, ZhuPenghao2024, Bhalla2022,wu_giant_2017,cook_design_2017,de_juan_quantized_2017, PhysRevB.61.5337,zhang_electrically_2018,low_topological_2015,ma_observation_2019,facio_strongly_2018,shvetsov_nonlinear_2019,dzsaber_giant_2021,kumar_room-temperature_2021,singh_engineering_2020,zeng_nonlinear_2021,du_band_2018,you_berry_2018,xiao_theory_2019,son_strain_2019,zhou_highly_2020,rostami_probing_2020,shao_nonlinear_2020,zhao_global_2017,  das_intrinsic_2023, Jankowski2024a, kaplan_unification_2024, Jankowski2025, xi_terahertz_2025, PhysRevLett.129.186801,wang_quantum-metric-induced_2023, doi:10.1021/acs.nanolett.4c05279, PhysRevB.105.155140}.
In strongly correlated systems, on the other hand, quantum geometry can guarantee finite superfluidity~\cite{torma_superconductivity_2022} and stabilize fractional Chern insulators~\cite{LIU2024515}.

The anomalous Hall effect establishes a direct connection between band topology and Berry curvature, being proportional to the integral of the latter over the Brillouin zone. In topological materials, concentrated Berry curvature originating from (near) band degeneracies, such as Weyl nodes or nodal planes, can strongly enhance the response~\cite{guo_moll_nat_phys_22, wilde_pfleiderer_news_view_nat_phys_22, huber_alpin_CoSi_PRL_22, zhong2016towards, Liang2016, wilde_Nature_21}.

In contrast, this connection is less understood in the nonlinear regime, where additional virtual interband transitions relate to state properties that are captured by more diverse quantum geometric invariants than just the Berry curvature~\cite{ahn_riemannian_2022,Bouhon2023,Jankowski2024,Jankowski2024a,avdoshkin_multi-state_2024,Jankowski2025,Mitscherling2025,fontana2025quantumgeometryelectricmagnetochiral}. These invariants are also generically large near (gapped) topological band crossings, and extend to regimes where conventional contributions are symmetry forbidden or strongly suppressed. For example, in time-reversal symmetric systems, where the linear anomalous Hall response vanishes, the Berry curvature dipole yields a lifetime-dependent nonlinear Hall contribution~\cite{sodemann_quantum_2015}.

However, the systematic investigation of nonlinear responses in terms of distinct quantum geometric invariants remains challenging. In particular, the lifetime-independent contribution, attributed to the interband quantum metric dipole (or Berry curvature polarizability), has been debated, with different semiclassical approaches leading to conflicting predictions~\cite{gao_field_2014,kaplan_unification_2024,das_intrinsic_2023,PhysRevB.110.245406}. This ambiguity is particularly consequential in $\mathcal{P}\mathcal{T}$-symmetric antiferromagnets, where lifetime-independent terms are expected to dominate, and has already led to contradictory experimental interpretations~\cite{doi:10.1126/science.adf1506,wang_quantum-metric-induced_2023}.

\begin{table*}
    \centering
    \begin{tabular}{lcccc}
        \hline\hline\\[-3mm]
         & \hspace{0.5cm}\makecell{Nonlinear Drude \\ weight (NLD)} \hspace{0.5cm}\phantom{a}& \hspace{0.5cm}\makecell{Berry curvature \\ dipole (BCD)} \hspace{0.5cm} \phantom{a} & \makecell{Interband quantum \\ metric dipole (interQMD)} & \makecell{Intraband quantum \\ metric dipole (intraQMD)} \\\hline\\[-2mm]
         Symbol & $\sigma^{a;bb}_{\mathrm{NLD}}(\bk)$ & $\sigma^{a;bb}_{\mathrm{BCD}}(\bk)$ & $\sigma^{a;bb}_{\mathrm{interQMD}}(\bk)$ & $\sigma^{a;bb}_{\mathrm{intraQMD}}(\bk)$ \\[2mm]\hline\\[-3mm]
         Full expression & $-\tau^2 \frac{v_n^{a}(\bk)}{m^{bb}_n(\bk)} \,N_n(\bk)$ & $\tau\,v_n^{b}(\bk)\, \Omega_{n}^{ab}(\bk)$ & $\sum_{m \neq n}2\frac{v_n^{b}(\bk)\,g_{mn}^{ab}(\bk)- v_n^{a}(\bk)\,g_{mn}^{bb}(\bk)}{\epsilon_{mn}(\bk)}$ & $-\frac{1}{2}\partial_{k_a}\,g^{bb}_n(\bk)$ \\[2mm]\hline\\[-2mm]
        Lifetime scaling & $\tau^2$ & $\tau$ & $\tau^0$ & $\tau^0$ \\[1mm]\hline\\[-3mm]
        \makecell{Quantum geometric \\ invariant} & $N_n(\bk)$ & $\Omega_{n}^{ab}(\bk)$ & $g_{mn}^{ab}(\bk)$ & $\partial_{k_a}\,g^{bb}_n(\bk)$ \\[2mm]
        & $\text{tr}\big[\hat P_n\big]$ &  $i\,\text{tr}\big[\hat P_n\,[\partial_{k_a} \hat P_n,\,\partial_{k_b}\hat P_n]\big]$ & $\frac{1}{2}\text{tr}\big[\partial_{k_a} \hat P_m\,\partial_{k_b} \hat P_n\big]$ & $\text{tr}\big[\partial_{k_a}\partial_{k_b} \hat{P}_n\,\partial_{k_b} \hat{P}_n\big]$ \\[1mm]\hline\\[-3mm]
        $\mathcal{P}$ & \xmark &\xmark  &\xmark  & \xmark \\
        $\mathcal{T}$ & \xmark & $\checkmark$ &  \xmark& \xmark \\
        $\mathcal{P}\mathcal{T}$ & $\checkmark$ & \xmark & $\checkmark$ & $\checkmark$ \\
        {$C_2^z\,,\ C_2^x\,,\ \mathcal{M}_x$} & \xmark & \xmark & \xmark &  \xmark\\
        $C_2^y\,,\ \mathcal{M}_y$ & $\checkmark$ & $\checkmark$ & $\checkmark$ &  $\checkmark$\\
        $C_3^z$& $\checkmark$ &  \xmark & \xmark & $\checkmark$\\
        \hline\hline
    \end{tabular}
    \caption{Full expressions for the leading-order contributions of distinct quantum geometric origin to the nonlinear conductivity, Eq.~\eqref{eq:finalformula}, together with their scaling in the quasiparticle lifetime $\tau$. The explicitly appearing quantum geometric invariants are the band degeneracy $N_n(\bk)$, Berry curvature $\Omega_{n}^{ab}(\bk)$, quantum metric $g_{mn}^{ab}(\bk)$, and quantum metric dipole $\partial_{k_a} g^{bb}_n(\bk)$. Symmetry conditions for all components of Eq.~\eqref{eq:finalformula} are shown for the representative case $\sigma^{x;yy}$. A check mark ($\checkmark$) indicates that a term is allowed under the corresponding symmetry. A complete table of nonzero tensor components, together with a magnetic space group analysis, is provided in the Supplemental Material~\cite{supplement}. $\mathcal{P}$, $\mathcal{T}$, $\mathcal{C}_n$, and $\mathcal{M}$ denote inversion, time-reversal, $n$-fold rotation, and mirror symmetries, respectively.}
    \label{table:summaryTerms}
\end{table*}

In this work, we present a comprehensive, fully gauge-invariant projector-based Green's-function framework that systematically decomposes the second-order nonlinear Hall effect (NLHE) into its quantum geometric contributions~\cite{PhysRevB.85.014435, Pozo2020, Graf2021, Mera2022, avdoshkin_extrinsic_2023, avdoshkin_multi-state_2024, Antebi2024, Avdoshkin2024, Mitscherling2025}. Our approach recovers the Berry curvature dipole and the interband quantum metric dipole (interQMD), confirmed by the results of a very recent reassessment~\cite{qiang2025clarificationquantummetricinducednonlineartransport} of deviating semiclassical results~\cite{gao_field_2014, kaplan_unification_2024, das_intrinsic_2023, PhysRevB.110.245406}, and reveals a third lifetime-independent contribution originating from the momentum derivative of the intraband quantum metric (intraQMD). This framework applies to both non-degenerate and degenerate bands, enabling analytical, gauge-invariant expressions for representative (gapped) Dirac/Weyl points, nodal lines, and nodal planes that serve as minimal models for the different contributions. By combining the analytical results with a comprehensive symmetry analysis for all magnetic space groups (MSG), we obtain stringent search and design criteria for materials with large intrinsic NLHE, as required for
high-frequency rectifiers and spintronic memory devices~{\cite{doi:10.1126/sciadv.aay2497, doi:10.1073/pnas.2100736118, wangIntrinsicNonlinearHall2023}}. The presented framework sets the stage for a systematic exploration of other nonlinear DC responses, 
advancing the ongoing research on nonlinear optical responses.

{\it Quantum geometric expansion of the NLHE.}---
We start by calculating the second-order DC conductivity tensor, relating the nonlinear response of the current density to the electric fields
via $j^a = \sum_{b,c}\sigma^{a;bc} E^b E^c$, with spatial indices $a$, $b$, $c$, for a general quadratic Hamiltonian 
\begin{align}
	\hat H = \sum_{\alpha,\beta} \hat c^\dagger_\alpha(\bk)\,H_{\alpha\beta}(\bk)\,\hat c^{}_{\beta}(\bk)  \, .
\end{align}
The Hamiltonian defines the Bloch Hamiltonian matrix $\hat H(\bk)=\big(H_{\alpha\beta}(\bk)\big)$ given in the ``orbital" basis, which may include spin, atomic orbitals, sublattice, or other degrees of freedom within the unit cell, with corresponding creation operators for fixed lattice momentum denoted by $\hat c^\dagger_\alpha(\bk)$. Using Green's function formalism, the nonlinear DC conductivity is given by~\cite{supplement,du_quantum_2021,michishita_effects_2021}
\begin{equation}\label{eq:KeldyshResultMain}
    \begin{aligned}
          &\sigma^{a;bc}=-\frac{e^3}{\hbar} \int_\text{BZ}\!\frac{d^d\bk}{(2\pi)^d}\int_{-\infty}^\infty \!\!\!\!\!d\epsilon\,\,\, \frac{\partial f(\epsilon)}{\partial \epsilon} \\ 
    &\times \text{Re}\,\text{tr}\Big[\hat A(\epsilon,\bk)\, \hat H^a(\bk)\, \frac{\partial \hat G^R(\epsilon,\bk)}{\partial \epsilon}\\&\times\Big( \hat H^b(\bk)\, \hat G^R(\epsilon,\bk)\,\hat H^c(\bk) +\frac{1}{2} \hat H^{bc}(\bk)+(b\leftrightarrow c)\Big)\Big]\, ,
    \end{aligned}
\end{equation}
involving momentum derivatives $\hat H^a(\bk) = \partial_{k_a}\hat H(\bk)$ and $\hat H^{ab}(\bk) = \partial_{k_a}\partial_{k_b}\hat H(\bk)$ of the Bloch Hamiltonian, the retarded Green's function $\hat G^R(\epsilon,\bk) = \big[\epsilon+\mu-\hat H(\bk)+i\Gamma\big]^{-1}$ with chemical potential $\mu$, the Fermi function distribution $f(\epsilon)$ at temperature $k_BT$, and the spectral function $\hat A(\epsilon,\bk) = -\pi \,\text{Im}\,\hat G^R(\epsilon,\bk)$. 
We introduce a phenomenological constant and nonzero imaginary part of the electron self-energy ${\Gamma>0}$~\cite{eberlein_fermi_2016, verret_phenomenological_2017, kontani_intrinsic_2007, tanaka_intrinsic_2008, mitscherling_longitudinal_2020, michishita_dissipation_2022}, capturing the electron lifetime $\tau=1/2\Gamma$. 
It corresponds to a broadening of the spectral density and can, e.g., be derived microscopically from white-noise disorder within a self-consistent Born approximation~\cite{rickayzen_greens_1980, Altland_Simons_2010}. A finite electron lifetime is required to regularize the conductivity in the DC limit. Further discussion of this approximation and a generalization to a band-dependent self-energy is provided in the Supplemental Material (SM)~\cite{supplement}.

The matrices within the trace in Eq.~\eqref{eq:KeldyshResultMain} generally do not commute, reflecting the nontrivial multiband nature of the system, which we characterize in terms of quantum geometry. In the following, we focus on $\sigma^{a;bb}$, which is most relevant for experiments, while the more general case $b \neq c$ is treated in the SM~\cite{supplement}. To reveal the relevant quantum geometric invariants for $\sigma^{a;bb}$, we diagonalize the Bloch Hamiltonian matrix, yielding
\begin{equation} \label{eq:projectordiag}
	  \hat H(\bk) = \sum_{n \in \text{bands}} E_n(\bk) \hat P_n(\bk) \, ,
\end{equation}
with dispersion $E_n(\bk)$ and orthogonal projector $\hat P_n(\bk)$ onto the (non-)degenerate bands. 

Importantly, all subsequent quantities, expressed in projectors, are manifestly gauge-invariant.
The nontrivial geometry arises from virtual interband transitions due to the coupling to the electric field and can be systematically characterized through a ``quantum geometric expansion" of $\sigma^{a;bb}$ obtained by inserting Eq.~\eqref{eq:projectordiag} into Eq.~\eqref{eq:KeldyshResultMain}. In this expansion, we separate band-dependent quantities, such as the direct band gap $\epsilon_{mn} = E_m - E_n$, the quasiparticle velocity $v^a_n = \partial_{k_a} E_n$, and the quasiparticle mass $m_n^{ab}=(\partial_{k_a} \partial_{k_b} E_n)^{-1}$, from quantum geometric invariants expressed as traces over projectors $\hat P_n$ and their derivatives $\partial_{k_a}\hat P_n$~\cite{Avdoshkin2024, Mitscherling2025}.

In order to determine the leading order contributions to the nonlinear DC response systematically, we combine the quantum geometric expansion with a perturbative expansion in $1/(\tau W)$ with bandwidth $W$ and keep the three dominant orders $\tau^{2}$, $\tau^{1}$, and $\tau^0$.
The straightforward but technical steps of the derivation are provided in the Supplemental Material~\cite{supplement}. In the zero-temperature limit, this yields the leading-order decomposition 
\begin{align}
   &\sigma^{a;bb} = \frac{e^3}{\hbar}\!\!\!\sum_{n\in\text{bands}}\int\!\!\!\frac{d^d\bk}{(2\pi)^d} \,\,\delta(E_n(\bk)-\mu) \Big[\sigma^{a;bb}_{\mathrm{NLD}}(\bk)\nonumber \\ &\hspace{8mm}+\sigma^{a;bb}_{\mathrm{BCD}}(\bk)+\sigma^{a;bb}_{\mathrm{interQMD}}(\bk)+\sigma^{a;bb}_{\mathrm{intraQMD}}(\bk)\Big]\,, \label{eq:finalformula}
\end{align}
for which we summarize the four distinct contributions, their scaling in the quasiparticle lifetime~$\tau$, and the relevant quantum geometric quantities in the manifestly gauge-invariant projector form in Tab.~\ref{table:summaryTerms}. These results can be extended to finite temperature. We identify four relevant quantum geometric invariants for the NLHE that cannot be further reduced up to partial integration in momentum: the band degeneracy $N_n$, Berry curvature $\Omega_{n}^{ab}$, two-state quantum metric $g_{mn}^{ab}$, and quantum metric dipole $\partial_{k_a} g^{bb}_n$. We fixed the ambiguity arising from partial integration in momentum by providing all expressions as Fermi-surface contributions.

{\it Distinct quantum geometric origins of the NLHE.}---
The nonlinear Drude weight (NLD), or Drude weight dipole, arises at order $\tau^2$ and only trivially depends on the quantum states via the band degeneracy $N_n = \text{tr}[\hat P_n]$~\cite{parker_diagrammatic_2019, kaplan_unifying_2023}. We find a further contribution proportional to $N_n$ at subleading order $\tau^0$ given by $\int_{\bk}  \sum_n \frac{1}{24}\delta'(E_n) \frac{v_n^{a}(\bk)}{m^{bb}_n(\bk)} N_n(\bk)$, as shown in the SM~\cite{supplement}.
The remaining three contributions are due to virtual interband transitions conveniently captured by the product of gauge-invariant interband transition operators $\hat e^a_{nm} = i\hat P_n\partial_{k_a}\hat P_m \hat P_m$ \cite{ahn_riemannian_2022, avdoshkin_multi-state_2024, Mitscherling2025}. We employ the defining properties of orthogonal projectors $\sum_{n}P_n=\mathbb{1}$ and $P_m P_n = \delta_{mn}P_m$ to systematically simplify the expressions by identifying the relevant quantum geometric invariants; see 
Tab.~\ref{table:summaryTerms}~\cite{Mitscherling2025, supplement}. At order $\tau^1$, we recover the well-established Berry curvature dipole (BCD)~\cite{sodemann_quantum_2015}. 

At order $\tau^0$, we find two distinct contributions. The first one involves the two-state quantum metric $g^{ab}_{mn}$, and is commonly referred to as (interband) quantum metric dipole (interQMD)~\cite{gao_field_2014}. While alternative approaches, such as the Luttinger-Kohn method~\cite{kaplan_unification_2024}, a Moyal product expansion~\cite{park2025quantumgeometrymoyalproduct}, or other semiclassical treatments~\cite{PhysRevB.110.174423, das_intrinsic_2023,PhysRevB.110.245406}, yield qualitatively similar terms, they differ in coefficients and symmetry behavior, particularly in $\mathcal{C}_3$-symmetric systems~\cite{gao_field_2014, kaplan_unification_2024}. These inconsistencies have led to conflicting interpretations of experiments~\cite{doi:10.1126/science.adf1506, wang_quantum-metric-induced_2023}. Our formulation, based solely on the Green's function expression Eq.~\eqref{eq:KeldyshResultMain} and a constant imaginary self-energy, agrees with the result of Gao \textit{et al.}~\cite{gao_field_2014}, and is consistent with the recent independent reinvestigation of the semiclassical theories~\cite{qiang2025clarificationquantummetricinducednonlineartransport}.

The second contribution at order $\tau^0$ involves the dipole of the {\it intraband quantum metric} $\partial_{k_a}g^{bb}_n$, which we label as intraband quantum metric dipole (intraQMD). It has not yet been identified in any aforementioned calculations~\cite{gao_field_2014, das_intrinsic_2023,kaplan_unification_2024, park2025quantumgeometrymoyalproduct} to the best of our knowledge. The quantum metric dipole arises from symmetrized and band-traced two-state quantum geometric connections $\text{tr}[\hat e_{nm}^b\partial_{k_a}\hat e_{mn}^c]$ and three-band contributions $\text{tr}\big[\hat e_{nm}^a \hat e_{ml}^b \hat e_{ln}^c\big]$ \cite{supplement, Mitscherling2025, mehraeen2025quantumresponsetheorymomentumspace}, showing its nontrivial interband origin. These contributions are present in Feynman diagrams containing only off-diagonal photon vertices, which distinguishes them from processes arising from extrinsic mechanisms such as skew scattering or side jumps~\cite{du_quantum_2021}. It originates from virtual interband processes captured naturally in the Green's function formalism but absent in semiclassical approaches, where quasiparticles are assigned delta-function spectral weight and lifetime effects enter only through a relaxation-time approximation~\cite{Rammer1986}. We discuss the difference to semiclassical methods further in the SM~\cite{supplement}. Note that the coefficient of the intraQMD is independent of the band structure.

Numerically, the separation of quantum geometry and band structure enables efficient evaluation in many cases~\cite{liu2023covariantderivativesberrytypequantities,PhysRevB.85.014435}. Besides the numerical differentiation schemes based on finite differences, derivatives of the projectors can be replaced by appropriate combinations of derivatives of the Bloch Hamiltonian matrix~\cite{supplement}.

{\it Distinguishing contributions by symmetries.}---
In the linear case, dissipative and non-dissipative (Hall) components correspond to the symmetric and antisymmetric parts of the conductivity tensor $\sigma^{ab}$ with $j^a = \sigma^{ab} E^b$. In the nonlinear case, this separation is more subtle and requires the full tensor $\sigma^{a;bc}$. Details are given in the SM~\cite{supplement}; importantly, all components of $\sigma^{a;bc}$ can be accessed experimentally~\cite{chichinadze2024observation}.

The different contributions can be distinguished by their transformation properties under common symmetries~\cite{supplement, kaplan_unification_2024, zhu_magnetic_2025, PhysRevB.109.085419}, as summarized in Tab.~\ref{table:summaryTerms}.  In the SM~\cite{supplement}, we provide the list of nonzero tensor components for each contribution. In particular, for antiferromagnetic order with the magnetic $\mathcal{P}\mathcal{T}$ symmetry, all terms except the BCD contribute. Moreover, $\mathcal{C}_3$ symmetry forbids the BCD and interQMD contributions, leaving the intraQMD as the only geometric response. Below, we will extend this magnetic symmetry analysis to identify material candidates.

To quantify the relevance of the different contributions beyond the symmetry analysis and establish their connection to band topology, we derive analytical results for the NLHE in several representative low-energy models.
We focus on non-centrosymmetric systems with broken time-reversal symmetry, thereby allowing for contributions beyond the BCD (see Table~\ref{table:summaryTerms}). The zero-, one-, or two-dimensional  degeneracies in momentum space can be locally approximated by low-energy Hamiltonians. In the following, we discuss a $\mathcal{P}\mathcal{T}$-symmetric 3D tilted Dirac point. In the SM~\cite{supplement}, we provide results for the two-dimensional (2D) tilted Dirac point, a 3D nodal line, and a 3D nodal plane, along with a description of the analytic methods employed. An overview of the discussed models is provided in Tab.~\ref{table:systems}.

\begin{figure}[t!]
    \centering
    \includegraphics[width=0.44\textwidth]{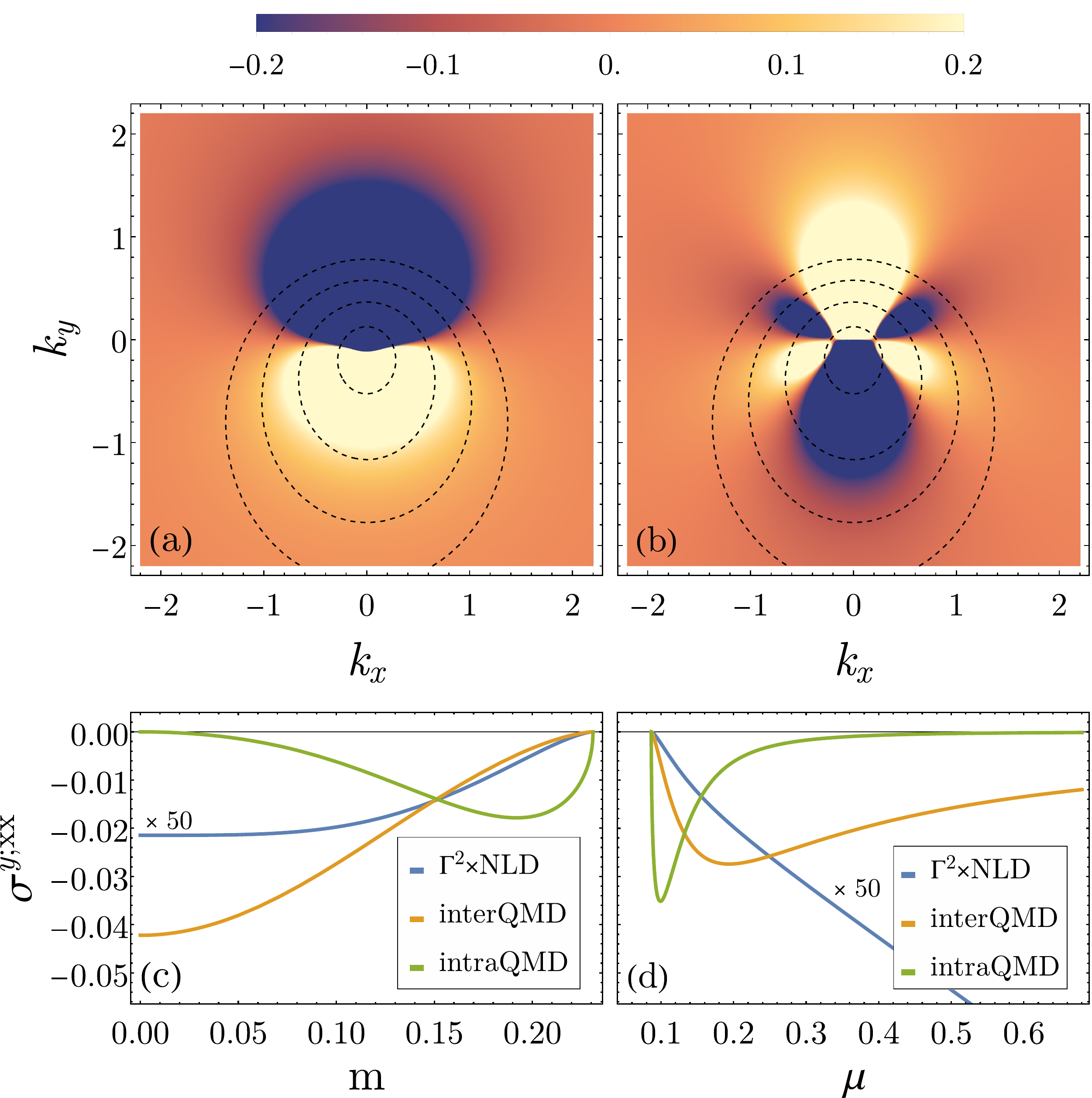}
    \caption{The (a) inter- and (b) intraQMD contributions before $\bk$-integration along the cut of momentum space $k_z=0$ for the 3D Dirac model defined in Eq.~\eqref{eq:model_3DDirac} with velocity $v=1$, tilt $t=0.5$, and mass $m=0.2$.  We give the dependence of the three nonzero contributions on (c) the gap $m$ for $\mu=0.2$ and (d) the chemical potential $\mu$ for $m=0.1$ of $\sigma^{y;xx}$.  In both cases, the chemical potential lies within the upper band.
    }
\label{fig:3DDiracPlot}
\end{figure}

{\it 3D tilted Dirac point.}---
To elucidate the behavior of QMD-driven contributions (inter- and intraQMD), we begin by investigating a four-band $\mathcal{P}\mathcal{T}$ symmetric 3D tilted Dirac point with uniform velocity $v$ described by the Hamiltonian
\begin{align} \label{eq:model_3DDirac}
    H = t k_y \sigma_0 \tau_0 +v\left(k_x \sigma_z \tau_x +k_y \sigma_0 \tau_y + k_z \sigma_x \tau_x \right) + m \sigma_y \tau_x\,,
\end{align}
with tensor product of Pauli matrices $\tau$ and $\sigma$. Antiferromagnetic systems with $\mathcal{P}\mathcal{T}$ symmetry have recently been studied to identify contributions to the NLHE beyond the BCD~\cite{PhysRevLett.127.277201,liu_shengyuan_PRL_21,doi:10.1126/science.adf1506}.  A mass term proportional to $m$ is introduced, which breaks $\mathcal{P}$ and $\mathcal{T}$ individually, while preserving $\mathcal{P}\mathcal{T}$. The tilt parameter $t\propto\mathds{1}$ also breaks inversion symmetry, affecting the Fermi surface but not the quantum geometry. Thus, to obtain contributions to the NLHE, we effectively compute skewed averages of the underlying quantum geometry of a Dirac point. This is illustrated in Fig.~\ref{fig:3DDiracPlot}(a,b), where we plot the inter- and intraQMD contributions prior to integration in a cross-section of $\bk$-space, along with the skewed Fermi surfaces. The model also possesses mirror symmetries $\mathcal{M}_x$, $\mathcal{M}_z$, which, as indicated in Table~\ref{table:summaryTerms}, imply that the only nonzero Hall conductivities are $\sigma^{y;xx}=\sigma^{y;zz}$. $\mathcal{P}\mathcal{T}$ symmetry enforces that the BCD contribution vanishes. The material \ce{CuMnAs} realizes such Dirac points near the Fermi energy and has previously been connected to the NLHE~\cite{tang_dirac_2016, PhysRevLett.127.277201}. Under pressure, \ce{Bi2CuO4} may also exhibit Dirac points near the Fermi surface in an antiferromagnetic phase~\cite{PhysRevB.96.121106}.

Note that the model consists of two doubly degenerate bands. The projector formalism is particularly convenient in this case, as it effectively reduces the system to a two-band model with projectors $\hat P_1(\bk)$ and $\hat P_2(\bk)$ projecting onto the entire degenerate eigenspaces of each band. Analytic expressions for the conductivities are obtained and presented in the SM~\cite{supplement}. All contributions except for the BCD are nonzero. The results are plotted in Fig.~\ref{fig:3DDiracPlot}(c,d) as functions of $m$ and $\mu$. At a fixed chemical potential $\mu=0.2$, the intraQMD contribution dominates for larger gap sizes. For a fixed gap $m=0.1$, the intraQMD and interQMD contributions are of equal importance when approaching the band edge. Particle-hole symmetry enforces that all contributions are odd functions of $\mu$~\cite{supplement}, which is confirmed by the analytic expressions. When the mass $m$ is set to 0, the Hamiltonian can be block-diagonalized into two gapless 3D Weyl points, with nonzero NLD and interQMD contributions.
The latter is proportional to 1/$\mu$, diverging at the band crossing point~\cite{supplement}.

\begin{table}[t!] 
\caption{\label{table:systems}
Contributions found for the minimal (tilted) Weyl/Dirac, nodal line, and nodal plane models discussed in the main text and the SM \cite{supplement}. Here, minimal indicates that terms of higher order in $k$ are neglected, which are typically
much smaller than the leading ones.}
\begin{ruledtabular}
\begin{tabular}{lcccc}
                       & NLD & BCD & interQMD & intraQMD \\ \hline
Gapped Dirac point in 3D     & $\checkmark$        & \xmark        & $\checkmark$             & $\checkmark$                                         \\ \hline
Gapless Weyl point in 3D         & $\checkmark$        & \xmark        & $\checkmark$             & \xmark                                        \\ \hline
Gapped Dirac point in 2D     & $\checkmark$        & $\checkmark$        & $\checkmark$             & $\checkmark$                                         \\ \hline
Gapless Dirac point in 2D & $\checkmark$        & \xmark  & $\checkmark$                   & $\checkmark$                                         \\ \hline
Nodal line in 3D     & $\checkmark$        & $\checkmark$        & $\checkmark$             & $\checkmark$                                         \\ \hline
Nodal plane in 3D           & $\checkmark$        & \xmark        & $\checkmark$             & $\checkmark$
\end{tabular}
\end{ruledtabular}
\end{table}

Extending the analysis to 3D nodal lines and planes completes the treatment of 0-, 1-, and 2-dimensional band degeneracies. For both the nodal line and plane, contributions increase as the band feature is made flatter, which can be understood in terms of an increased density of states. While for nodal lines QMD-driven contributions are strongly enhanced for a vanishing mass gap, the intraQMD is found to be proportional to the gap in the minimal nodal-plane model.~\cite{supplement, PhysRevB.111.L081115}.

{\it Candidate materials.}---
To identify materials exhibiting an NLHE driven by quantum geometric invariants beyond the BCD, we derive symmetry constraints for all magnetic space groups (MSGs), resulting in four distinct scenarios, each associated with candidate materials drawn from MAGNDATA~\cite{gallego_magndata_2016,gallego_magndata_2016-1}. Complete MSG listings and additional candidates are provided in the SM~\cite{supplement}. Representative examples include \ce{Yb3Pt4}, which realizes a case where, within a given plane, the BCD and interQMD are forbidden while the intraQMD remains symmetry-allowed. Another example is \ce{MnNb3S6}, a metallic ferromagnetic system with nodal-plane–enhanced quantum geometry, in which the BCD is forbidden, but both interQMD and intraQMD are allowed.

{\it Discussion.}---
Within a fully gauge-invariant approach, we identified three quantum geometric contributions to the second-order nonlinear Hall effect, including a previously unrecognized intraband quantum metric dipole (intraQMD). Our approach reproduces the established semiclassical contributions (NLD and BCD) while resolving prior discrepancies surrounding the interband quantum metric dipole (interQMD). A magnetic symmetry analysis revealed nodal-plane materials and antiferromagnetic Dirac systems as promising candidates for quantum-metric-dipole-driven NLHE, complementing the Berry-curvature–driven response.

Disorder and interactions are expected to generate additional quantum-geometric structure in nonlinear Hall responses. Our quantum-geometric Green's-function expansion provides a framework to incorporate microscopic scattering and assess whether further invariants emerge. Temperature-dependent measurements may distinguish contributions through their characteristic scattering-rate scaling, including the nonlinear Drude and quantum metric dipole terms or other extrinsic effects~\cite{wang_quantum-metric-induced_2023,PhysRevLett.129.186801}.

An even richer dependence on quantum geometric invariants beyond the quantum metric and Berry curvature can be expected for disorder- and interaction-induced nonlinear Hall responses when combining microscopic impurity or electron-phonon scattering with nontrivial Bloch states. While numerical progress has been made for minimal models, such as the 2D Dirac point~\cite{du_quantum_2021}, a complete understanding remains an open challenge. The presented quantum geometric expansion provides a suitable framework, starting from a perturbative Green's function approach, to enable studies of how contributions are modified and whether additional geometric invariants emerge. Temperature-dependent experiments may help separate contributions via the distinct scaling in the scattering rate, such as the nonlinear Drude and quantum metric dipole contributions, or new extrinsic effects~\cite{wang_quantum-metric-induced_2023, PhysRevLett.129.186801}.
 
The presented techniques can also be straightforwardly extended to further DC responses, such as the spin Hall, Nernst, and DC Kerr effects in the nonlinear regime. Additionally, our scattering rate expansion \cite{supplement} also enables access to higher-order effects such as third-order or magneto-nonlinear Hall responses~\cite{gao_field_2014}. A systematic understanding of the quantum geometric structure of nonlinear responses may reveal unexpected links between observables, clarify the role of emerging geometric invariants~\cite{ahn_riemannian_2022, avdoshkin_extrinsic_2023, Bouhon2023, Avdoshkin2024, Jankowski2024, Jankowski2024a, avdoshkin_multi-state_2024, Mitscherling2025, Jankowski2025}, and guide the design of nonlinear phenomena in quantum materials.

{\it Note added.--} During the refereeing process, a related pre-print appeared~\cite{guo2025dissipationshapedquantumgeometrynonlinear}, in which the nonlinear conductivity was calculated using the density-matrix formalism for a system coupled to a featureless bath and subsequently decomposed into its quantum geometric contributions. Their five contributions are fully consistent with ours.

\vspace{1ex}
{\it Acknowledgments.--}
We thank Pavel~Ostrovsky for deriving Eq.~\eqref{eq:KeldyshResultMain} and for enlightening us
about the role of disorder on the NLHE.
The authors are also grateful to
Tobias Holder, Moritz Hirschmann, and Raymond Wiedmann for  helpful discussions.
Y.U., L.C. and A.P.S. are funded by the Deutsche Forschungsgemeinschaft (DFG, German Research Foundation) – TRR 360 – 492547816.
J.M. was, in part, supported by the Deutsche Forschungsgemeinschaft under Grant cluster of excellence ct.qmat (EXC 2147, Project No. 390858490). 

\let\oldaddcontentsline\addcontentsline
\renewcommand{\addcontentsline}[3]{}
\bibliography{bibliography_prl_re}
\let\addcontentsline\oldaddcontentsline

%%%%% Supplemental Material %%%%%%%%%%%%%%%%%%%%%

\widetext
\pagebreak
\begin{center}
{\bf \huge Supplemental Material}
\end{center}

\thispagestyle{empty}
\tableofcontents

\renewcommand{\theequation}{S\arabic{equation}}

\renewcommand{\thefigure}{S\arabic{figure}}
\renewcommand{\thetable}{S\arabic{table}}
\renewcommand{\thesection}{S\arabic{section}}

\setcounter{equation}{0}
\setcounter{figure}{0}
\setcounter{table}{0}
\setcounter{section}{0}

\pagebreak
\setcounter{page}{1}

\section{NLHE in Keldysh formalism and the scattering rate} \label{app:Keldysh}
The nonlinear Hall conductivity can be calculated using the Keldysh formalism; the corresponding result was provided to us by Pavel Ostrovsky, also shown in the main  text:
\begin{equation}\label{suppeq:KeldyshResultMain}
    \begin{aligned}
          &\sigma^{a;bc}=-\frac{e^3}{\hbar} \int_\text{BZ}\!\frac{d^d\bk}{(2\pi)^d}\int_{-\infty}^\infty \!\!\!\!\!d\epsilon\,\,\, \frac{\partial f(\epsilon)}{\partial \epsilon}\, \text{Re}\,\text{tr}\Big[\hat A(\epsilon,\bk)\, \hat H^a(\bk)\, \frac{\partial \hat G^R(\epsilon,\bk)}{\partial \epsilon}\,\Big( \hat H^b(\bk)\, \hat G^R(\epsilon,\bk)\,\hat H^c(\bk) +\frac{1}{2} \hat H^{bc}(\bk)+(b\leftrightarrow c)\Big)\Big]\, ,
    \end{aligned}
\end{equation}
involving momentum derivatives $\hat H^a(\bk) = \partial_{k^a}\hat H(\bk)$ and $\hat H^{ab}(\bk) = \partial_{k^a}\partial_{k^b}\hat H(\bk)$ of the Bloch Hamiltonian, the retarded Green's function $\hat G^R(\epsilon,\bk) = \big[\epsilon+\mu-\hat H(\bk)+i\Gamma\big]^{-1}$ with chemical potential $\mu$, the Fermi function distribution $f(\epsilon)$ at temperature $T$, and the spectral function $\hat A(\epsilon,\bk) = -\pi \,\text{Im}\,\hat G^R(\epsilon,\bk)$. This result is consistent with previous literature results~\cite{du_quantum_2021}, and also with Ref.~\cite{michishita_effects_2021} up to a prefactor of 2. We attribute this discrepancy to a missing factor of $1/2$ in their derivation. In both cases, the Fermi sea contribution vanishes, as can be confirmed by explicit calculation. In our convention, $e>0$.

The conductivity of a material is rendered finite by various scattering processes (arising from impurities, disorder, etc.). 
The constant phenomenological scattering rate that we include in our Green's functions accounts for these, and is comparable but not equivalent to the relaxation-time approximation employed in semiclassical calculations~\cite{kaplan_unifying_2023}. Indeed, as discussed in the main text, they do not capture the intraQMD contribution~\cite{kaplan_unification_2024, gao_field_2014, das_intrinsic_2023, park2025quantumgeometrymoyalproduct}. Previously, some Green's function calculations began with expressions for the conductivity at optical frequencies $\omega_i$ in clean systems and introduced lifetimes via the substitution ${\omega_i \rightarrow \omega_i + i \Gamma}$~\cite{holder_consequences_2020, oiwa_systematic_2022}. As we now show, this procedure yields different results. For example, for Green's functions that depend on two frequencies,
\begin{align}
    \mathcal{G}_n^{R}(\epsilon+\omega_1+\omega_2,\bk) = \frac{1}{\epsilon+\omega_1+\omega_2-E_n(\bk)+ i0^+} \rightarrow \frac{1}{\epsilon+\omega_1+\omega_2-E_n(\bk)+ 2i\Gamma}\,,
\end{align}
while a Green's function of the form $\mathcal{G}_n^{R}(\epsilon+\omega_1,\bk)$ would only be assigned a self-energy of $i\Gamma$. In the DC limit, the substitution ${\omega \rightarrow \omega + i \Gamma}$ thus leads to retarded and advanced Green's functions $\mathcal{G}_n^{R/A}(\epsilon, \bk)$ evaluated at the same energy $\epsilon$ and momentum $\bk$ but with different self-energies. In our formalism, both become $({\epsilon - E_n(\bk) + i\Gamma})^{-1}$ at $\omega_i = 0$. Rather than introducing self-energies via frequency substitutions, we assume that the system possesses an intrinsic scattering rate, which we approximate to be constant.

As mentioned in the main text, the constant scattering rate we use can be derived microscopically in a self-consistent Born approximation that includes the effects of white-noise disorder~\cite{rickayzen_greens_1980, Altland_Simons_2010}. The perturbative expansion to obtain this result is controlled by $\Gamma$ being much smaller than the Fermi momentum $\bk_F$ \cite{rickayzen_greens_1980, Altland_Simons_2010}.
In a complete treatment of disorder for a specific system, one would need to calculate a generally momentum-dependent self-energy as well as vertex corrections by explicitly including interactions with scattering potentials. Crossing diagrams---generally higher-order in $\Gamma$ than their non-crossing counterparts---may also play a role, as in the linear-response case~\cite{ado_anomalous_2015, rickayzen_greens_1980}. The constant $\Gamma$ we consider corresponds to calculating only the non-interacting diagrams, but with a self-energy included in the Green's functions. Our approach is designed to remain as agnostic as possible to the microscopic details of the system, isolating quantum geometric contributions to the intrinsic nonlinear Hall conductivity. The incorporation of aforementioned corrections is left for future work. 

We discuss the perturbative expansion in the scattering rate $\Gamma$ in Sec.~\ref{suppsec:cleanlimit} and the implications of a band-dependent $\Gamma$ in Sec.~\ref{suppsec:banddependentGamma}.

\section{Geometric decomposition of the nonlinear conductivity tensor}

In this section, we demonstrate how the nonlinear conductivity given in Eq.~\eqref{suppeq:KeldyshResultMain} is decomposed into its quantum geometric components. The summary of steps performed is as follows: We begin by expressing the Hamiltonian (and thus also the Green's functions) in terms of band projectors \cite{Mitscherling2025}. The obtained traces of projectors and their derivatives are then simplified into a minimal number of independent quantum geometric invariants, such as quantum metric and Berry curvature. The coefficients in front of each of these components are simplified and expanded in the scattering rate $\Gamma$. Finally, we combine all contributions and use integration by parts to streamline the result. 

\subsection{Introduction of band projectors and quantum geometric decomposition} \label{suppsec:insertingProjectors}

\subsubsection{Insertion of band projectors}

We employ the fact that the Bloch Hamiltonian can be decomposed in terms of its band dispersion $E_n(\bk)$ and the corresponding projection $\hat P_n(\bk)$ to the eigenspace, 
\begin{align}
    \hat{H}(\bk) = \sum_n E_n(\bk)  \hat{P}_n(\bk) \, ,
\end{align}
where we consider orthogonal projectors $\hat P_n(\bk)\hat P_m(\bk) = \delta_{nm} \hat P_n(\bk)$. To be concrete, the projectors take the form $\hat P_n(\bk) = \sum_\alpha |u_n^\alpha(\bk)\rangle\langle u_n^\alpha(\bk)|$ with Bloch functions $|u_n^\alpha(\bk)\rangle$, where $\alpha$ spans the possibly degenerate eigenspace of eigenvalue $E_n$. Note that the band projector is explicitly $U(N)$ invariant for $N$-degenerate bands \cite{Mera2022}. Due to the form of $\hat P_n$ all subsequent decompositions are explicitly gauge-invariant by construction. 
The retarded and advanced Green's function yield the form
\begin{align}
    G^{R/A}(\epsilon, \bk)  =  \frac{1}{\epsilon - \hat{H}(\bk) \pm i\Gamma} = \sum_n \frac{ \hat{P}_n(\bk)}{\epsilon - E_n(\bk) \pm i\Gamma}\, .
\end{align}
Inserting the projector expansion for the Green's functions into Eq.~\eqref{suppeq:KeldyshResultMain} results in 
\begin{align} \label{eq:SuppconductivityAnyT}
    \sigma^{a; b c}=& -\frac{e^3}{\hbar} \int\!\!\frac{d^d\bk}{(2\pi)^d}\int\!\! d\epsilon \,\frac{\partial f(\epsilon)}{\partial \epsilon}\Bigg[
    \sum_{m,n}w^{nm}(\epsilon)\,\text{tr}\big[\hat{P}_m (\partial^{a}\hat{H}) \hat{P}_n (\partial^{b}\partial^{c}\hat{H})\big]\nonumber\\
    & +\frac{1}{2}\sum_{l,m,n}u^{lmn}(\epsilon)\,\Big(\text{tr}\big[\hat{P}_l(\partial^{a}\hat{H})\hat{P}_m (\partial^{b}\hat{H})\hat{P}_n (\partial^{ c }\hat{H})\big]+\text{tr}\big[\hat{P}_l(\partial^{a}\hat{H})\hat{P}_m (\partial^{c}\hat{H})\hat{P}_n (\partial^{ b }\hat{H})\big]\Big)\Bigg]\,,
\end{align}
where we drop the explicit reference to the momentum dependence for shorter notation. Note the explicit $b\leftrightarrow c$ symmetry. We keep the momentum derivatives of the Bloch Hamiltonian explicit at that point. The functions $w^{nm}(\epsilon)$ and $u^{lmn}(\epsilon)$, which also depend on the energies of the bands, are given by
\begin{alignat}{3}
    &w^{nm}(\epsilon) &&= -\frac{1}{2} \left(A_n(\epsilon) G^A_m(\epsilon)^2+A_m(\epsilon) G^R_n(\epsilon )^2\right)\nonumber\\\label{eq:suppwDef}
    & &&=-\frac{\Gamma }{2 \pi  (E_m+i \Gamma )^2 \left(\Gamma ^2+E_n^2\right)}-\frac{\Gamma }{2 \pi  \left(\Gamma ^2+E_m^2\right) (E_n-i \Gamma )^2}\,, \\[2mm]
    &u^{lmn}(\epsilon) &&= -A_m(\epsilon ) G^A_l(\epsilon )^2 G^A_n(\epsilon)-A_l(\epsilon ) G^R_m(\epsilon )^2
   G^R_n(\epsilon )\nonumber \\ \label{eq:suppuDef}
   & &&=\frac{\Gamma }{\pi  \left(\Gamma ^2+E_l^2\right) (E_m-i \Gamma )^2 (E_n-i \Gamma )}+\frac{\Gamma }{\pi  (E_l+i \Gamma )^2 \left(\Gamma ^2+E_m^2\right) (E_n+i \Gamma )}\,,
\end{alignat}
where we used the spectral function
\begin{align}
    A_n(\epsilon) = -\frac{1}{2\pi i}\left(G^R(\epsilon)-G^A(\epsilon)\right)=\frac{\Gamma }{\pi  \left(\Gamma ^2+(\epsilon -E_n)^2\right)}\,.
\end{align}
At zero temperature $\frac{\partial f(\epsilon)}{\partial \epsilon}= -\delta(\epsilon)$ such that
\begin{align} \label{eq:SuppconductivityT0}
    \sigma^{a; b c}=& \frac{e^3}{\hbar} \int \!\!\frac{d^d\bk}{(2\pi)^d}\Bigg[
    \sum_{m,n}w^{nm}\,\text{tr}\big[\hat{P}_m (\partial^{a}\hat{H}) \hat{P}_n (\partial^{b}\partial^{c}\hat{H})\big]\nonumber\\
    & +\frac{1}{2}\sum_{l,m,n}u^{lmn}\,\Big(\text{tr}\big[\hat{P}_l(\partial^{a}\hat{H})\hat{P}_m (\partial^{b}\hat{H})\hat{P}_n (\partial^{ c }\hat{H})\big]+\text{tr}\big[\hat{P}_l(\partial^{a}\hat{H})\hat{P}_m (\partial^{c}\hat{H})\hat{P}_n (\partial^{ b }\hat{H})\big]\Big)\Bigg]\,,
\end{align}
where $w^{nm}= w^{nm}(0)$ and $u^{lmn} = u^{lmn}(0)$.

\subsubsection{Summary of quantum geometric invariants}

We briefly summarize and set the notation for the quantum geometric quantities in their projector form that we will use hereafter \cite{avdoshkin_multi-state_2024, Mitscherling2025}. The single- and multi-band quantum geometric tensor are defined as
\begin{align} \label{eq:QGdefQGTsingleband}
    &Q_{n}^{ab} = \text{tr}\big[\hat{P}_n(\partial^a \hat{P}_n)(\partial^b \hat{P}_n)\big] = g_n^{ab}-\frac{i}{2}\Omega_n^{ab}\,, \\
    \label{eq:QGdefQGT}
    &Q_{mn}^{ab} = \text{tr}\big[\hat{P}_n(\partial^a \hat{P}_m)(\partial^b \hat{P}_n)\big]\,.
\end{align}
where $g_n^{ab}$ is the (real) single-band quantum metric and $\Omega_n^{ab}$ is the Berry curvature. The two-state quantum metric is related to the multi-band quantum geometric tensor by $g_{mn}^{ab} =\operatorname{Re} Q_{mn}^{ab}$.
The (two-band) quantum geometric connection $C^{a;bc}_{mn}$ is defined as
\begin{align} \label{eq:QGdefConnection}
    C^{a;bc}_{mn} = \text{tr}\Big[P_n (\partial^b \hat{P}_m) \big((\partial^a\partial^c \hat{P}_n)+(\partial^a \hat{P}_m)(\partial^c \hat{P}_n)\big)\Big]\,.
\end{align}
There is also a single-band connection, or skewness tensor $Q_{n}^{ a ; b  c }$
\begin{align} \label{eq:QGdefSkewness}
    Q_{n}^{ a ; b  c } = \text{tr}\big[\hat{P}_n(\partial^ a  \hat{P}_n)(\partial^ b  \partial^ c  \hat{P}_n)\big] = Q_{n}^{ a ; c  b }\,.
\end{align}

\subsubsection{Quantum geometric decomposition}

We now turn to decomposing the two types of traces appearing in Eq.~\eqref{eq:SuppconductivityT0}. The first derivative of the Bloch Hamiltonian decomposes as \cite{Mitscherling2025} 
\begin{align} \label{eq:dHformulaProj}
    \hat{P}_m \big(\partial^{a} \hat H\big) \hat{P}_n = \delta_{mn}\, E^a_n\,\hat{P}_n - \epsilon_{mn} \hat{P}_m (\partial^{a}\hat{P}_n)\hat{P}_n 
\end{align}
with $E^a_n = \partial^a E_n$ and $\epsilon_{mn}=E_m-E_n$. Moreover, $\hat{P}_m (\partial^{a}\hat{P}_n)\hat{P}_n=-\hat{P}_m (\partial^{a}\hat{P}_m)\hat{P}_n$. 
Using this decomposition, the trace involving three first derivatives of the Bloch Hamiltonian is expanded into 
\begin{align}
\text{tr}[\hat{P}_l(\partial^{a}\hat{H})\hat{P}_m (\partial^{b}\hat{H})\hat{P}_n (\partial^{ c }\hat{H})]
    =& \,\text{tr}\big[\hat{P}_l(\partial^{a}\hat{H})\hat{P}_m \hat{P}_m (\partial^{b}\hat{H})\hat{P}_n \hat{P}_n (\partial^{ c }\hat{H}) \hat{P}_l\big] \\
    =& \,\delta_{lm}\delta_{mn}\delta_{nl} \,E^a_m E^b_nE^c_l \,\text{tr}\big[\hat{P}_n \hat{P}_m \hat{P}_l\big] \nonumber \\\nonumber
    & + \delta_{lm}\,\epsilon_{mn}\epsilon_{nl}\, E^a_m\, \text{tr}\big[\hat{P}_m \hat{P}_m (\partial^{b}\hat{P}_n)\hat{P}_n \hat{P}_n (\partial^{ c }\hat{P}_l)\hat{P}_l\big]\\\nonumber
    & + \delta_{mn}\,\epsilon_{lm}\epsilon_{nl}\, E^b_n \, \text{tr}\big[\hat{P}_l (\partial^{a}\hat{P}_m)\hat{P}_m \hat{P}_n \hat{P}_n (\partial^{ c }\hat{P}_l)\hat{P}_l\big]\\\nonumber
    & + \delta_{nl}\,\epsilon_{lm}\epsilon_{mn}\, E^c_l\, \text{tr}\big[\hat{P}_l (\partial^{a}\hat{P}_m)\hat{P}_m \hat{P}_m (\partial^{b}\hat{P}_n)\hat{P}_n \hat{P}_l\big] \\
    &- \epsilon_{lm}\epsilon_{mn}\epsilon_{nl}\,\text{tr}\big[\hat{P}_l(\partial^{a}\hat{P}_m)\hat{P}_m \hat{P}_m (\partial^{b}\hat{P}_n)\hat{P}_n \hat{P}_n (\partial^{ c }\hat{P}_l)\hat{P}_l\big] \\
    =&\  \delta_{lm}\delta_{mn}\delta_{nl} \, E^a_m E^b_n E^c_l \,\text{tr}\big[\hat{P}_m\big] \nonumber \\\nonumber
    & - \delta_{lm}\,\epsilon_{mn}^2\, E^a_m \,\text{tr}\big[\hat{P}_m (\partial^{b}\hat{P}_n)\hat{P}_n (\partial^{ c }\hat{P}_l)\big]\\\nonumber
    & - \delta_{mn}\,\epsilon_{nl}^2 \, E^b_n \,\text{tr}\big[\hat{P}_l (\partial^{a}\hat{P}_m)\hat{P}_m (\partial^{ c }\hat{P}_l)\big]\\\nonumber
    & - \delta_{nl}\,\epsilon_{lm}^2 \,E^c_l \, \text{tr}\big[\hat{P}_l (\partial^{a}\hat{P}_m)\hat{P}_m (\partial^{b}\hat{P}_n)\big] \\
    &- \epsilon_{lm}\epsilon_{mn}\epsilon_{nl}\,\text{tr}\big[\hat{P}_l(\partial^{a}\hat{P}_m)\hat{P}_m (\partial^{b}\hat{P}_n)\hat{P}_n (\partial^{ c }\hat{P}_l)\big] \,.
\end{align}
Terms with two velocities, such as $E^a_m E^b_n$, vanish due to $\epsilon_{nn}=0$. Using that $\epsilon_{nm}\neq 0 \Rightarrow n\neq m$ and $\hat P_n (\partial^ a  \hat{P}_m)\hat{P}_m = \hat{P}_n (\partial^ a  \hat{P}_m) $ for $n\neq m$, we obtain
\begin{align}
&\text{tr}\big[\hat{P}_l(\partial^{a}\hat{H})\hat{P}_m (\partial^{b}\hat{H})\hat{P}_n (\partial^{ c }\hat{H})\big] \nonumber \\
    &=\,\delta_{lm}\delta_{mn}\delta_{nl} \,E^a_m E^b_n E^c_l \,\text{tr}\big[\hat{P}_l\big] \nonumber \\\nonumber
    & - \delta_{lm}\,\epsilon_{mn}^2\,E^a_m \,\text{tr}\big[\hat{P}_m(\partial^{ b }\hat{P}_n)(\partial^{ c }\hat{P}_m)\big] - \delta_{mn}\,\epsilon_{nl}^2\, E^b_n\, \text{tr}\big[\hat{P}_l(\partial^{ a }\hat{P}_m)(\partial^{ c }\hat{P}_l)\big] - \delta_{nl}\,\epsilon_{lm}^2\, E^c_l\, \text{tr}\big[\hat{P}_n(\partial^{ a }\hat{P}_m)(\partial^{ b }\hat{P}_n)\big] \\
    &- \epsilon_{lm}\epsilon_{mn}\epsilon_{nl}\,\text{tr}\big[\hat{P}_l(\partial^{ a }\hat{P}_m)(\partial^{ b }\hat{P}_n) (\partial^{ c }\hat{P}_l)\big]\,. 
\end{align}
We can now identify the two-state quantum geometric tensor given in Eq.~\eqref{eq:QGdefQGT} in the second line, arriving at
\begin{align} \label{eq:tripleHTraceGeometricDecomp}
\text{tr}\big[\hat{P}_l(\partial^{a}\hat{H})\hat{P}_m (\partial^{b}\hat{H})\hat{P}_n (\partial^{ c }\hat{H})\big]
    &=\delta_{lm}\delta_{mn}\delta_{nl} \, E_m^ a  E_n^{ b } E_l^{ c }  \,\text{tr}\big[\hat{P}_l\big] \nonumber \\\nonumber
    & - \delta_{lm} \, \epsilon_{nm}^2 \, E_m^{ a } \, Q_{nm}^{ b  c }- \delta_{mn}\,\epsilon_{ml}^2 \, E_m^{ b } Q_{ml}^{ a  c }- \delta_{nl}\,\epsilon_{ml}^2 \, E_l^{ c } \, Q_{ml}^{ a  b } \\
    &- \epsilon_{lm}\epsilon_{mn}\epsilon_{nl}\,\text{tr}\big[\hat{P}_l(\partial^{ a }\hat{P}_m)(\partial^{ b }\hat{P}_n) (\partial^{ c }\hat{P}_l)\big]\,. 
\end{align}
The second trace in Eq.~\eqref{eq:SuppconductivityT0} is more elaborate as it involves the second derivative of the Bloch Hamiltonian. To simplify the expression, we use the (hermitian) gauge-invariant interband transition rate in projector form \cite{ahn_riemannian_2022, Mitscherling2025}
\begin{align}
    \hat{e}^a_{m n}=i  \hat{P}_m (\partial^a  \hat{P}_n)  \hat{P}_n,
\end{align}
such that the first derivative of the Bloch Hamiltonian yields the form
\begin{align} \label{eq:SuppDelHamEFormula}
    \hat{P}_m \big(\partial^a \hat H\big) \hat{P}_n = \delta_{mn} \, E_n^{a} \, \hat{P}_n + i \,\epsilon_{mn}\,\hat e^{a}_{mn}\,.
\end{align}
We use this decomposition in combination with the completeness of the band projectors $\sum_l \hat P_l(\bk) = \mathds{1}$ to obtain
\begin{align}
    \hat{P}_m \big(\partial^{ b} \partial^{ c }\hat{H}\big) \hat{P}_n &= \sum_{l_1,l_2}\hat{P}_m \,\big(\partial^{ b }\big[\hat{P}_{l_1}(\partial^{ c }\hat{H}) \hat{P}_{l_2}\big]\big)\,\hat{P}_n \\
    &=\sum_{l_1,l_2} \hat{P}_m \big(\partial^{ b }\big[\delta_{l_1,l_2}E_{l_1}^{ c }\hat{P}_{l_1}\big]\big)\hat{P}_n + i\sum_{l_1,l_2} \hat{P}_m \big(\partial^{ b }\big[\epsilon_{l_1 l_2}\hat e^{ c }_{l_1 l_2}\big]\big)\hat{P}_n \\
    &=\sum_{l_1} E_{l_1}^{ b  c }\hat{P}_m\hat{P}_{l_1}\hat{P}_n + \sum_{l_1} E_{l_1}^{ c } \hat{P}_m (\partial^{ b }\hat{P}_{l_1})\hat{P}_n + i\sum_{l_1,l_2}\epsilon_{l_1 l_2}^{ b } \hat{P}^{}_m \hat e^{ c }_{l_1 l_2}\hat{P}^{}_n+ i\sum_{l_1,l_2}\epsilon_{l_1 l_2} \hat{P}^{}_m (\partial^{ b }\hat e^{ c }_{l_1 l_2})\hat{P}^{}_n \\
    &= \delta_{mn} E_{n}^{ b  c }\hat{P}_{n} + i\epsilon_{m n}^{ c } \hat e^{ b }_{m n} + i\epsilon_{m n}^{ b } \hat e^{ c }_{m n}+ i\sum_{l_1,l_2}\epsilon_{l_1 l_2} \hat{P}_m (\partial^{ b } \hat e^{ c }_{l_1 l_2})\hat{P}_n\,.
    \label{eqn:secondDerivHPre}
\end{align}
using $\hat P_m(\partial^b\hat P_l)\hat P_n = 0$ for $l\neq n,m$. We introduced the short notation $E^{bc}_n = \partial^b\partial^c E_n$ and $\epsilon^b_{mn} = \partial^b \epsilon_{mn}$. The last term needs further consideration. We note that it vanishes if none of the indices $l_1$ and $ l_2$ are $ m$ or $ n$, since 
\begin{align}
    \hat{P}^{}_m (\partial^{ b }\hat e^{ c }_{l_1l_2})\hat{P}^{}_n &=i \hat{P}^{}_m(\partial^{ b }\hat{P}_{l_1}) (\partial^{ c }\hat{P}^{}_{l_2}) \hat{P}_{l_2} \hat P^{}_n + i \hat{P}^{}_m \hat{P}^{}_{l_1} (\partial^{ b }\partial^{ c }\hat{P}^{}_{l_2}) \hat{P}^{}_{l_2} \hat{P}^{}_n + i \hat{P}^{}_m \hat{P}^{}_{l_1} (\partial^{ c }\hat{P}^{}_{l_2}) (\partial^{ b }\hat{P}^{}_{l_2}) \hat{P}^{}_n\\
    &=i \delta_{nl_2} \hat{P}^{}_m(\partial^{ b }\hat{P}^{}_{l_1}) (\partial^{ c }\hat{P}^{}_n) \hat P^{}_n + i \delta_{ml_1}\delta_{nl_2}\hat{P}^{}_m (\partial^{ b }\partial^{ c }\hat{P}^{}_n) \hat{P}^{}_n + i \delta_{ml_1} \hat{P}^{}_m (\partial^{ c }\hat{P}^{}_{l_2}) (\partial^{ b }\hat{P}^{}_{l_2}) \hat{P}^{}_n\,.
    \label{eqn:secondDerivHDecomp}
\end{align}
Using $l_1\neq l_2$ due to $\epsilon_{l_1l_2}$, we obtain
\begin{align}
   \sum_{l_1,l_2}\epsilon_{l_1 l_2} \hat{P}_m (\partial^{ b }\hat e^{ c }_{l_1 l_2})\hat{P}_n &= \epsilon_{mn} \hat{P}_m (\partial^{ b }\hat e^{ c }_{mn})\hat{P}_n+\epsilon_{nm} \hat{P}_m (\partial^{ b }\hat e^{ c }_{nm})\hat{P}_n\nonumber \\ &\,\,\,\,+\sum_{l\neq n,m}\epsilon_{lm} \hat{P}_m (\partial^{ b }\hat e^{ c }_{lm})\hat{P}_n + \sum_{l\neq n,m}\epsilon_{ln} \hat{P}_m (\partial^{ b }\hat e^{ c }_{ln})\hat{P}_n \nonumber \\ &\,\,\,\,+\sum_{l\neq n,m}\epsilon_{ml} \hat{P}_m (\partial^{ b }\hat e^{ c }_{ml})\hat{P}_n +\sum_{l\neq n,m}\epsilon_{nl} \hat{P}_m (\partial^{ b }\hat e^{ c }_{nl})\hat{P}_n \\[2mm]
   &= \epsilon_{mn} \hat{P}_m (\partial^{ b }\hat e^{ c }_{mn})\hat{P}_n + \sum_{l\neq n,m}\epsilon_{ml} \hat{P}_m (\partial^{ b }\hat e^{ c }_{ml})\hat{P}_n +\sum_{l\neq n,m}\epsilon_{ln} \hat{P}_m (\partial^{ b }\hat e^{ c }_{ln})\hat{P}_n  \, ,
\end{align}
where we first considered the cases with either $l_1$ or $l_2$ are equal to $n$ or $m$ and took all constraints of Eq.~\eqref{eqn:secondDerivHDecomp} into account in the second step. Now using Eq.~\eqref{eqn:secondDerivHDecomp} and $l\neq m,n$, the contributions in the sums can be simplified into
\begin{align}
    \hat{P}_m (\partial^{ b }\hat e^{ c }_{ml})\hat{P}_n = i  \hat{P}^{}_m (\partial^{ c }\hat{P}^{}_{l}) (\partial^{ b }\hat{P}^{}_{l}) \hat{P}^{}_n = i  \hat{P}^{}_m (\partial^{ c }\hat{P}^{}_{l}) \hat{P}_l (\partial^{ b }\hat{P}^{}_{l}) \hat{P}^{}_n = -i  \hat{P}^{}_m (\partial^{ c }\hat{P}^{}_{l}) \hat{P}_l (\partial^{ b }\hat{P}^{}_{n}) \hat{P}^{}_n = i \hat e^c_{ml}\hat e^b_{ln}\,,
\end{align}
and
\begin{align}
    \hat{P}_m (\partial^{ b }\hat e^{ c }_{ln})\hat{P}_n = i  \hat{P}^{}_m (\partial^{ b }\hat{P}^{}_{l}) (\partial^{ c }\hat{P}^{}_{n}) \hat{P}^{}_n = i  \hat{P}^{}_m (\partial^{ b }\hat{P}^{}_{l}) \hat{P}_l (\partial^{ c }\hat{P}^{}_{n}) \hat{P}^{}_n = -i \hat e^b_{ml}\hat e^c_{ln}\,.
\end{align}
Introducing the covariant derivative of the transition dipole moment defined by $\nabla^a \hat{e}_{mn}^b \equiv \hat{P}_m (\partial^a \hat{e}_{mn}^b) \hat{P}_n$ and combining the result with the other contributions in Eq.~\eqref{eqn:secondDerivHPre}, we conclude
\begin{align} \label{eq:suppPddHP}
    \hat{P}_m \big(\partial^b\partial^c\hat H\big)\hat{P}_n &= \delta_{mn} E_{n}^{ b  c }\hat{P}_{n} + i\epsilon_{m n}^{ c } \hat e^{ b }_{m n} + i\epsilon_{m n}^{ b } \hat e^{ c }_{m n} +i\epsilon_{m n} \nabla^{ b }\hat e^{ c }_{m n} - \sum_{l\neq m,n}\epsilon_{m l} \hat e^c_{ml}\hat e^b_{ln} + \sum_{l\neq m,n}\epsilon_{l n} \hat e^b_{ml}\hat e^c_{ln}\,,
\end{align}
For completeness, we also provide an alternate form, symmetrized in the indices $b$ and $c$. First note that 
\begin{align}
    &-\sum_{l\neq n,m} \epsilon_{ml}\hat e^c_{ml}\hat e^b_{ln} + \sum_{l\neq n,m} \epsilon_{ln}\hat e^b_{ml}\hat e^c_{ln} = \hat P_m(\partial^c\hat P_m)\big(\hat H-E_m\big)(\partial^b\hat P_n)\hat P_n + \hat P_m(\partial^b\hat P_m)\big(\hat H-E_n\big)(\partial^c\hat P_n)\hat P_n\,,
\end{align}
using the projector decomposition of the Bloch Hamiltonian. Also symmetrizing $i\epsilon_{mn}\nabla^b \hat{e}_{mn}^c$, one obtains
\begin{align}
    \hat{P}_m \big(\partial^b\partial^c\hat H\big)\hat{P}_n &= \delta_{mn} E_{n}^{ b  c }\hat{P}_{n} + i\epsilon_{m n}^{ c } \hat e^{ b }_{m n} + i\epsilon_{m n}^{ b } \hat e^{ c }_{m n} \nonumber \\ &+\frac{i}{2}\epsilon_{m n} \nabla^{ b }\hat e^{ c }_{m n}+\frac{i}{2}\epsilon_{m n} \nabla^{ c }\hat e^{ b }_{m n} \nonumber \\ &\hat P_m(\partial^c\hat P_m)\big(\hat H-\frac{1}{2}\big[E_n+E_m\big]\big)(\partial^b\hat P_n)\hat P_n + \hat P_m(\partial^b\hat P_m)\big(\hat H-\frac{1}{2}\big[E_n+E_m\big]\big)(\partial^c\hat P_n)\hat P_n\,,
\end{align}
generalizing a gauge-invariant separation for single-degenerate bands \cite{Mitscherling2021}.

To proceed with the calculation at hand, we now combine Eq.~\eqref{eq:suppPddHP} with Eq.~\eqref{eq:SuppDelHamEFormula}, to calculate the other trace appearing in the conductivity Eq.~\eqref{eq:SuppconductivityT0} :
\begin{align}
    \text{tr}[\hat{P}_m(\partial^{ a }\hat{H})\hat{P}_n (\partial^{ b }\partial^{ c }\hat{H})]
=\,&  \delta_{mn} E^{ a }_n E_{n}^{ b  c }\text{tr}[\hat{P}_n]\nonumber \\\nonumber
&- \delta_{mn} E^{ a }_n\sum_{l\neq m,n}\epsilon_{n l} \text{tr}[\hat{P}_n\hat e^c_{nl}\hat e^b_{ln}]+ \delta_{mn} E^{ a }_n\sum_{l\neq m,n}\epsilon_{l n} \text{tr}[\hat{P}_n\hat e^b_{nl}\hat e^c_{ln}] \\\nonumber
&- \epsilon_{mn}\epsilon_{nm}^{ c }\text{tr}[\hat e^{ a }_{mn} \hat e^{ b }_{nm}]- \epsilon_{mn}\epsilon_{nm}^{ b }\text{tr}[\hat e^{ a }_{mn} \hat e^{ c }_{nm}]\\\nonumber
&-\epsilon_{mn}\epsilon_{nm}\text{tr}[\hat e^{ a }_{mn} \nabla^{ b }\hat e^{ c }_{nm}] \\
&- i \epsilon_{mn}\sum_{l\neq m,n}\epsilon_{n l}\text{tr}[\hat e^{ a }_{mn}\hat e^c_{nl}\hat e^b_{lm}]+i\epsilon_{mn}\sum_{l\neq m,n}\epsilon_{lm}\text{tr}[\hat e^{ a }_{mn}\hat e^b_{nl}\hat e^c_{lm}]\,.
\end{align}
Now $\hat{P}_n \hat e^b_{nl}=\hat e^b_{nl}$, and we also identify the quantum geometric tensor Eq.~\eqref{eq:QGdefQGT}, expressed as $Q^{ab}_{nm} = -\operatorname{tr}[\hat e_{mn}^{a}\hat e_{nm}^{b}]$:
\begin{align}
    \text{tr}[\hat{P}_m(\partial^{ a }\hat{H})\hat{P}_n (\partial^{ b }\partial^{ c }\hat{H})]
=\,&  \delta_{mn} E^{ a }_n E_{n}^{ b  c }\text{tr}[\hat{P}_n]\nonumber \\\nonumber
&+ \delta_{mn} E^{ a }_n\sum_{l\neq m,n}\epsilon_{n l} Q_{ln}^{cb}- \delta_{mn} E^{ a }_n\sum_{l\neq m,n}\epsilon_{l n} Q_{ln}^{bc} \\\nonumber
&+ \epsilon_{mn}\epsilon_{nm}^{ c }Q_{nm}^{ab}+ \epsilon_{mn}\epsilon_{nm}^{ b }Q_{nm}^{ac}\\\nonumber
&-\epsilon_{mn}\epsilon_{nm}\text{tr}[\hat e^{ a }_{mn} \nabla^{ b }\hat e^{ c }_{nm}] \\
&- i \epsilon_{mn}\sum_{l\neq m,n}\epsilon_{n l}\text{tr}[\hat e^{ a }_{mn}\hat e^c_{nl}\hat e^b_{lm}]+i\epsilon_{mn}\sum_{l\neq m,n}\epsilon_{lm}\text{tr}[\hat e^{ a }_{mn}\hat e^b_{nl}\hat e^c_{lm}]\,.
\end{align}
The quantum geometric connection Eq.~\eqref{eq:QGdefConnection} is also given by ${C^{a;bc}_{mn} = - \text{tr}[\hat{e}_{nm}^b \nabla^a \hat{e}_{mn}^c]}$ for $n\neq m$. We also reorder indices in the $\epsilon$:
\begin{align}
    \text{tr}[\hat{P}_m(\partial^{ a }\hat{H})\hat{P}_n (\partial^{ b }\partial^{ c }\hat{H})]
=\, &  \delta_{mn} E^{ a }_n E_{n}^{ b  c }\text{tr}[\hat{P}_n]\nonumber \\\nonumber
& - \delta_{mn} E^{ a }_n\sum_{l\neq m,n}\epsilon_{l n} \left(Q_{ln}^{bc}+Q_{ln}^{cb}\right) \\\nonumber
&- \epsilon_{nm}\epsilon_{nm}^{ c }Q_{nm}^{ab}- \epsilon_{nm}\epsilon_{nm}^{ b }Q_{nm}^{ac}\\\nonumber
&-\epsilon_{nm}^2C_{nm}^{ b ; a  c } \\
&+ i \epsilon_{nm}\sum_{l\neq m,n}\epsilon_{n l}\text{tr}[\hat e^{ a }_{mn}\hat e^c_{nl}\hat e^b_{lm}]-i\epsilon_{nm}\sum_{l\neq m,n}\epsilon_{lm}\text{tr}[\hat e^{ a }_{mn}\hat e^b_{nl}\hat e^c_{lm}]\,.
\end{align}
Finally, we symmetrize in $b,\ c$, since the initial trace has this symmetry:
\begin{align} \label{eq:doubleDerHamiltonianTraceGeometric}
\text{tr}[\hat{P}_m(\partial^{ a }\hat{H})\hat{P}_n (\partial^{ b }\partial^{ c }\hat{H})]=\,&  \delta_{mn} E^{ a }_n E_{n}^{ b  c }\text{tr}[\hat{P}_n]\nonumber\\\nonumber
&-2\delta_{mn} E^{ a }_n\sum_{l\neq n}\epsilon_{ln} Q_{(ln)}^{ b  c }\\\nonumber
&- \epsilon_{nm}\epsilon_{nm}^{ c }Q_{nm}^{ a  b }- \epsilon_{nm}\epsilon_{nm}^{ b }Q_{nm}^{ a  c }\\\nonumber
&-\frac{1}{2}\epsilon_{nm}^2C_{nm}^{ b ; a  c }-\frac{1}{2}\epsilon_{nm}^2C_{nm}^{ c ; a  b } \\
&+  \frac{i}{2}\epsilon_{nm}\sum_{l\neq m,n}(\epsilon_{n l}-\epsilon_{lm})(\text{tr}[\hat e^{ a }_{mn}\hat e^{ c }_{nl}\hat e^{ b }_{lm}]+\text{tr}[\hat e^{ a }_{mn}\hat e^{ b }_{nl}\hat e^{ c }_{lm}])\,.
\end{align}
Here, $(a_1\ldots a_n)$ indicates symmetrization of a tensor:
\begin{equation}
   T_{\left(i_1 i_2 \cdots i_n\right)}=\frac{1}{n!} \sum_{\sigma \in \mathfrak{S}_n} T_{i_{\sigma 1} i_{\sigma 2} \cdots i_{\sigma n}}\, ,
\end{equation}
 where the summation extends over the symmetric group on n symbols, and a conventional prefactor of $1/(n!)$ was included. We define antisymmetrization, denoted by $T_{[a_1\ldots a_n]}$, similarly:
 \begin{equation}
   T_{\left[i_1 i_2 \cdots i_n\right]}=\frac{1}{n!} \sum_{\sigma \in \mathfrak{S}_n} \mathrm{sgn}(\sigma) T_{i_{\sigma 1} i_{\sigma 2} \cdots i_{\sigma n}}\, ,
\end{equation}
where $\mathrm{sgn}(\sigma)$ denotes the signature of the permutation $\sigma$.
Note also that $Q_{(mn)}^{ b  c }= Q_{(mn)}^{ (b  c) } = Q_{mn}^{(bc)}$.

\subsection{Collecting contributions to quantum geometric quantities} 
\label{suppsec:GeomContrib}

Having expanded the traces containing the Hamiltonian into quantum geometric quantities in the previous section, we now return to Eq.~\eqref{eq:SuppconductivityT0} and collect the coefficient in front of each independent geometric quantity (those that appeared in Eqs.~\eqref{eq:tripleHTraceGeometricDecomp}, \eqref{eq:doubleDerHamiltonianTraceGeometric}) in the conductivity. Recall that the functions $w^{nm}$, $u^{lmn}$ are defined in Eqs.~\eqref{eq:suppwDef}, \eqref{eq:suppuDef}.
To streamline the calculation and clarify notation, we absorb the prefactor and momentum integral \( \frac{e^3}{\hbar} \int \frac{d^d\bk}{(2\pi)^d} \) into a compact notation \( \int_{\bk} \), and decompose the conductivity into five distinct contributions:
\begin{align}
    \sigma^{a;bc} &= \underbrace{\int_{\bk} (\ldots)\, \text{tr}[\hat{P}_n]}_{\mathlarger{A}} + \underbrace{\int_{\bk} (\ldots)\,\left(Q_{nm}^{ a  b }+Q_{nm}^{ a c }\right)}_{\mathlarger{B}}+ \underbrace{\int_{\bk} (\ldots)\,Q_{nm}^{ b  c }}_{\mathlarger{C}} \\
    &\qquad + \underbrace{\int_{\bk} (\ldots)\,\left(C_{nm}^{ b ; a c }+C_{nm}^{c ; a  b }\right)}_{\mathlarger{D}} + \underbrace{\int_{\bk} (\ldots)\left(\text{tr}[e^{ a }_{mn}e^{ b }_{nl}e^{c }_{lm}]+\text{tr}[e^{ a }_{mn}e^{c }_{nl}e^{ b }_{lm}]\right)}_{\mathlarger{E}}\,,
\end{align}
where each term represents an integral over a band-structure-dependent prefactor (indicated by $(\ldots)$), multiplied by a distinct quantum geometric invariant. We used the $b\leftrightarrow c$ symmetry to collect the quantum geometric tensors $Q_{nm}^{ a  b },\, Q_{nm}^{ a c }$, the connections $C_{nm}^{ b ; a c }$,\, $C_{nm}^{c ; a  b }$, and the three-band traces $\text{tr}[e^{ a }_{mn}e^{ b }_{nl}e^{c }_{lm}]$,\, $\text{tr}[e^{ a }_{mn}e^{c }_{nl}e^{ b }_{lm}]$ into one term each. We collect the quantities $A$, $B$, $C$, $D$, and $E$ in what follows. 
We continue treating the general case $b\neq c$, which will reduce to the result shown in the main text upon setting $c=b$.

\subsubsection{Contribution A: trivial geometry \texorpdfstring{$\text{tr}[\hat{P}_n]$}{}}

All three traces contribute a diagonal (in the band indices) term:
\begin{align}
A = \int_{\bk}\left(\sum_n \frac{u^{nnn}}{2}E^{ a }_n E_{n}^{ b }E_n^{c }\text{tr}[\hat{P}_n]+\sum_n \frac{u^{nnn}}{2}E^{ a }_n E_{n}^{c }E_n^{ b }\text{tr}[\hat{P}_n]+
\sum_n w^{nn}E^{ a }_n E_{n}^{ b c }\text{tr}[\hat{P}_n]\right)\,.
\end{align}
By integrating by parts in momentum, we may combine these contributions. The main technique is to use momentum derivatives of the energy as internal derivatives (since $u^{nnn},w^{nn}$ are functions of $E_n$), i.e.,
 \begin{align}
     \partial^a f(E_n) = f'(E_n) E_n^a\,.
 \end{align}
 Denoting indefinite integrals in $E_n$ by $\int\ldots dE_n$, the above can be manipulated into
\begin{align}
A &= \int_{\bk}\sum_n \left[u^{nnn}E^{ a }_n E_{n}^{ b }E_n^{c }+
w^{nn}E^{ a }_n E_{n}^{ b c }\right]\text{tr}[\hat{P}_n]\\ &= \int_{\bk}\sum_n \left[-\left(\int u^{nnn}dE_n\right)  \partial^a\left(E_{n}^{ b }E_n^{c }\right)
- \left(\int w^{nn} dE_n\right) E_{n}^{a b c }\right]\text{tr}[\hat{P}_n] \\
&=\int_{\bk}\sum_n \left[-\left(\int u^{nnn}dE_n\right) \left(E_n^{c }E_{n}^{ ab }+E_{n}^{ b }E_n^{ac }\right)
- \left(\int w^{nn} dE_n\right) E_{n}^{a b c }\right]\text{tr}[\hat{P}_n]\\\label{eq:trivialGeometryCoeffBeforeExpansion}
&=\int_{\bk}\sum_n \left[\left(\iint u^{nnn}dE_n dE_n\right) 
- \left(\int w^{nn} dE_n\right) \right]E_{n}^{a b c }\,\text{tr}[\hat{P}_n]\,.
\end{align}

\subsubsection{Contribution B: quantum geometric tensors  \texorpdfstring{$Q_{nm}^{ a  b }$ and $Q_{nm}^{ a c }$}{}}
Again all traces contribute a term proportional to $Q_{nm}^{ a  b }$ (or equivalently to $Q_{nm}^{ a c }$):
\begin{align}
     B= \int_{\bk}\Biggl(-\sum_{n \neq m}w^{nm}\epsilon_{nm}\epsilon_{nm}^{c }Q_{nm}^{ a  b } -\sum_{n \neq m}\left(\frac{u^{m n m}}{2}E_m^{c }+ \frac{u^{m n n}}{2} E_n^{c }\right)\epsilon_{nm}^2 Q_{nm}^{ a  b }\Biggr)+(b\leftrightarrow c)\,.
\end{align}
It is convenient to separate $Q_{nm}^{ a  b }$ into its symmetric and antisymmetric part in $n,m$, which also correspond to the real and imaginary part respectively: 
\begin{align}\nonumber
 B= \int_{\bk}\Biggl(&-\sum_{n \neq m}  w^{nm}\epsilon_{nm}\epsilon_{nm}^{c }Q_{(nm)}^{ a  b } -\sum_{n \neq m}\left(\frac{u^{m n m}}{2}E_m^{c }+ \frac{u^{m n n}}{2} E_n^{c }\right)\epsilon_{nm}^2 Q_{(nm)}^{ a  b } \\
    & \quad -\sum_{n \neq m}  w^{nm}\epsilon_{nm}\epsilon_{nm}^{c }Q_{[nm]}^{ a  b } -\sum_{n \neq m}\left(\frac{u^{m n m}}{2}E_m^{c }+ \frac{u^{m n n}}{2} E_n^{c }\right)\epsilon_{nm}^2 Q_{[nm]}^{ a  b }\Biggr)+(b\leftrightarrow c)\,.
\end{align}
Expanding $\epsilon_{nm}^c = E_n^c-E_m^c$:
\begin{align}\nonumber
    B= \int_{\bk}\Biggl(&-\sum_{n \neq m}  w^{nm}\epsilon_{nm}E_n^{c }Q_{(nm)}^{ a  b } -\sum_{n \neq m}\left(\frac{u^{n m n}}{2}E_n^{c }+ \frac{u^{m n n}}{2} E_n^{c }\right)\epsilon_{nm}^2 Q_{(nm)}^{ a  b } \\\nonumber
    &\quad+\sum_{n \neq m} w^{nm}\epsilon_{nm}E_m^{c }Q_{(nm)}^{ a  b } -\sum_{n \neq m}\left(-\frac{u^{n m n}}{2}E_n^{c }+ \frac{u^{m n n}}{2} E_n^{c }\right)\epsilon_{nm}^2 Q_{[nm]}^{ a  b } \\
    &\quad+\sum_{n \neq m} - w^{nm}\epsilon_{nm}E_n^{c }Q_{[nm]}^{ a  b } +\sum_{n \neq m}w^{nm}\epsilon_{nm}E_m^{c }Q_{[nm]}^{ a  b }\Biggr)+(b\leftrightarrow c)\,.
\end{align}
 We continue by relabeling sums so that all $E^c$ are in band $n$:
\begin{align}\nonumber
    B= \int_{\bk}\Biggl(-\sum_{n \neq m}&  w^{nm}\epsilon_{nm}E_n^{c }Q_{(nm)}^{ a  b } -\sum_{n \neq m}\left(\frac{u^{n m n}}{2}E_n^{c }+ \frac{u^{m n n}}{2} E_n^{c }\right)\epsilon_{nm}^2 Q_{(nm)}^{ a  b } \\\nonumber
    &\quad-\sum_{n \neq m} w^{mn}\epsilon_{nm}E_n^{c }Q_{(nm)}^{ a  b } -\sum_{n \neq m}\left(-\frac{u^{n m n}}{2}E_n^{c }+ \frac{u^{m n n}}{2} E_n^{c }\right)\epsilon_{nm}^2 Q_{[nm]}^{ a  b } \\
    &\quad-\sum_{n \neq m} w^{nm}\epsilon_{nm}E_n^{c }Q_{[nm]}^{ a  b } +\sum_{n \neq m}w^{mn}\epsilon_{nm}E_n^{c }Q_{[nm]}^{ a  b }\Biggr)+(b\leftrightarrow c)\,.
\end{align}
Collecting identical contributions finally gives
\begin{align}\label{eq:QabCoeffBeforeExpansion}
    B= \int_{\bk}\Biggl(&-\sum_{n \neq m}\left(2w^{(nm)}+\frac{u^{n m n}}{2}\epsilon_{nm}+ \frac{u^{m n n}}{2} \epsilon_{nm}\right)\epsilon_{nm} E_n^{c }Q_{(nm)}^{ a  b }\\
    &\quad-\sum_{n \neq m}\left(2w^{[nm]}-\frac{u^{n m n}}{2}\epsilon_{nm}+ \frac{u^{m n n}}{2} \epsilon_{nm}\right)\epsilon_{nm} E_n^{c }Q_{[nm]}^{ a  b }\Biggr)+(b\leftrightarrow c) \,.
\end{align}

\subsubsection{Contribution C: quantum geometric tensor \texorpdfstring{$Q_{nm}^{ b c }$}{}}

The last quantum geometric tensor contribution is $Q_{nm}^{ b c }$. Collecting all coefficients yields
\begin{align}
C= \int_{\bk}\Biggl(\sum_{n \neq m} &2w^{nn}\epsilon_{nm}E_n^{ a }Q_{(nm)}^{ b c } -\sum_{n \neq m}\left(\frac{u^{m m n}}{2}E_m^{ a }+ \frac{u^{n n m}}{2} E_n^{ a }\right)\epsilon_{nm}^2 Q_{nm}^{ b c }\Biggr)\,.
\end{align}
As previously, we separate into symmetric and antisymmetric components:
\begin{align}
   C= \int_{\bk}\Biggl(& \sum_{n \neq m} 2w^{nn}\epsilon_{nm}E_n^{ a }Q_{(nm)}^{ b c } -\sum_{n \neq m}\left(\frac{u^{n n m}}{2}E_n^{ a }+ \frac{u^{n n m}}{2} E_n^{ a }\right)\epsilon_{nm}^2 Q_{(nm)}^{ b c } \nonumber \\
   &\quad-\sum_{n \neq m}\left(-\frac{u^{n n m}}{2}E_n^{ a }+ \frac{u^{n n m}}{2} E_n^{ a }\right)\epsilon_{nm}^2 Q_{[nm]}^{ b c }\Biggr)\,.
\end{align}
The third sum is evidently 0, and the result simplifies to
\begin{align}\label{eq:QbcCoeffBeforeExpansion}
    C&= \int_{\bk}\Biggl(\sum_{n \neq m} 2w^{nn}\epsilon_{nm}E_n^{ a }Q_{(nm)}^{ b c } -\sum_{n \neq m}u^{n n m} E_n^{ a }\epsilon_{nm}^2 Q_{(nm)}^{ b c }\Biggr) \\
    &= \int_{\bk}\Biggl(\sum_{n \neq m}\bigl(2w^{nn} - u^{n n m}\epsilon_{nm}\bigr)\epsilon_{nm} E_n^{ a }Q_{(nm)}^{ b c }\Biggr)\,.
\end{align}

\subsubsection{Contribution D: connections \texorpdfstring{$C_{nm}^{ b ; a c }$, $C_{nm}^{c ; a  b }$}{}}
The total contribution from connections is 
\begin{align} \label{eq:ConnectionCoeffBeforeExpansion}
D= \int_{\bk}\Biggl(-\sum_{m,n} \frac{1}{2}w^{nm}\epsilon_{nm}^2C_{nm}^{ b ; a c } - \sum_{m,n} \frac{1}{2}w^{nm}\epsilon_{nm}^2C_{nm}^{c ; a  b }\Biggr)\,.
\end{align}

\subsubsection{Contribution E: three-band traces \texorpdfstring{$\text{tr}[e^{ a }_{mn}e^{ b }_{nl}e^{c }_{lm}]$, $\text{tr}[e^{ a }_{mn}e^{c }_{nl}e^{ b }_{lm}]$}{}}

The contribution to the three-band traces $\text{tr}[e^{ a }_{mn}e^{ b }_{nl}e^{c }_{lm}]$, $\text{tr}[e^{ a }_{mn}e^{ c }_{nl}e^{b }_{lm}]$ is given by
\begin{align}
 E= \int_{\bk}\Biggl(\sum_{m,n} \frac{i}{2}&w^{nm}\epsilon_{nm}\sum_{l\neq m,n}(\epsilon_{n l}-\epsilon_{lm})\text{tr}[e^{ a }_{mn}e^{ b }_{nl}e^{c }_{lm}]-i\sum_{l,m,n}\frac{u^{m n l}}{2}\epsilon_{mn}\epsilon_{nl}\epsilon_{lm}\text{tr}[\hat{e}_{mn}^{ a }\hat{e}_{nl}^{ b }\hat{e}_{lm}^{c }]\Biggr)+(b\leftrightarrow c)\,, 
\end{align}
which we combine into a single sum:
\begin{align}
   E= \int_{\bk}\Biggl( \frac{i}{2}\sum_{(l, m,n)}\biggl(w^{nm}\epsilon_{nm}(\epsilon_{n l}-\epsilon_{lm})-u^{m n l}\epsilon_{mn}\epsilon_{nl}\epsilon_{lm}\biggr)\bigl(\text{tr}[e^{ a }_{mn}e^{ b }_{nl}e^{c }_{lm}]+\text{tr}[e^{ a }_{mn}e^{c }_{nl}e^{ b }_{lm}]\bigr)\,.
\end{align}
We indicate summing only over different band indices with the notation $(l,m,n)$ (i.{}e.{} $l\neq m$, $l\neq n$, $m\neq n$). 
Now
\begin{align}
    \text{tr}[e^{ a }_{mn}e^{ b }_{nl}e^{c }_{lm}]^* = \text{tr}[e^{c }_{ml}e^{ b }_{ln}e^{ a }_{nm}]= \text{tr}[e^{ a }_{nm} e^{c }_{ml}e^{ b }_{ln}]\,,
\end{align}
effectively permuting $b,c$ and $n,m$ at the same time. The contribution may then be written as
\begin{align}
   E= \int_{\bk}\Biggl( \frac{i}{2}\sum_{(l, m,n)}\biggl(w^{nm}\epsilon_{nm}(\epsilon_{n l}-\epsilon_{lm})-u^{m n l}\epsilon_{mn}\epsilon_{nl}\epsilon_{lm}\biggr)\bigl(\text{tr}[e^{ a }_{mn}e^{ b }_{nl}e^{c }_{lm}]+\text{tr}[e^{ a }_{nm}e^{b }_{ml}e^{ c }_{ln}]^*\bigr)\Biggr)\,.
\end{align}
We now separate the functions $w,u$ into symmetric and antisymmetric parts in $n,m$:
\begin{align}\nonumber
    E= \int_{\bk}\Biggl(&\frac{i}{2}\sum_{(l, m,n)}\biggl(w^{(nm)}\epsilon_{nm}(\epsilon_{n l}-\epsilon_{lm})-u^{(m n) l}\epsilon_{mn}\epsilon_{nl}\epsilon_{lm}\biggr)\bigl(\text{tr}[e^{ a }_{mn}e^{ b }_{nl}e^{c }_{lm}]+\text{tr}[e^{ a }_{nm}e^{b }_{ml}e^{ c }_{ln}]^*\bigr)\\
    &\quad+\frac{i}{2}\sum_{(l, m,n)}\biggl(w^{[nm]}\epsilon_{nm}(\epsilon_{n l}-\epsilon_{lm})-u^{[m n] l}\epsilon_{mn}\epsilon_{nl}\epsilon_{lm}\biggr)\bigl(\text{tr}[e^{ a }_{mn}e^{ b }_{nl}e^{c }_{lm}]+\text{tr}[e^{ a }_{nm}e^{b }_{ml}e^{ c }_{ln}]^*\bigr)\Biggr)\,.
\end{align}
Taking the antisymmetry of the $\epsilon$ combinations multiplying $w,u$ into consideration, we reindex sums to permute the indices $n,m$ of $\text{tr}[e^{ a }_{nm}e^{b }_{ml}e^{ c }_{ln}]^*$:
\begin{align}\nonumber
    E= \int_{\bk}\Biggl(&\frac{i}{2}\sum_{(l, m,n)}\biggl(w^{(nm)}\epsilon_{nm}(\epsilon_{n l}-\epsilon_{lm})-u^{(m n) l}\epsilon_{mn}\epsilon_{nl}\epsilon_{lm}\biggr)\bigl(\text{tr}[e^{ a }_{mn}e^{ b }_{nl}e^{c }_{lm}]-\text{tr}[e^{ a }_{mn}e^{b }_{nl}e^{ c }_{lm}]^*\bigr)\\
    &\quad+\frac{i}{2}\sum_{(l, m,n)}\biggl(w^{[nm]}\epsilon_{nm}(\epsilon_{n l}-\epsilon_{lm})-u^{[m n] l}\epsilon_{mn}\epsilon_{nl}\epsilon_{lm}\biggr)\bigl(\text{tr}[e^{ a }_{mn}e^{ b }_{nl}e^{c }_{lm}]+\text{tr}[e^{ a }_{mn}e^{b }_{nl}e^{ c }_{lm}]^*\bigr)\Biggr)\,.
\end{align}
Identifying the real and imaginary parts of $\text{tr}[e^{ a }_{mn}e^{ b }_{nl}e^{c }_{lm}]$, we finally obtain
\begin{align}\nonumber
E= \int_{\bk}\Biggl(&-\sum_{(l, m,n)}\biggl(w^{(nm)}\epsilon_{nm}(\epsilon_{n l}-\epsilon_{lm})-u^{(m n) l}\epsilon_{mn}\epsilon_{nl}\epsilon_{lm}\biggr)\text{Im}\,\text{tr}[e^{ a }_{mn}e^{ b }_{nl}e^{c }_{lm}]  \\\label{eq:TripleTraceCoeffBeforeExpansion}
    &\quad+i\sum_{(l, m,n)}\biggl(w^{[nm]}\epsilon_{nm}(\epsilon_{n l}-\epsilon_{lm})-u^{[m n] l}\epsilon_{mn}\epsilon_{nl}\epsilon_{lm}\biggr)\,\text{Re}\,\text{tr}[e^{ a }_{mn}e^{ b }_{nl}e^{c }_{lm}]\Biggr)\,.
\end{align}

\subsection{How to perform the \texorpdfstring{$\Gamma$}{Gamma} expansion}\label{suppsec:cleanlimit}

Now that we have collected the coefficient in front of each geometric contribution, we describe the way the $\Gamma$ expansion mentioned in the main text is done. Indeed, the functions in Eqs.~\eqref{eq:trivialGeometryCoeffBeforeExpansion}, \eqref{eq:QabCoeffBeforeExpansion}, \eqref{eq:QbcCoeffBeforeExpansion}, \eqref{eq:ConnectionCoeffBeforeExpansion}, \eqref{eq:TripleTraceCoeffBeforeExpansion} are in general complicated functions of the band energies and $\Gamma$. As an example, the coefficient of trivial geometry is given by
\begin{align} \label{eq:suppGammaExpansionTrivialExample0}
    \frac{3 E_n^3 \Gamma+5 E_n \Gamma^3+3 E_n^4 \arctan\left(\frac{E_n}{\Gamma}\right)+6 E_n^2 \Gamma^2 \arctan\left(\frac{E_n}{\Gamma}\right)+3 \Gamma^4 \arctan\left(\frac{E_n}{\Gamma}\right)}{12 \pi  \Gamma^2 \left(E_n^2+\Gamma^2\right)^2}\,.
\end{align}
Other coefficients are in general more complicated, and depend on the energies of multiple bands. However, it is physically reasonable to assume that $\Gamma$ is smaller than the bandwidth, and we would like to make use of this fact to ``expand'' these functions in $\Gamma$. Schematically, the previous calculations reduce the conductivity to a sum of terms of the type
\begin{align} \label{eq:suppsecGammaExpansionSchematic1}
    \propto \int \frac{d^d\bk}{(2\pi)^d} f(\Gamma, E_n,E_m,\ldots) (\textrm{quantum geometric quantity})\,,
\end{align}
where $f$ is one of the coefficients found in the previous subsection, and we explicitly highlighted its $\Gamma$ dependence. We want to Taylor expand this expression (the result after integration) in $\Gamma$ around zero by replacing $f$ by a suitable simpler function, to collect the (perturbatively) most significant contributions. As we show now it is, however, incorrect to simply Taylor expand $f$ in $\Gamma$. 

To illustrate the problem we begin with the simple example of a Lorentzian:
\begin{align}
	f(x, \Gamma)=\frac{\Gamma}{\pi(\Gamma^2 + x^2)}\,.
\end{align}
A Taylor expansion $\forall x \neq 0$ yields that the first contribution is of order $\mathcal{O}(\Gamma^1)$. However, for $x=0$ the function reduces to $1/(\pi\Gamma)$, which has a pole in $\Gamma$ around zero (in other words it is of order $\mathcal{O}(\Gamma^{-1})$). The resolution of this conundrum resides in taking the limit in the \textit{distributional} sense. Indeed, it is a well-known fact that 
\begin{align}
	\frac{\Gamma}{\pi(\Gamma^2 + x^2)} \xrightarrow{\Gamma \rightarrow 0} \delta(x)
\end{align}
in the space of distributions. Since we are calculating the conductivity by integrating over momentum space (effectively also integrating over energies $\xi_n$), this distributional expansion is the appropriate one to expand the final result in orders of $\Gamma$.

The coefficients Eqs.~\eqref{eq:trivialGeometryCoeffBeforeExpansion}, \eqref{eq:QabCoeffBeforeExpansion}, \eqref{eq:QbcCoeffBeforeExpansion}, \eqref{eq:ConnectionCoeffBeforeExpansion}, \eqref{eq:TripleTraceCoeffBeforeExpansion} of the geometric quantities are more complicated than Lorentzians, and a systematic approach is required. 

Conveniently, the distributional limit above is a continuous limit of functions in Fourier space. Indeed, the Fourier transform of the Lorentzian above (in $x$) is given by
\begin{align}
    f(x,\Gamma) \xrightarrow[]{\mathcal{F}} \widehat{f}(\xi,\Gamma) = \frac{1}{\sqrt{2\pi}}\left(e^{\Gamma \xi} \theta(-\xi) +e^{-\Gamma \xi} \theta(\xi) \right)\, .
\end{align}
Now the limit $\Gamma\rightarrow 0$ (or Taylor expansion in $\Gamma$) is defined everywhere, and performing the inverse Fourier transform yields 
\begin{align}
    \widehat{f}(\xi,\Gamma) \xrightarrow[]{\Gamma \rightarrow 0} \frac{1}{\sqrt{2\pi}} \xleftrightarrow[]{\mathcal{F}^{-1}} \delta(x)\,,
\end{align}
that is, if we perform the inverse Fourier transform of the limit we now recover a Dirac delta function. This is the method we use for the functions appearing in 
Eq.~\eqref{suppsec:GeomContrib}: We 
\begin{enumerate}
    \item Fourier transform in all relevant energies appearing (there can be multiple if multiple bands are involved)
    \item Taylor expand in $\Gamma$ up to order $\mathcal{O}\left(\Gamma^0\right)$ (check if this is defined everywhere in Fourier space)
    \item perform the inverse Fourier transform back.
\end{enumerate}
For this paper we choose to expand in $\Gamma$ up to order $\mathcal{O}\left(\Gamma^0\right)$, but this can be chosen entirely differently in another context. In particular, there is no constraint to collecting terms of order $\mathcal{O}\left(\Gamma^1\right)$ or higher. Clearly we are able to recover a delta distribution from a Lorentzian, but, in general, what types of functions (or distributions) can we recover using this method? First, we require our input functions $f(\Gamma, E_n,E_m,\ldots)$ to have a Fourier transform. Note that many functions that don't have a well-defined Fourier transform (like Heaviside's step function) do have a well-defined \emph{distributional} Fourier transform. In our case, a sufficient definition of being well-defined is that the inverse Fourier transform of the Fourier transform recovers the original function. The second condition is that the Taylor expansion in $\Gamma$ must be defined everywhere in Fourier space.  Effectively, this procedure transforms a complicated function $f(\Gamma, E_n,E_m,\ldots)$ into a simpler function $f_s(\Gamma, E_n,E_m,\ldots)$ such that the schematic expression Eq.~\eqref{eq:suppsecGammaExpansionSchematic1} is expanded:
\begin{align}
    \int \frac{d^d\bk}{(2\pi)^d} f(\Gamma, E_n,E_m,\ldots) (\textrm{QG quantity}) = \int \frac{d^d\bk}{(2\pi)^d} f_s(\Gamma, E_n,E_m,\ldots) (\textrm{QG quantity}) + \mathcal{O}\left(\Gamma^1\right)\,.
\end{align}
As an example, for Eq.~\eqref{eq:suppGammaExpansionTrivialExample0} (the explicit form of the coefficient of trivial geometry Eq.~\eqref{eq:trivialGeometryCoeffBeforeExpansion}), these steps work out to be
\begin{enumerate}
    \item $\frac{i e^{\Gamma \xi _n} \left(\Gamma^2 \xi _n^2-3 \Gamma \xi _n+3\right) \theta \left(-\xi _n\right)}{12 \sqrt{2 \pi } \Gamma^2 \xi _n}+\frac{i e^{\Gamma \left(-\xi _n\right)} \left(\Gamma^2 \xi _n^2+3 \Gamma \xi _n+3\right) \theta \left(\xi _n\right)}{12 \sqrt{2 \pi } \Gamma^2 \xi _n}\,$,
    \item $-\frac{i \left(\Gamma^2 \xi _n^2-6\right)}{24 \sqrt{2 \pi } \Gamma^2 \xi _n}\,$,
    \item $\frac{1-2 \theta (-E_n)}{8 \Gamma^2}+\frac{1}{24} \delta '(E_n)\,$,
\end{enumerate}
where $\xi_n$ denotes the Fourier space variable corresponding to $E_n$. For the other cases, the explicit functions and their Fourier transforms become too complicated to present explicitly, but, as shown in the next section, the final results are indeed simple. When the function depends on multiple energies $E_n,\ E_m,\ldots$, each one needs to be Fourier transformed, introducing Fourier space variables $\xi_n,\ \xi_m,\ldots$. Explicitly, we perform Fourier transforms using \textsc{mathematica}~\cite{Mathematica}. 

\subsection{Performing the \texorpdfstring{$\Gamma$}{Gamma} expansion} \label{suppsec:DoingGammaExpansion}

Having found a procedure for the expansion in $\Gamma$, we now apply it to each coefficient of a geometric quantity Eqs.~\eqref{eq:trivialGeometryCoeffBeforeExpansion}, \eqref{eq:QabCoeffBeforeExpansion}, \eqref{eq:QbcCoeffBeforeExpansion}, \eqref{eq:ConnectionCoeffBeforeExpansion}, \eqref{eq:TripleTraceCoeffBeforeExpansion}, keeping terms of order $\mathcal{O}(\Gamma^{-2})$, $\mathcal{O}(\Gamma^{-1})$ and $\mathcal{O}(\Gamma^{0})$. The $\Gamma$-expanded versions of contributions A–E are denoted by $\bar{A}$ through $\bar{E}$.

\subsubsection{Contribution \texorpdfstring{$\bar{A}$}{A}: trivial geometry \texorpdfstring{$\text{tr}[\hat{P}_n]$}{}}

As shown in the previous subsection, in the $\Gamma$ expansion contribution A (shown in Eq.~\eqref{eq:trivialGeometryCoeffBeforeExpansion}) is simply
\begin{align}
  \bar{A} &= \int_{\bk}  \sum_n \frac{1-2 \theta (-E_n)}{8 \Gamma^2} E^{ a  b c }_n \text{tr}[\hat{P}_n] = -\frac{1}{4 \Gamma ^2}\int_{\bk} \sum_n\theta (-E_n)  E^{ a  b c }_n \text{tr}[\hat{P}_n]\,.
\end{align}
We neglected a part proportional to $\delta '(E_n)$, as it is two orders higher in $\Gamma$ than the first and will not mix with any other geometric quantity later on. The equality follows from the momentum integration over $\bk$, since $\int \frac{d^d\bk}{(2\pi)^d} E^{ a  b c }_n = 0$. 

If we kept the part proportional to $\Gamma^0$, it would provide a contribution given by
\begin{align}
    \bar{A}_{\Gamma^0} &= \int_{\bk}  \sum_n \frac{1}{24}\delta'(E_n) E^{ a  b c }_n \text{tr}[\hat{P}_n] \\
    &= -\int_{\bk}  \sum_n \frac{1}{24}\delta''(E_n)E^{c}_n E^{ a  b}_n \text{tr}[\hat{P}_n] \\
    &= \int_{\bk}  \sum_n \frac{1}{24}\delta^{(3)}(E_n)E^{ a}_n E_n^b  E^{c}_n \text{tr}[\hat{P}_n] +\int_{\bk}  \sum_n \frac{1}{24}\delta''(E_n)E^{ a}_n E^{bc}_n  \text{tr}[\hat{P}_n] \\
    &= \int_{\bk}  \sum_n \frac{1}{24}\delta^{(3)}(E_n)E^{ a}_n E_n^b  E^{c}_n \text{tr}[\hat{P}_n] +\int_{\bk}  \sum_n \frac{1}{24}(\partial^a\delta'(E_n)) E^{bc}_n  \text{tr}[\hat{P}_n] \\
    &= \int_{\bk}  \sum_n \frac{1}{24}\delta^{(3)}(E_n)E^{ a}_n E_n^b  E^{c}_n \text{tr}[\hat{P}_n] - \bar{A}_{\Gamma^0}\,,
\end{align}
from which we conclude that this contribution could also be written as
\begin{align}
    \bar{A}_{\Gamma^0} = \int_{\bk}  \sum_n \frac{1}{48}\delta^{(3)}(E_n)E^{ a}_n E_n^b  E^{c}_n\,.
\end{align}

\subsubsection{Contribution \texorpdfstring{$\bar{B}$}{B}: quantum geometric tensors  \texorpdfstring{$Q_{nm}^{ a  b }$}{} and \texorpdfstring{$Q_{nm}^{ a c }$}{}}

In the $\Gamma$ expansion, contribution B (shown in Eq.~\eqref{eq:QabCoeffBeforeExpansion}) reduces to
\begin{align}
 \bar{B} &= \int_{\bk}\Biggl(\,\sum_{m\neq n} \left(\frac{\delta (E_n)}{ E_m}+\frac{1}{2}\delta '(E_n)\right)E_n^{ c }Q_{(nm)}^{ a  b }+\sum_{m\neq n} \frac{i}{2\Gamma} \delta(E_n)E_n^{ c }Q_{[nm]}^{ a  b }\Biggr)+(b\leftrightarrow c)\,.
\end{align}

\subsubsection{Contribution \texorpdfstring{$\bar{C}$}{C}: quantum geometric tensor \texorpdfstring{$Q_{nm}^{ b c }$}{}}

In the $\Gamma$ expansion, contribution C (shown in Eq.~\eqref{eq:QbcCoeffBeforeExpansion}) becomes
\begin{align}
 \bar{C} &= \int_{\bk}\sum_{m\neq n} \left(-\frac{2\delta (E_n)}{E_m}-\delta '(E_n)\right) E^{ a }_n Q_{(nm)}^{ b  c }\,.
\end{align}

\subsubsection{Contribution \texorpdfstring{$\bar{D}$}{D}: connections \texorpdfstring{$C_{nm}^{ b ; a c }$, $C_{nm}^{c ; a  b }$}{}}

In the $\Gamma$ expansion, contribution D (shown in Eq.~\eqref{eq:ConnectionCoeffBeforeExpansion}) simplifies to
\begin{align} \label{eq:ConnectionCoeffAfterExpansion}
\bar{D} &= \int_{\bk}\sum_{m,n} \frac{1}{4}\bigl(\delta(E_m)+\delta(E_n)\bigr)\left(C_{(nm)}^{ b ; a c }+C_{(nm)}^{c ; a  b }\right) \\
&= \int_{\bk}\sum_{m,n} \frac{1}{2}\delta(E_n)\left(C_{(nm)}^{ b ; a c }+C_{(nm)}^{c ; a  b }\right)\,.
\end{align}
We utilized the $n,m$ symmetry of $\delta(E_m)+\delta(E_n)$ in the last equality.

\subsubsection{Contribution \texorpdfstring{$\bar{E}$}{E}: to three-band traces \texorpdfstring{$\text{tr}[e^{ a }_{mn}e^{ b }_{nl}e^{c }_{lm}]$, $\text{tr}[e^{ a }_{mn}e^{c }_{nl}e^{ b }_{lm}]$}{}}

In the $\Gamma$ expansion of the final contribution, E (shown in Eq.~\eqref{eq:TripleTraceCoeffBeforeExpansion}), only the term proportional to $\text{Im}\,\text{tr}[e^{ a }_{mn}e^{ b }_{nl}e^{ c }_{lm}]$ survives, and we obtain
\begin{align}
\bar{E} = \int_{\bk}&\sum_{(l, m,n)} \left(\frac{\delta (E_n)}{2}-\frac{\delta (E_m)}{2}\right)\text{Im}\,\text{tr}[e^{ a }_{mn}e^{ b }_{nl}e^{ c }_{lm}]\,.
\end{align}
Now that the coefficient does not depend on the $l$ band, simplifications are possible. Indeed, explicitly writing out the trace yields
\begin{align}
\bar{E} = \int_{\bk}&\sum_{(l, m,n)} \left(\frac{\delta (E_n)}{2}-\frac{\delta (E_m)}{2}\right)\text{Im}\left\{-i\text{tr}[\hat{P}_m(\partial^ a  \hat{P}_n)\hat{P}_n (\partial^ b  \hat{P}_l)\hat{P}_l (\partial^ c  \hat{P}_m)]\right\}\,,
\end{align}
which we split into
\begin{align}\label{eq:tripleTraceafterExpansionIntermediate1}
\bar{E} = \int_{\bk}\Biggl(&-\sum_{(l, m,n)} \frac{\delta (E_n)}{2}\,\text{Re}\left\{\text{tr}[\hat{P}_m(\partial^ a  \hat{P}_n)\hat{P}_n (\partial^ b  \hat{P}_l)\hat{P}_l (\partial^ c  \hat{P}_m)]\right\}  \\
&\quad+ \sum_{(l, m,n)} \frac{\delta (E_m)}{2}\,\text{Re}\left\{\text{tr}[\hat{P}_m(\partial^ a  \hat{P}_n)\hat{P}_n (\partial^ b  \hat{P}_l)\hat{P}_l (\partial^ c  \hat{P}_m)]\right\}\Biggr)\,.
\end{align}
Now by definition of projectors,
\begin{align}
    \sum_n \hat{P}_n = \mathbb{1} \Rightarrow \sum_n \hat{P}_n^2 = \mathbb{1}\,.
\end{align}
Applying a derivative yields
\begin{align}
    \sum_n (\partial^ a  \hat{P}_n) \hat{P}_n = -\sum_n \hat{P}_n(\partial^ a  \hat{P}_n)\,.
\end{align}
Using this identity to reduce Eq.~\eqref{eq:tripleTraceafterExpansionIntermediate1} yields (remember $l,m,n$ are necessarily all different in this sum)
\begin{align}
 \bar{E} &= \int_{\bk}\sum_{n} \frac{\delta (E_n)}{2}\,\text{Re}\left\{\sum_{m\neq n}\text{tr}[\hat{P}_m(\partial^ a  \hat{P}_n)\hat{P}_n (\partial^ b  \hat{P}_n)(\partial^ c  \hat{P}_m)]\right\} \\
&\qquad- \int_{\bk}\sum_{m\neq n}\frac{\delta (E_m)}{2}\,\text{Re}\left\{\sum_{n}\text{tr}[\hat{P}_m(\partial^ a  \hat{P}_n)\hat{P}_n (\partial^ b  \hat{P}_n) (\partial^ c  \hat{P}_m)]\right\}\,,
\end{align}
which, since $\hat{P}_m(\partial^ a  \hat{P}_n)\hat{P}_n = \hat{P}_m(\partial^ a  \hat{P}_n)$ for $m\neq n$, reduces to
\begin{align}
   \bar{E} = \int_{\bk} \Biggl(& \sum_{n\neq m} \frac{\delta (E_n)}{2}\,\text{Re}\left\{\text{tr}[\hat{P}_m(\partial^ a  \hat{P}_n) (\partial^ b  \hat{P}_n)(\partial^ c  \hat{P}_m)]\right\} -\sum_{n \neq m}\frac{\delta (E_m)}{2}\,\text{Re}\left\{\text{tr}[\hat{P}_m(\partial^ a  \hat{P}_n) (\partial^ b  \hat{P}_n) (\partial^ c  \hat{P}_m)]\right\}\Biggr)\,.
\end{align}
We now relabel the second sum and combine both parts:
\begin{align}
    \bar{E} = \int_{\bk}\sum_{n\neq m}& \frac{\delta (E_n)}{2}\,\text{Re}\left\{\text{tr}[\hat{P}_m(\partial^ a  \hat{P}_n) (\partial^ b  \hat{P}_n)(\partial^ c  \hat{P}_m)]-\text{tr}[\hat{P}_n(\partial^ a  \hat{P}_m) (\partial^ b  \hat{P}_m) (\partial^ c  \hat{P}_n)]\right\}\,.
\end{align}
Now $\text{Re}\left\{\text{tr}[\hat{P}_m(\partial^ a  \hat{P}_n) (\partial^ b  \hat{P}_n)(\partial^ c  \hat{P}_m)]\right\} = \text{Re}\left\{\text{tr}[\hat{P}_m(\partial^ a  \hat{P}_n) (\partial^ b  \hat{P}_n)(\partial^ c  \hat{P}_m)]^\dagger\right\}$, which we use to transform the first trace:
\begin{align}
\bar{E} &= \int_{\bk} \sum_{n\neq m}\frac{\delta (E_n)}{2}\,\text{Re}\left\{\text{tr}[(\partial^ c  \hat{P}_m)(\partial^ b  \hat{P}_n)(\partial^ a  \hat{P}_n)\hat{P}_m]-\text{tr}[\hat{P}_n(\partial^ a  \hat{P}_m) (\partial^ b  \hat{P}_m) (\partial^ c  \hat{P}_n)]\right\} \nonumber \\
 &=-\int_{\bk}\sum_{n\neq m} \frac{\delta (E_n)}{2}\,\text{Re}\left\{\text{tr}[ \hat{P}_n(\partial^ a  \hat{P}_m)(\partial^ c  \hat{P}_m)(\partial^ b  \hat{P}_n)]+\text{tr}[\hat{P}_n(\partial^ a  \hat{P}_m) (\partial^ b  \hat{P}_m) (\partial^ c  \hat{P}_n)]\right\}\,.
\end{align}
It is proven in Eq.~(33) of Ref.~\cite{Mitscherling2025} that
\begin{align}
    \sum_{m \neq n} C^{a ; b c}_{m n}=-Q^{b ; a c}_n+\text{tr}\left[\hat{P}_n\left(\sum_{m \neq n} (\partial^b \hat{P}_m) (\partial^a \hat{P}_m)\right) (\partial^c \hat{P}_n)\right]\, ,
\end{align}
where $Q_{n}^{ a ; b  c }$ is the skewness tensor Eq.~\eqref{eq:QGdefSkewness}.

Using this equation and $\text{Re}\,C_{nm}^{ a ; b  c } =C_{(nm)}^{ a ; b  c } $, the contribution finally simplifies to
\begin{align}
   \bar{E} = -\int_{\bk}\Biggl( \sum_{n\neq m}\frac{\delta (E_n)}{2}C_{(nm)}^{ c ; a  b } +\sum_{n\neq m}\frac{\delta (E_n)}{2}C_{(nm)}^{ b ; a  c }  +\sum_{n}\delta (E_n)\,\text{Re}\,Q_{n}^{ a ; b  c }\Biggr)\,.
\end{align}
Interestingly, these connections exactly cancel with the connections explicitly obtained before in Eq.~\eqref{eq:ConnectionCoeffAfterExpansion}.

\subsection{Combining all contributions} \label{suppsec:finalresult}

We may now combine all previously isolated contributions $\bar{A}$ through $\bar{E}$, and use integration by parts and quantum geometric identities to simplify the total result. The starting point is the sum of all results of the previous section, given by
\begin{equation}
\begin{aligned}
    \sigma^{a;bc} &= \frac{e^3}{\hbar}\int \frac{d^d\bk}{(2\pi)^d} \Biggl( -\sum_n \theta(-E_n) \frac{1}{4\Gamma^2} E^{ a  b  c }_n \operatorname{tr}[\hat{P}_n] \\
    &\quad+ \sum_{m\neq n} \left(-\frac{2\delta (E_n)}{E_m}-\delta '(E_n)\right) E^{ a }_n Q_{(nm)}^{ b  c } \\
    &\quad+\sum_{m\neq n} \left(\frac{\delta (E_n)}{E_m}+\frac{1}{2}\delta '(E_n)\right)E_n^{ c }Q_{(nm)}^{ a  b }+\sum_{m\neq n} \frac{i}{2\Gamma} \delta(E_n)E_n^{ c }Q_{[nm]}^{ a  b }\\
    &\quad+\sum_{m\neq n} \left(\frac{\delta (E_n)}{ E_m}+\frac{1}{2}\delta '(E_n)\right)E_n^{ b }Q_{(nm)}^{ a  c }+\sum_{m\neq n} \frac{i}{2\Gamma} \delta(E_n)E_n^{ b }Q_{[nm]}^{ a  c } \\
    &\quad- \sum_{n}\delta (E_n)\,\operatorname{Re}Q_{n}^{ a ; b  c } \Biggr)\,.
\end{aligned}
\end{equation}
Where possible, we will use the factor of $E_n^{ a }$ to perform integration by parts since $\delta'(E_n)E_n^a=\partial_a \delta(E_n)$. Symmetric and antisymmetric parts are identified as real and imaginary, respectively. We begin by splitting up sums to simplify contributions individually
\begin{equation}
\begin{aligned}
    \sigma^{a;bc} &= \frac{e^3}{\hbar}\int \frac{d^d\bk}{(2\pi)^d} \Biggl( -\sum_n \theta(-E_n) \frac{1}{4\Gamma^2} E^{ a  b  c }_n\, \operatorname{tr}[\hat{P}_n] \\
    &\quad- \sum_{m\neq n} \frac{2\delta (E_n)}{E_m} E^{ a }_n \,\operatorname{Re}Q_{nm}^{ b  c }- \sum_{m\neq n} \delta '(E_n) E^{ a }_n \,\operatorname{Re}Q_{nm}^{ b  c } \\
    &\quad+\sum_{m\neq n} \frac{\delta (E_n)}{ E_m}E_n^{ c }\,\operatorname{Re}Q_{nm}^{ a  b }+\sum_{m\neq n} \frac{1}{2}\delta '(E_n)E_n^{ c }\,\operatorname{Re}Q_{nm}^{ a  b }-\sum_{m\neq n} \frac{1}{2\Gamma} \delta(E_n)E_n^{ c }\,\operatorname{Im}Q_{nm}^{ a  b }\\
    &\quad+\sum_{m\neq n} \frac{\delta (E_n)}{E_m}E_n^{ b }\,\operatorname{Re}Q_{nm}^{ a  c }+\sum_{m\neq n} \frac{1}{2}\delta '(E_n)E_n^{ b }\,\operatorname{Re}Q_{nm}^{ a  c }-\sum_{m\neq n} \frac{1}{2\Gamma} \delta(E_n)E_n^{ b }\,\operatorname{Im}Q_{nm}^{ a  c } \\
    &\quad- \sum_{n}\delta (E_n)\,\operatorname{Re}Q_{n}^{ a ; b  c } \Biggr)\,.
\end{aligned}
\end{equation}
We begin integrating by parts using $E_n^a$ as an internal derivative, and additionally use $\frac{\delta(E_n)}{E_m} = \frac{\delta(E_n)}{E_m-E_n}$:
\begin{equation}
\begin{aligned}
    \sigma^{a;bc} &= \frac{e^3}{\hbar}\int \frac{d^d\bk}{(2\pi)^d} \Biggl( -\sum_n \theta(-E_n) \frac{1}{4\Gamma^2} E^{ a  b  c }_n\, \operatorname{tr}[\hat{P}_n] \\
    &\quad- \sum_{m\neq n} 2 \delta (E_n) E_n^a\frac{\operatorname{Re}Q_{nm}^{ b  c }}{E_m-E_n}  + \sum_{m\neq n} \delta(E_n)  \operatorname{Re}\partial^{ a }Q_{nm}^{ b  c } \\
    &\quad+\sum_{m\neq n} \delta (E_n)E_n^c \frac{\operatorname{Re}Q_{nm}^{ a  b }}{E_m-E_n}-\sum_{m\neq n} \frac{1}{2}\delta(E_n)\,\operatorname{Re}\partial^{ c }Q_{nm}^{ a  b }-\sum_{m\neq n} \frac{1}{2\Gamma} \delta(E_n)E_n^{ c }\,\operatorname{Im}Q_{nm}^{ a  b }\\
    &\quad+\sum_{m\neq n} \delta (E_n)E_n^b \frac{\operatorname{Re}Q_{nm}^{ a  c }}{E_m-E_n}-\sum_{m\neq n} \frac{1}{2}\delta(E_n)\,\operatorname{Re}\partial^{ b }Q_{nm}^{ a  c }-\sum_{m\neq n} \frac{1}{2\Gamma} \delta(E_n)E_n^{ b }\,\operatorname{Im}Q_{nm}^{ a  c } \\
    &\quad- \sum_{n}\delta (E_n)\,\operatorname{Re}Q_{n}^{ a ; b  c } \Biggr)\,.
\end{aligned}
\end{equation}
Now there are convenient relations between the single-band and multiband QGTs (defined in Eqs.~\eqref{eq:QGdefQGTsingleband}, \eqref{eq:QGdefQGT})
\begin{align} \label{eq:QSumRules}
        \sum_{m\neq n} \operatorname{Re} Q_{nm}^{ a  b } &= \sum_{m\neq n} \operatorname{Re} (Q_{mn}^{ a  b })^\dagger = -\operatorname{Re} Q_{n}^{ a  b }\,, \\
        \sum_{m\neq n} \operatorname{Im} Q_{nm}^{ a  b } &= \sum_{m\neq n} \operatorname{Im} (Q_{mn}^{ a  b })^\dagger =\operatorname{Im} Q_{n}^{ a  b }\,.
\end{align}
The sums are only over $m$ in this case. We may then transform the conductivity to 
\begin{equation}
\begin{aligned}
    \sigma^{a;bc} &= \frac{e^3}{\hbar}\int \frac{d^d\bk}{(2\pi)^d} \Biggl( -\sum_n \theta(-E_n) \frac{1}{4\Gamma^2} E^{ a  b  c }_n\, \operatorname{tr}[\hat{P}_n] \\
    &\quad- \sum_{m\neq n}2 \delta (E_n) E_n^a\frac{\operatorname{Re}Q_{mn}^{ b  c }}{E_m-E_n}  - \sum_{m\neq n} \delta(E_n)  \operatorname{Re}\partial^{ a }Q_{n}^{ b  c } \\
    &\quad+\sum_{m\neq n} \delta (E_n)E_n^c \frac{\operatorname{Re}Q_{mn}^{ a  b }}{E_m-E_n}+\sum_{n} \frac{1}{2}\delta(E_n)\,\operatorname{Re}\partial^{ c }Q_{n}^{ a  b }-\sum_{n} \frac{1}{2\Gamma} \delta(E_n)E_n^{ c }\operatorname{Im}Q_{n}^{ a  b }\\
    &\quad+\sum_{m\neq n} \delta (E_n)E_n^b \frac{\operatorname{Re}Q_{mn}^{ a  c }}{E_m-E_n}+\sum_{n} \frac{1}{2}\delta(E_n)\,\operatorname{Re}\partial^{ b }Q_{n}^{ a  c }-\sum_{n} \frac{1}{2\Gamma} \delta(E_n)E_n^{ b }\operatorname{Im}Q_{n}^{ a  c } \\
    &\quad- \sum_{n}\delta (E_n)\,\operatorname{Re}Q_{n}^{ a ; b  c } \Biggr)\,.
\end{aligned}
\end{equation}
Using the identity
\begin{align} \label{eq:QGTdipoleAsSkewness}
    \partial^ a  Q_n^{ b c} = (Q_n^{c; a  b })^\dagger+Q_n^{ b ; a c} \Rightarrow \operatorname{Re}\partial^ a  Q_n^{ b c} = \operatorname{Re}Q_n^{c; a  b }+\operatorname{Re} Q_n^{ b ; a c}
\end{align}
leads to the following relation, reminiscent of Christoffel symbols in terms of the metric:
\begin{align}
    Q_{n}^{ a ; b  c } = \frac{1}{2}\left(\partial^{ b }Q_{n}^{ a  c }+\partial^{ c }Q_{n}^{ a  b }-\partial^{ a }Q_{n}^{ b  c }\right)\,.
\end{align}
We insert this relation for the last term:
\begin{equation}
\begin{aligned}
    \sigma^{a;bc} &= \frac{e^3}{\hbar}\int \frac{d^d\bk}{(2\pi)^d} \Biggl( -\sum_n \theta(-E_n) \frac{1}{4\Gamma^2} E^{ a  b  c }_n\, \operatorname{tr}[\hat{P}_n] \\
    &\quad- \sum_{m\neq n}2 \delta (E_n) E_n^a\frac{\operatorname{Re}Q_{mn}^{ b  c }}{E_m-E_n} \\
    &\quad+\sum_{m\neq n} \delta (E_n)E_n^c \frac{\operatorname{Re}Q_{mn}^{ a  b }}{E_m-E_n}-\sum_{n} \frac{1}{2\Gamma} \delta(E_n)E_n^{ c }\operatorname{Im}Q_{n}^{ a  b }\\
    &\quad+\sum_{m\neq n} \delta (E_n)E_n^b \frac{\operatorname{Re}Q_{mn}^{ a  c }}{E_m-E_n}-\sum_{n} \frac{1}{2\Gamma} \delta(E_n)E_n^{ b }\operatorname{Im}Q_{n}^{ a  c } \\
    &\quad- \sum_{n}\frac{1}{2}\delta (E_n)\,\operatorname{Re}\partial^{a}Q_n^{bc} \Biggr)\,.
\end{aligned}
\end{equation}
As is apparent, the first term (the NLD contribution) is completely symmetric in $a,b,c$. Accordingly, it is classified as an Ohmic (as opposed to Hall, i.e.\ non-dissipative) contribution according to~\cite{OhmicAndHall}. We may now reorder terms and integrate the first term by parts to isolate a Fermi surface delta function everywhere:
\begin{equation}
\begin{aligned}
   \sigma^{a;bc} &=\frac{e^3}{\hbar}\int \frac{d^d\bk}{(2\pi)^d} \sum_n \delta(E_n)\Biggl( - \frac{1}{(2\Gamma)^2} E^a E^{b  c}_n \,\operatorname{tr}[\hat{P}_n] \\
    &\quad - \frac{1}{2\Gamma} \left(E_n^{ c }\operatorname{Im}Q_{n}^{ a  b }+E_n^{ b }\operatorname{Im}Q_{n}^{ a  c }\right)\\
    &\quad- \sum_{m\neq n}\left(2 E_n^a\frac{\operatorname{Re}Q_{mn}^{ b  c }}{E_m-E_n} -E_n^c \frac{\operatorname{Re}Q_{mn}^{ a  b }}{E_m-E_n}- E_n^b \frac{\operatorname{Re}Q_{mn}^{ a  c }}{E_m-E_n}\right)\\
    &\quad- \frac{1}{2}  \operatorname{Re}\partial^{ a }Q_{n}^{ b  c }\Biggr)\,.
\end{aligned}
\end{equation}
Using the decomposition of $Q_n^{ a  b }$ into the familiar quantum metric and Berry curvature Eq.~\eqref{eq:QGdefQGTsingleband}
\begin{align}
    Q_n^{ a  b } = g_n^{ a  b } -\frac{i}{2}\Omega_n^{ a  b }\,,
\end{align}
as well as the (symmetric) two-state metric $\operatorname{Re} Q_{mn}^{ab} = g_{mn}^{ab}$, the conductivity is finally given by
\begin{equation} \label{eq:suppfinalresult}
\begin{aligned}
    \sigma^{a;bc} &=\frac{e^3}{\hbar}\int \frac{d^d\bk}{(2\pi)^d} \sum_n \delta(E_n)\Biggl( - \frac{1}{(2\Gamma)^2} E^a_n E^{b  c}_n\, \operatorname{tr}[\hat{P}_n] \\
    &\quad + \frac{1}{4\Gamma} \left(E_n^{ c }\Omega_{n}^{ a  b }+E_n^{ b }\Omega_{n}^{ a  c }\right)\\
    &\quad- \sum_{m\neq n}\left(2 E_n^a\frac{g_{mn}^{ b  c }}{E_m-E_n} -E_n^c \frac{g_{mn}^{ a  b }}{E_m-E_n}- E_n^b \frac{g_{mn}^{ a  c }}{E_m-E_n}\right)\\
    &\quad- \frac{1}{2}  \partial^{ a }g_{n}^{ b  c }\Biggr)\,.
\end{aligned}
\end{equation}
Setting $b=c$, this expression yields the NLD, BCD, interQMD, and intraQMD contributions in the main text. The third line is the interQMD term, found with varying coefficients throughout the literature~\cite{gao_field_2014, das_intrinsic_2023,kaplan_unification_2024, park2025quantumgeometrymoyalproduct}, depending on the approach used and how the scattering is introduced (see Sec.~\ref{app:Keldysh}). In Sec.~\ref{suppsec:banddependentGamma}, we discuss what would have changed if $\Gamma$ had been band-dependent, and whether such an approach is consistent.

\section{Discussion on our treatment of finite quasiparticle lifetimes}
In this supplement, we discuss whether the treatment shown here could be extended to a band-dependent scattering rate, and then describe the differences in our treatment compared to existing literature on the nonlinear Hall effect. 
\subsection{A note on band-dependent scattering rates} \label{suppsec:banddependentGamma}
To model a band-dependent scattering rate, in Sec.~\ref{suppsec:insertingProjectors} we could have introduced a scattering rate $\Gamma$ that is diagonal in bands, with Green's functions given by
\begin{align}
    G^{R/A}(\epsilon,\bk) =  \frac{1}{\epsilon - \hat{H}(\bk) \pm i\hat{\Gamma}} = \sum_n \frac{ \hat{P}_n(\bk)}{\epsilon - E_n(\bk) \pm i\Gamma_n}\,.
\end{align}
This was also motivated by the attempt to match approaches used in previous semiclassical results, where different lifetimes for on- and off-shell particles were used~\cite{kaplan_unifying_2023, holder_consequences_2020}. The expansion in $\Gamma$ works almost identically,  setting $\Gamma_n = c_n \Gamma$ for some $c_n$ and then expanding in $\Gamma$. The final result of this approach is given by
\begin{align}
\begin{aligned}
    \sigma^{a;bc} &=\frac{e^3}{\hbar}\int \frac{d^d\bk}{(2\pi)^d} \sum_n \delta(E_n)\Biggl( - \frac{1}{(2\Gamma_n)^2} E^a_n E^{b  c}_n \operatorname{tr}[\hat{P}_n] \\
    &\quad + \frac{1}{4\Gamma_n} \left(E_n^{ c }\Omega_{n}^{ a  b }+E_n^{ b }\Omega_{n}^{ a  c }\right)\\
    &\quad- \sum_{m\neq n}\left(\frac{(\Gamma_n+\Gamma_m)^2}{2\Gamma_n^2} E_n^a\frac{g_{mn}^{ b  c }}{E_m-E_n} -\frac{\Gamma_m}{\Gamma_n}E_n^c \frac{g_{mn}^{ a  b }}{E_m-E_n}- \frac{\Gamma_m}{\Gamma_n}E_n^b \frac{g_{mn}^{ a  c }}{E_m-E_n}\right)\\
    &\quad+ \sum_{m\neq n} \frac{\Gamma_m}{2\Gamma_n}  \partial^{ a }g_{mn}^{ b  c }\Biggr)\,.
\end{aligned}
\end{align}
Setting $\Gamma_n = \Gamma$ recovers the main result Eq.~\eqref{eq:suppfinalresult}. Since $n$ is the on-shell band while $m$ is (where it appears) necessarily off-shell, varying lifetimes could be introduced for each one, inspired by previous work~\cite{kaplan_unifying_2023, holder_consequences_2020}. Indeed, setting the off-shell self-energies to 0, we find a vanishing intraQMD and modified interQMD. Ultimately, however, this approach does not allow matching results found with other methods.

Additionally, a band-dependent scattering rate is in general momentum-dependent in the orbital basis, since it is diagonal in the band basis. Such a self-energy necessarily breaks the Ward identity, and already at the linear level the well-known TKNN formula is not recovered~\cite{PhysRevLett.49.405}. Indeed, the antisymmetric contribution is found to be
\begin{equation}
    \sigma^{[a;b]} = -\frac{e^2}{\hbar}\int \frac{d^d\bk}{(2\pi)^d} \sum_n \theta(-E_n) \frac{\Gamma_m}{\Gamma_n}\operatorname{Im} Q_{nm}^{ab}\,,
\end{equation}
which only recovers the TKNN formula when $\Gamma_m = \Gamma_n$. To take a non-constant scattering rate $\Gamma$ into account consistently, a full perturbative treatment of disorder including vertex corrections (and possibly crossing diagrams) should be done, which we leave to potential future work.

\subsection{Finite quasiparticle lifetime and interband effects}
As discussed in the main text, our result in Eq.~\eqref{eq:suppfinalresult} agrees with earlier semiclassical predictions for the Berry curvature dipole (BCD) and the (band-normalized) interband quantum metric dipole (interQMD). However, to the best of our knowledge, the intraband quantum metric dipole (intraQMD) has not been captured within a semiclassical framework. Finding similar but not identical formulas by different approaches is commonly observed in the nonlinear regime~\cite{jiang_holder_review_arxiv_25}. The discrepancy can be traced to differences in how virtual interband processes and scattering effects are treated, depending on the underlying formalism --- such as Boltzmann transport or density matrix methods. As shown in Sec.~\ref{suppsec:banddependentGamma}, assuming an infinite lifetime of remote states using a phenomenological two-lifetime approach \cite{kaplan_unifying_2023, holder_consequences_2020, jiang_holder_review_arxiv_25} within our derivation, we find a vanishing intraQMD and strongly modified interQMD. This suggests a physical interpretation of this contribution as a virtual hopping to a remote band followed by a decay.

Interestingly, the Boltzmann equation can be derived from Green's function formalism by assuming that quasiparticles have an infinite lifetime (i.e., a delta-function spectral function)~\cite{Rammer1986}. This implies that the Boltzmann equation cannot account for the physical process mentioned above, explaining why it was not previously obtained. Furthermore, this  suggests that the additional intraQMD term may arise in a semiclassical theory by lifting this assumption within a multiband setting. As mentioned in the main text and visible in the derivation of Eq.~\eqref{eq:suppfinalresult},  the intraQMD originates from a combination of two- and three-band processes, highlighting how the treatment of finite lifetimes directly affects the interband contributions obtained. 

Recent progress has been made here by employing Wigner-Moyal approaches~\cite{mitscherling2025orbitalwignerfunctionsquantum, park2025quantumgeometrymoyalproduct}, where the intraQMD contribution may arise after including a suitable collision integral. The theory of DC response functions in dispersive multiband systems would greatly benefit from a careful reevaluation of both semiclassical and fully quantum approaches, systematically investigating the implications of nontrivial quantum geometry within a fully gauge-invariant framework, as presented here. It may also provide further insights on unconventional transport in flatband systems, where similar discrepancies have been discussed recently \cite{Mitscherling2022, Huhtinen2023, Antebi2024, Chen2025}.

\section{Symmetry decomposition of the nonlinear conductivity tensor and candidate materials} \label{suppsec:symmetries}

\subsection{Ohmic and Hall contributions}
The conductivity tensor can be separated into a fully symmetric Ohmic (dissipative) part and a non-dissipative Hall part containing the remaining components~\cite{OhmicAndHall}. Using the full expression for the conductivity tensor Eq.~\eqref{eq:suppfinalresult}, we identify the NLD contribution as purely Ohmic, the BCD and interQMD contributions as purely Hall, and the intraQMD contribution as containing both Ohmic and Hall components.
% \sout{Interestingly, the intraQMD contribution separates into its Ohmic and Hall components as ${\sigma^{a;bc}_{\mathrm{intraQMD}}(\bk)=-\frac{3}{2}\partial{\vphantom{\Gamma}}^{(a}\,g{\vphantom{\Gamma}}^{bc)}_n(\bk)+\frac{1}{2}\Gamma^{abc}_n(\bk)}$, where $\Gamma^{abc}_n(\bk)=\frac{1}{2}\left(\partial^b\, g^{ac}_n + \partial^c\, g^{ab}_n-\partial^a \, g^{bc}_n\right)$ are the Christoffel symbols related to the quantum metric}~\cite{PhysRevResearch.4.013217,fontana2025quantumgeometryelectricmagnetochiral}.} 
The Ohmic contributions can, in general, be observed in both longitudinal and transverse components of the conductivity, while Hall contributions are purely transverse.
Moreover, through the use of angle-dependent measurements and specific sample geometry, it is experimentally possible to access all components of the full tensor $\sigma^{a;bc}$, allowing for the separation of Ohmic and Hall contributions~\cite{chichinadze2024observation}.

\subsection{Symmetry transformation of band projectors and quantum geometric invariants}

A Bloch Hamiltonian $\hat H(\bk)$ is symmetric under a transformation $\mathcal{U}$ if
\begin{align} \label{eq:suppdefSymmetricHam}
   \mathcal{U}^\dagger\, \hat H\big(D_U\bk\big) \,\mathcal{U} = \hat H(\bk) \,.
\end{align}
Here, $D_U$ is a representation of $\mathcal{U}$ in momentum space, which belongs to $O(3)$. Importantly, $\mathcal{U}$ may be either unitary or antiunitary. We assume that $\mathcal{U}$ is $\bk$-independent, up to a possible $\bk$-dependent overall phase. For a lattice Hamiltonian, affine symmetries (consisting of a linear part and a translation) and time-reversal $\mathcal{T}$ can always be represented in a $\bk$-independent way for an adequate Fourier basis choice~\cite{PhysRevB.92.195116, Vanderbilt_2018}. Indeed, in the so-called convention I Bloch-like basis functions are defined by~\cite{Vanderbilt_2018}
\begin{align}
    \left|\chi_j^{\mathbf{k}}\right\rangle=\sum_{\mathbf{R}} e^{i \mathbf{k} \cdot\left(\mathbf{R}+\boldsymbol{\tau}_j\right)}\left|\phi_{\mathbf{R} j}\right\rangle\,,
\end{align}
where $\mathbf{R}$ is a lattice vector which specifies the unit cell, and the index $j$ runs over the orbitals, located at $\boldsymbol{\tau}_j$, in the cell. The wavefunction of the tight-binding orbital $j$ in the unit cell located at $\mathbf{R}$ is denoted by $\left|\phi_{\mathbf{R} j}\right\rangle$. Affine symmetries and $\mathcal{T}$ can then be represented in a $\bk$-independent way up to a $\bk$-dependent phase~\cite{PhysRevB.92.195116}. These symmetries suffice to generate all magnetic space groups. Furthermore, only this Fourier convention is consistent with coupling to the electromagnetic potential $\mathbf{A}$ via~\cite{mitscherling_longitudinal_2020}
\begin{align}
    \hat H(\bk) \rightarrow \hat H(\bk - q\mathbf{A})\,.
\end{align}
If $|\psi_n(\bk)\rangle$ is an eigenstate of $\hat H(\bk)$, we see from Eq.~\eqref{eq:suppdefSymmetricHam} that
\begin{equation}
    \hat H\big(D_U\bk\big)\,\mathcal{U}\,|\psi_n(\bk)\rangle =  \mathcal{U}\, \hat H(\bk) \,|\psi_n(\bk)\rangle = E_n \,\mathcal{U} \,|\psi_n(\bk)\rangle \,.
\end{equation}
Thus, we find
\begin{equation}
  |\psi_n(D_U \bk)\rangle =  \mathcal{U} |\psi_n(\bk)\rangle\,,
\end{equation}
up to gauge choice of eigenbasis. This implies that the projectors transform as
\begin{equation} \label{eq:suppprojTransformunderSymmetry}
  \hat P_n( D_U\bk)=|\psi_n(D_U \bk)\rangle \langle \psi_n(D_U \bk)| =  \mathcal{U}\, \hat P_n(\bk)\, \mathcal{U}^\dagger \,.
\end{equation}
Note that any $\bk$-dependent overall phase in $\mathcal{U}$ cancels in the projector, so that $\mathcal{U}$ can be assumed to be momentum independent without loss of generality. The transformation of the band projectors under symmetries Eq.~\eqref{eq:suppprojTransformunderSymmetry} allows us to deduce transformation rules for quantum geometric invariants. We illustrate this for the multiband QGT Eq.~\eqref{eq:QGdefQGT}
\begin{align}
    Q_{mn}^{ab}(\bk) = \text{tr}\Big[\hat{P}_n(\bk)\big[(\partial^a \hat{P}_m)(\bk)\big]\big[(\partial^b \hat{P}_n)(\bk)\big]\Big]\,.
\end{align}
Its real part is the two-state quantum metric $g_{nm}^{ab} = \operatorname{Re}Q_{nm}^{ab}$,  see Eq.~\eqref{eq:QGdefQGT}. We start by considering a unitary $\mathcal{U}$. We evaluate it at $D_U \bk$ leading to
\begin{align}
    Q_{mn}^{ab}(D_U \bk) = \text{tr}\Big[\hat{P}_n(D_U \bk)\big[(\partial^a \hat{P}_m)(D_U \bk)\big]\big[(\partial^b \hat{P}_n)(D_U \bk)\big]\Big]\,.
\end{align}
The chain rule implies $\partial^{a'}\big(\hat{P}_n(D_U \bk)\big) = \big(\partial^{b}\hat{P}_n\big)(D_U \bk)\,\partial^{a'}\big(D_U \bk\big)^b = D_U^{ba'}(\partial^{b}\hat{P}_n)(D_U \bk) =\bigl(D_U^{-1}\bigr)^{a'b}\big(\partial^{b}\hat{P}_n\big)(D_U \bk)$ leading to $\big(\partial^a \hat{P}_n\big)(D_U \bk) = D_U^{aa'}\partial^{a'}\big(\hat{P}_n(D_U \bk)\big)$, where we sum over repeated indices. Inserting Eq.~\eqref{eq:suppprojTransformunderSymmetry} leads to
\begin{align}
    Q_{mn}^{ab}(D_U \bk) &= D_U^{aa'}D_U^{bb'}\text{tr}\Big[\hat{P}_n(D_U \bk)\big[\partial^{a'} \big(\hat{P}_m(D_U \bk)\big)\big]\big[\partial^{b'} \big(\hat{P}_n(D_U \bk)\big)\big]\Big]\\
    &= D_U^{aa'}\,D_U^{bb'}\,\text{tr}\Big[\big(\mathcal{U}\,\hat P_n(\bk)\, \mathcal{U}^\dagger\big)\,\big[\partial^{a'} \big(\mathcal{U} \,\hat P_m(\bk)\, \mathcal{U}^\dagger\big)\big]\big[\partial^{b'} \big(\mathcal{U}\, \hat P_n(\bk)\, \mathcal{U}^\dagger\big)\big]\Big]\\
    &= D_U^{aa'}\,D_U^{bb'}\,\text{tr}\big[\hat P_n(\bk) \big(\partial^{a'} \hat P_m(\bk)\big)\big(\partial^{b'} \,\hat P_n(\bk)\big)\big]\\
    &= D_U^{aa'}\,D_U^{bb'}\,Q_{mn}^{a'b'}(\bk) \qquad \textrm{(unitary transformation)}\,,
    \label{eqn:unitary}
\end{align}
using the $\bk$-independence of $\mathcal{U}$ and $\mathcal{U}^\dagger\mathcal{U}=\mathbb{1}$. The multiband QGT thus transforms as a tensor. Any higher-order trace of projectors, i.e., higher-order quantum geometric invariants such as the quantum geometric connection, will also transform as tensors. In particular, $Q_n^{ab} = Q_{nn}^{ab}$ will equally transform in the same way.

If the symmetry transformation $\mathcal{U}$ is antiunitary, we have $\mathcal{U}=\mathcal{D}\mathcal{K}$, where $\mathcal{D}$ is unitary and $\mathcal{K}$ represents complex conjugation. Then $|\psi_n(D_U \bk)\rangle =  \mathcal{D} \mathcal{K}|\psi_n(\bk)\rangle=\mathcal{D} \big(|\psi_n(\bk)\rangle)^*$, leading to 
\begin{equation}
  \hat P_n( D_U\bk)=|\psi_n(D_U \bk)\rangle \langle \psi_n(D_U \bk)| =  \mathcal{D} \left(\hat P_n(\bk)\right)^* \mathcal{D}^\dagger\,.
\end{equation}
Note that $D_U$ does include the effect of time-reversal on momentum. We obtain
\begin{align}
    \label{eqn:antiunitary}
    Q_{mn}^{ab}(D_U \bk) &= D_U^{aa'}D_U^{bb'}\left(Q_{mn}^{a'b'}(\bk)\right)^* \qquad \textrm{(antiunitary transformation)}\,.
\end{align}
Antiunitary transformations imply that the Berry curvature $\Omega_n^{ab}=-\text{Im}\,Q_{nn}^{ab}$ yields an additional sign flip, while the two-state quantum metric $g_{nm}^{ab}=\text{Re}\,Q_{nm}^{ab}$ does not.

\subsection{Symmetry transformation of nonlinear conductivity tensor}

The implications of the symmetry transformations given via Eqs.~\eqref{eqn:unitary} and \eqref{eqn:antiunitary} on the conductivity tensor $\sigma^{a;bc}$ are now almost immediate. The band energies are invariant under symmetry transformations: $E_n(D_U \bk)=E_n(\bk)$. Similarly to projectors, the velocity and second derivative of the energy then transform as
\begin{gather}
    v^a(D_U\bk) = (\partial^a E)(D_U\bk) = D_U^{aa'}\partial^{a'}\big(E(D_U \bk)\big) = D_U^{aa'}\partial^{a'}E(\bk) = D_U^{aa'}v^{a'}(\bk)\,,\\
    (\partial^a \partial^b E)(D_U\bk) = D_U^{aa'}D_U^{bb'}(\partial^a \partial^b E(\bk))\,.
\end{gather}
Starting with Eq.~\eqref{eq:suppfinalresult} for the conductivity with general indices $a,b,c$ and performing a change of variables $\bk \rightarrow D_U\bk$, the conductivity transforms as
\begin{align} 
   \sigma^{a;bc} &= \frac{e^3}{\hbar}\!\!\int\!\!\!\frac{d^d\bk}{(2\pi)^d}\sum_n \delta(E_n(\bk)-\mu) \big(\sigma^{a;bc}_{\mathrm{NLD}}(\bk)+\sigma^{a;bc}_{\mathrm{BCD}}(\bk)+\sigma^{a;bc}_{\mathrm{interQMD}}(\bk)+\sigma^{a;bc}_{\mathrm{intraQMD}}(\bk)\big)\\
   &=\frac{e^3}{\hbar}\!\!\int\!\!\!\frac{d^d\bk}{(2\pi)^d}\sum_n \delta(E_n(D_U\bk)-\mu) \big(\sigma^{a;bc}_{\mathrm{NLD}}(D_U\bk)+\sigma^{a;bc}_{\mathrm{BCD}}(D_U\bk)+\sigma^{a;bc}_{\mathrm{interQMD}}(D_U\bk)+\sigma^{a;bc}_{\mathrm{intraQMD}}(D_U\bk)\big)\\
   &=D_U^{aa'}D_U^{bb'}D_U^{cc'}\frac{e^3}{\hbar}\!\!\int\!\!\!\frac{d^d\bk}{(2\pi)^d}\sum_n \delta(E_n(\bk)-\mu) \big(\sigma^{a';b'c'}_{\mathrm{NLD}}(\bk)+\eta \sigma^{a';b'c'}_{\mathrm{BCD}}(\bk)+\sigma^{a';b'c'}_{\mathrm{interQMD}}(\bk)+\sigma^{a';b'c'}_{\mathrm{intraQMD}}(\bk)\big)\,,
\end{align}
where $\eta=1$ ($\eta=-1$) for unitary (antiunitary) transformations.

The equation above reduces to the constraint
\begin{equation}\label{eq:conductivitySymconstraint}
   \sigma^{a;bc}=  D_U^{aa'}D_U^{bb'}D_U^{cc'}\sigma^{a';b'c'}_{\mathrm{NLD}}+\eta D_U^{aa'}D_U^{bb'}D_U^{cc'}\sigma^{a';b'c'}_{\mathrm{BCD}}+D_U^{aa'}D_U^{bb'}D_U^{cc'}\sigma^{a';b'c'}_{\mathrm{interQMD}}+D_U^{aa'}D_U^{bb'}D_U^{cc'}\sigma^{a';b'c'}_{\mathrm{intraQMD}} \,.
\end{equation}
This tensorial equation yields a system of 27 individual equations (one for each of the tensor components).

Additionally, further constraints for each of the contributions appear due to their specific form (see Eq.~\eqref{eq:suppfinalresult}). Indeed,
\begin{align} \label{eq:conductivitiesAdditionalconstraints}
    \sigma^{a;bc}_{\mathrm{NLD}} &= \sigma^{(a;bc)}_{\mathrm{NLD}}\,, \\
    \sigma^{a;aa}_{\mathrm{BCD}} &= \sigma^{a;aa}_{\mathrm{interQMD}} = 0\,.
\end{align}
$(\ldots)$ indicates symmetrization here. Finally, the entire tensor is symmetric in its last two indices: $\sigma^{a;bc} = \sigma^{a;cb}$, which reduces the number of independent components from 27 to 18.

We begin by listing the result of these constraints for a few common symmetries in Table \ref{table:suppSymmetries}. We follow the syntax
\begin{equation} \label{eq:sigmatensor}
    \begin{pmatrix}
\sigma^{x;xx} & \sigma^{x;yy} & \sigma^{x;zz}\\
\sigma^{y;xx} & \sigma^{y;yy} & \sigma^{y;zz}\\
\sigma^{z;xx} & \sigma^{z;yy} & \sigma^{z;zz}
\end{pmatrix}\,,
\end{equation}
Terms with electric-field direction, such as $\sigma^{x;xy}$, are also constrained. Here, we continue to focus on the experimentally most accessible case $\sigma^{a;bb}$.
indicating by a zero that a symmetry prevents a certain conductivity, and maximally reduce the number of independent components of the conductivity tensor appearing.

\begin{table}[hb]
\caption{\label{table:suppSymmetries} Observable components of the conductivity tensor $\sigma^{a;bc}$, separated into the 4 components NLD, BCD, interQMD and intraQMD (see \ref{suppsec:symmetries}). We also show the representation $D_U$ of the symmetry used. See Eq.~\ref{eq:sigmatensor} for the definition of matrix components.}
\begin{center}
\resizebox{\columnwidth}{!}{%
\begin{tabular}{|Sc|Sc|Sc|Sc|Sc|Sc|}
\hline
Name              & $D_U$ & $\sigma_{\mathrm{NLD}}$ & $\sigma_{\mathrm{BCD}}$ & $\sigma_{\mathrm{interQMD}}$ & $\sigma_{\mathrm{intraQMD}}$ \\ \hline
$\mathcal{C}_2^z$ & $\left(\begin{array}{ccc} -1 & 0 & 0 \\ 0 & -1 & 0 \\ 0 & 0 & 1 \\\end{array}\right)$ & $\left(\begin{array}{ccc} 0 & 0 & 0 \\ 0 & 0 & 0 \\ \sigma^{z;xx}\
 
  & \sigma^{z;yy} & \sigma^{z;zz} \\\end{array}\right)$ & $\left(\begin{array}{ccc} 0 & 0 & 0 \\ 0 & 0 & 0 \\ \sigma^{z;xx} & \sigma^{z;yy} & 0 \\\end{array}\right)$ &\
 
  $\left(\begin{array}{ccc} 0 & 0 & 0 \\ 0 & 0 & 0 \\ \sigma^{z;xx} & \sigma^{z;yy} & 0 \\\end{array}\right)$ & $\left(\begin{array}{ccc} 0 & 0 & 0 \\ 0 & 0 & 0 \\\
 
  \sigma^{z;xx} & \sigma^{z;yy} & \sigma^{z;zz} \\\end{array}\right)$ \\ \hline
  $\mathcal{C}_2^z\mathcal{T}$ & $\left(\begin{array}{ccc} 1 & 0 & 0 \\ 0 & 1 & 0 \\ 0 & 0 & -1 \\\end{array}\right)$ & $\left(\begin{array}{ccc} \sigma^{x;xx} & \sigma^{x;yy}\
 
  & \sigma^{x;zz} \\ \sigma^{y;xx} & \sigma^{y;yy} & \sigma^{y;zz} \\ 0 & 0 & 0 \\\end{array}\right)$ & $\left(\begin{array}{ccc} 0 & 0 & 0 \\ 0 & 0 & 0 \\ \sigma^{z;xx} &\
 
  \sigma^{z;yy} & 0 \\\end{array}\right)$ & $\left(\begin{array}{ccc} 0 & \sigma^{x;yy} & \sigma^{x;zz} \\ \sigma^{y;xx} & 0 & \sigma^{y;zz} \\ 0 & 0 & 0\
 
  \\\end{array}\right)$ & $\left(\begin{array}{ccc} \sigma^{x;xx} & \sigma^{x;yy} & \sigma^{x;zz} \\ \sigma^{y;xx} & \sigma^{y;yy} & \sigma^{y;zz} \\ 0 & 0 & 0\
 
  \\\end{array}\right)$ \\ \hline
  $\mathcal{C}_3^z$ & $\left(\begin{array}{ccc} -\frac{1}{2} & -\frac{\sqrt{3}}{2} & 0 \\ \frac{\sqrt{3}}{2} & -\frac{1}{2} & 0 \\ 0 & 0 & 1 \\\end{array}\right)$ &\
 
  $\left(\begin{array}{ccc} -\sigma^{x;yy} & \sigma^{x;yy} & 0 \\ -\sigma^{y;yy} & \sigma^{y;yy} & 0 \\ \sigma^{z;yy} & \sigma^{z;yy} & \sigma^{z;zz} \\\end{array}\right)$ &\
 
  $\left(\begin{array}{ccc} 0 & 0 & 0 \\ 0 & 0 & 0 \\ \sigma^{z;yy} & \sigma^{z;yy} & 0 \\\end{array}\right)$ & $\left(\begin{array}{ccc} 0 & 0 & 0 \\ 0 & 0 & 0 \\\
 
  \sigma^{z;yy} & \sigma^{z;yy} & 0 \\\end{array}\right)$ & $\left(\begin{array}{ccc} -\sigma^{x;yy} & \sigma^{x;yy} & 0 \\ -\sigma^{y;yy} & \sigma^{y;yy} & 0 \\\
 
  \sigma^{z;yy} & \sigma^{z;yy} & \sigma^{z;zz} \\\end{array}\right)$ \\ \hline
  $\mathcal{C}_3^z\mathcal{T}$ & $\left(\begin{array}{ccc} \frac{1}{2} & \frac{\sqrt{3}}{2} & 0 \\ -\frac{\sqrt{3}}{2} & \frac{1}{2} & 0 \\ 0 & 0 & -1 \\\end{array}\right)$ &\
 
  $\left(\begin{array}{ccc} 0 & 0 & 0 \\ 0 & 0 & 0 \\ 0 & 0 & 0 \\\end{array}\right)$ & $\left(\begin{array}{ccc} 0 & 0 & 0 \\ 0 & 0 & 0 \\ \sigma^{z;yy} & \sigma^{z;yy} & 0\
 
  \\\end{array}\right)$ & $\left(\begin{array}{ccc} 0 & 0 & 0 \\ 0 & 0 & 0 \\ 0 & 0 & 0 \\\end{array}\right)$ & $\left(\begin{array}{ccc} 0 & 0 & 0 \\ 0 & 0 & 0 \\ 0 & 0 & 0\
 
  \\\end{array}\right)$ \\ \hline
  $\mathcal{C}_4^z$ & $\left(\begin{array}{ccc} 0 & -1 & 0 \\ 1 & 0 & 0 \\ 0 & 0 & 1 \\\end{array}\right)$ & $\left(\begin{array}{ccc} 0 & 0 & 0 \\ 0 & 0 & 0 \\ \sigma^{z;yy} &\
 
  \sigma^{z;yy} & \sigma^{z;zz} \\\end{array}\right)$ & $\left(\begin{array}{ccc} 0 & 0 & 0 \\ 0 & 0 & 0 \\ \sigma^{z;yy} & \sigma^{z;yy} & 0 \\\end{array}\right)$ &\
 
  $\left(\begin{array}{ccc} 0 & 0 & 0 \\ 0 & 0 & 0 \\ \sigma^{z;yy} & \sigma^{z;yy} & 0 \\\end{array}\right)$ & $\left(\begin{array}{ccc} 0 & 0 & 0 \\ 0 & 0 & 0 \\\
 
  \sigma^{z;yy} & \sigma^{z;yy} & \sigma^{z;zz} \\\end{array}\right)$ \\ \hline
  $\mathcal{C}_4^z\mathcal{T}$ & $\left(\begin{array}{ccc} 0 & 1 & 0 \\ -1 & 0 & 0 \\ 0 & 0 & -1 \\\end{array}\right)$ & $\left(\begin{array}{ccc} 0 & 0 & 0 \\ 0 & 0 & 0 \\\
 
  -\sigma^{z;yy} & \sigma^{z;yy} & 0 \\\end{array}\right)$ & $\left(\begin{array}{ccc} 0 & 0 & 0 \\ 0 & 0 & 0 \\ \sigma^{z;yy} & \sigma^{z;yy} & 0 \\\end{array}\right)$ &\
 
  $\left(\begin{array}{ccc} 0 & 0 & 0 \\ 0 & 0 & 0 \\ -\sigma^{z;yy} & \sigma^{z;yy} & 0 \\\end{array}\right)$ & $\left(\begin{array}{ccc} 0 & 0 & 0 \\ 0 & 0 & 0 \\\
 
  -\sigma^{z;yy} & \sigma^{z;yy} & 0 \\\end{array}\right)$ \\ \hline
  $\mathcal{M}_x$ & $\left(\begin{array}{ccc} -1 & 0 & 0 \\ 0 & 1 & 0 \\ 0 & 0 & 1 \\\end{array}\right)$ & $\left(\begin{array}{ccc} 0 & 0 & 0 \\ \sigma^{y;xx} & \sigma^{y;yy}\
 
  & \sigma^{y;zz} \\ \sigma^{z;xx} & \sigma^{z;yy} & \sigma^{z;zz} \\\end{array}\right)$ & $\left(\begin{array}{ccc} 0 & 0 & 0 \\ \sigma^{y;xx} & 0 & \sigma^{y;zz} \\\
 
  \sigma^{z;xx} & \sigma^{z;yy} & 0 \\\end{array}\right)$ & $\left(\begin{array}{ccc} 0 & 0 & 0 \\ \sigma^{y;xx} & 0 & \sigma^{y;zz} \\ \sigma^{z;xx} & \sigma^{z;yy} & 0\
 
  \\\end{array}\right)$ & $\left(\begin{array}{ccc} 0 & 0 & 0 \\ \sigma^{y;xx} & \sigma^{y;yy} & \sigma^{y;zz} \\ \sigma^{z;xx} & \sigma^{z;yy} & \sigma^{z;zz}\
 
  \\\end{array}\right)$ \\ \hline
  $\mathcal{M}_x\mathcal{T}$ & $\left(\begin{array}{ccc} 1 & 0 & 0 \\ 0 & -1 & 0 \\ 0 & 0 & -1 \\\end{array}\right)$ & $\left(\begin{array}{ccc} \sigma^{x;xx} & \sigma^{x;yy} &\
 
  \sigma^{x;zz} \\ 0 & 0 & 0 \\ 0 & 0 & 0 \\\end{array}\right)$ & $\left(\begin{array}{ccc} 0 & 0 & 0 \\ \sigma^{y;xx} & 0 & \sigma^{y;zz} \\ \sigma^{z;xx} & \sigma^{z;yy} &\
 
  0 \\\end{array}\right)$ & $\left(\begin{array}{ccc} 0 & \sigma^{x;yy} & \sigma^{x;zz} \\ 0 & 0 & 0 \\ 0 & 0 & 0 \\\end{array}\right)$ & $\left(\begin{array}{ccc}\
 
  \sigma^{x;xx} & \sigma^{x;yy} & \sigma^{x;zz} \\ 0 & 0 & 0 \\ 0 & 0 & 0 \\\end{array}\right)$ \\ \hline
  $\mathcal{P}\mathcal{T}$ & $\left(\begin{array}{ccc} 1 & 0 & 0 \\ 0 & 1 & 0 \\ 0 & 0 & 1 \\\end{array}\right)$ & $\left(\begin{array}{ccc} \sigma^{x;xx} & \sigma^{x;yy} &\
 
  \sigma^{x;zz} \\ \sigma^{y;xx} & \sigma^{y;yy} & \sigma^{y;zz} \\ \sigma^{z;xx} & \sigma^{z;yy} & \sigma^{z;zz} \\\end{array}\right)$ & $\left(\begin{array}{ccc} 0 & 0 & 0\
 
  \\ 0 & 0 & 0 \\ 0 & 0 & 0 \\\end{array}\right)$ & $\left(\begin{array}{ccc} 0 & \sigma^{x;yy} & \sigma^{x;zz} \\ \sigma^{y;xx} & 0 & \sigma^{y;zz} \\ \sigma^{z;xx} &\
 
  \sigma^{z;yy} & 0 \\\end{array}\right)$ & $\left(\begin{array}{ccc} \sigma^{x;xx} & \sigma^{x;yy} & \sigma^{x;zz} \\ \sigma^{y;xx} & \sigma^{y;yy} & \sigma^{y;zz} \\\
 
  \sigma^{z;xx} & \sigma^{z;yy} & \sigma^{z;zz} \\\end{array}\right)$ \\ \hline
\end{tabular}
}
\end{center}
\end{table}

In magnetic space groups (MSGs), however, multiple symmetries have to be taken into account simultaneously. We do this algorithmically, using data from~\cite{isomag}. A similar treatment was done in Ref.~\cite{zhu_magnetic_2025} for spin space groups, but with a different formula for the nonlinear conductivity, which changes results. Indeed, as mentioned above, the interQMD contribution is purely transverse, imposing further constraints. For each MSG, we simultaneously impose the constraints enforced by each symmetry to obtain the nonzero tensor components of $\sigma^{a;bc}$ for each of the four contributions. This allows us to sort through all 1651 MSGs, imposing certain conditions, of which we present relevant choices below. 

The first three cases are designed to isolate contributions from NLD, inter- and intraQMD together, or intraQMD and NLD alone. In the fourth case, we additionally impose a MSG which enforces a nodal plane~\cite{wilde_Nature_21}. In all cases shown below, the NLD contribution has the same constraints as the intraQMD contribution. Finally, the MAGNDATA database~\cite{gallego_magndata_2016, gallego_magndata_2016-1} is used to find experimentally verified materials belonging to the obtained MSGs. In general, the database does not provide information as to whether a material is metallic or an insulator (and thus does or does not have an NLHE). Therefore, in this SM we show all materials from the database regardless of whether they are insulating or not.
In the following, we indicate by $\sigma_{\ldots}\neq 0$ that a certain contribution is not constrained to be zero by magnetic symmetries.

\subsection{Magnetic space groups with no BCD}
We begin by simply imposing that the BCD is zero. To avoid MSGs which have no nonlinear conductivity at all, we also impose a nonzero intraQMD:
\begin{align}
    \sigma_{\mathrm{BCD}}=0,\quad \sigma_{\mathrm{intraQMD}}\neq 0\,.
\end{align}
208 MSGs, shown in Table~\ref{supptableSymms1}, are compatible with this constraint. In all cases the NLD is also nonzero, and in most cases the interQMD is as well (see below). Many materials (175 as of writing) in the MAGNDATA~\cite{gallego_magndata_2016, gallego_magndata_2016-1} database have phases belonging to one of these MSGs. Some examples are \ce{EuTiO3}, \ce{MnTiO3}, \ce{Mn2Au}, \ce{MnPS3}, \ce{Co4Nb2O9}, \ce{Ca2MnO4}, \ce{CuMnAs}, \ce{MgMnO3}, \ce{EuMnSb2}, \ce{Fe4Nb2O9}, \ce{KFeO2}, \ce{Co4Ta2O9}, \ce{KFeS2}, \ce{MnNb3S6},
  \ce{VNb3S6}, \ce{MnPd2}, \ce{MnNb2O6}, \ce{MnTa2O6}, \ce{SrMnO3}, \ce{ThMn2}, \ce{FeSn2}, \ce{FeGe2} and \ce{CoNb3S6}. Recently, the material \ce{MnBi2Te4} was connected to NLHE contributions beyond the BCD~\cite{doi:10.1126/science.adf1506,wang_quantum-metric-induced_2023}.  This material, however, breaks inversion symmetry only in the thin even-layered case~\cite{doi:10.1126/sciadv.aaw5685}.
\begin{table}[htb]
    \caption{List of MSGs enforcing the conditions $\sigma_{\mathrm{BCD}}=0,$ and $\sigma_{\mathrm{intraQMD}}\neq 0$ on the nonlinear conductivity.}
    \label{supptableSymms1}
\addtolength{\tabcolsep}{5pt}  
    \begin{center}
    \begin{tabular}{ccccccccc} \hline
         2.6 & 10.44 & 10.45 & 11.52 & 11.53 & 12.60 & 12.61 & 13.67 & 13.68\\14.77 & 14.78 & 15.87 & 15.88 & 16.3 & 17.9 & 17.10 & 18.18 & 18.19\\19.27 & 20.33 & 20.34 & 21.40 & 21.41 & 22.47 & 23.51 & 24.55 & 47.251\\48.259 & 49.267 & 49.268 & 50.279 & 50.280 & 51.291 & 51.292 & 51.293 & 52.307\\52.308 & 52.309 & 53.323 & 53.324 & 53.325 & 54.339 & 54.340 & 54.341 & 55.355\\55.356 & 56.367 & 56.368 & 57.379 & 57.380 & 57.381 & 58.395 & 58.396 & 59.407\\59.408 & 60.419 & 60.420 & 60.421 & 61.435 & 62.443 & 62.444 & 62.445 & 63.459\\63.460 & 63.461 & 64.471 & 64.472 & 64.473 & 65.483 & 65.484 & 66.493 & 66.494\\67.503 & 67.504 & 68.513 & 68.514 & 69.523 & 70.529 & 71.535 & 72.541 & 72.542\\73.550 & 74.556 & 74.557 & 83.46 & 83.47 & 84.54 & 84.55 & 85.62 & 85.63\\86.70 & 86.71 & 87.78 & 87.79 & 88.84 & 88.85 & 89.90 & 89.91 & 90.98\\90.99 & 91.106 & 91.107 & 92.114 & 92.115 & 93.122 & 93.123 & 94.130 & 94.131\\95.138 & 95.139 & 96.146 & 96.147 & 97.154 & 97.155 & 98.160 & 98.161 & 111.253\\111.255 & 112.261 & 112.263 & 113.269 & 113.271 & 114.277 & 114.279 & 121.329 & 121.331\\122.335 & 122.337 & 123.341 & 123.346 & 124.353 & 124.358 & 125.365 & 125.370 & 126.377\\126.382 & 127.389 & 127.394 & 128.401 & 128.406 & 129.413 & 129.418 & 130.425 & 130.430\\131.437 & 131.442 & 132.449 & 132.454 & 133.461 & 133.466 & 134.473 & 134.478 & 135.485\\135.490 & 136.497 & 136.502 & 137.509 & 137.514 & 138.521 & 138.526 & 139.533 & 139.538\\140.543 & 140.548 & 141.553 & 141.558 & 142.563 & 142.568 & 147.15 & 148.19 & 149.21\\149.23 & 150.27 & 151.29 & 151.31 & 152.35 & 153.37 & 153.39 & 154.43 & 155.47\\162.75 & 162.76 & 163.81 & 163.82 & 164.87 & 165.93 & 166.99 & 167.105 & 174.133\\174.135 & 175.139 & 175.140 & 176.145 & 176.146 & 177.153 & 178.159 & 179.165 & 180.171\\181.177 & 182.183 & 187.212 & 188.218 & 189.223 & 190.229 & 191.235 & 192.245 & 193.255\\ \hline
    \end{tabular}
\end{center}
\addtolength{\tabcolsep}{-5pt} 
\end{table}

\subsection{Magnetic space groups with no BCD and no interQMD}
\noindent To isolate the new intraQMD contribution, one can impose the stronger condition
\begin{align}
    \sigma_{\mathrm{BCD}}=0,\quad \sigma_{\mathrm{interQMD}}=0,\quad \sigma_{\mathrm{intraQMD}}\neq 0\,.
\end{align}
Only the following MSGs are compatible:
        \noindent 149.21, 151.29, 153.37, 162.76, 163.82, 174.133, 175.139, 176.145.

  In all cases, $\sigma_{\mathrm{NLD}}=0$ is not constrained to be zero. The only material candidates from the MAGNDATA database~\cite{gallego_magndata_2016, gallego_magndata_2016-1} that have phases belonging to one of these MSGs are      \ce{U14Au51} (in MSG 175.139) and \ce{K2Mn3(VO4)2CO3} (in MSG 176.145).

\subsection{Magnetic space groups with no BCD and no interQMD in a plane}
For the intraQMD (and NLD) contributions to be theoretically distinguishable, the previous condition can be relaxed to be true only in a plane:  
\begin{align}
    \sigma_{\mathrm{BCD}}=0\mathrm{\ in\ } xy \mathrm{\ plane},\quad \sigma_{\mathrm{interQMD}}=0\mathrm{\ in\ } xy \mathrm{\ plane},\quad \sigma_{\mathrm{intraQMD}}\neq 0\mathrm{\ in\ } xy \mathrm{\ plane}\,.
\end{align}
Now the following longer list of MSGs is compatible:
\noindent 147.15, 148.19, 149.21, 149.23, 151.29, 151.31, 153.37, 153.39, \
162.75, 162.76, 163.81, 163.82, 174.133, 175.139, 176.145.

Material candidates from the MAGNDATA database~\cite{gallego_magndata_2016, gallego_magndata_2016-1} with phases in these MSGs are  \ce{[Na(OH2)3]Mn(NCS)3}  (in MSG 147.15),
\ce{MnTiO3}, \ce{MnGeO3}, \ce{MgMnO3}, \ce{CoTe6O13}, \ce{Yb3Pt4} (all in MSG 148.19), \ce{U14Au51} (in MSG 175.139), and \ce{K2Mn3(VO4)2CO3} (in MSG 176.145).

Constraining to the $xz$ or $yz$ plane instead does not yield any new MSGs. \ce{U14Au51} and \ce{K2Mn3(VO4)2CO3} verify the constraint in all planes.

\subsection{Nodal planes and QMD-driven NLHE}
In~\cite{wilde_Nature_21}, a list of MSGs that enforce nodal planes is provided. Of these groups, the following enforce $\sigma_{\mathrm{BCD}}=0$ while being compatible with $\sigma_{\mathrm{intraQMD}}\neq 0$: 17.10, 18.18, 18.19, 19.27, 20.34, 90.98, 90.99, 92.114, 92.115, 94.130, 94.131, 96.146, 96.147, 113.269, 113.271, 114.277, 114.279.

All of them also do not constrain $\sigma_{\mathrm{interQMD}}$ to be 0 along a direction in which $\sigma_{\mathrm{intraQMD}}$ is nonzero. Material candidates from the MAGNDATA database~\cite{gallego_magndata_2016, gallego_magndata_2016-1} that belong to these MSGs are 
\ce{Ca2CoSi2O7} (in MSG 18.19), \ce{Cu3Mo2O9}, \ce{BaCrF5}, \ce{TbFeO3}, \ce{C7H14NFeCl4} (all in MSG 19.27), \ce{MnNb3S6}, \ce{SrMnO3}, \ce{RbNiCl3}, \ce{CsNiCl3} (all in MSG 20.34), \ce{Nd5Si4}, and \ce{Yb2Ge2O7} (both in MSG 92.114).

\section{Particle-hole and \texorpdfstring{$\mathcal{C}\mathcal{P}\mathcal{T}$}{CPT} Symmetry}
In this section, we discuss the implications of particle-hole (PH) $\mathcal{C}$ as well as $\mathcal{C}\mathcal{P}\mathcal{T}$ symmetry on the contributions to the nonlinear conductivity. 
Topological semimetals, with band crossings  close to the Fermi level, in general, exhibit
an approximate PH and/or 
$\mathcal{C}\mathcal{P}\mathcal{T}$ symmetry. We discuss below that the nonlinear Hall conductivity of these materials must then be odd in $\mu$ with respect to the charge neutrality point,
except for the BCD in $\mathcal{C}\mathcal{P}\mathcal{T}$ symmetric systems, which is even.

\subsection{PH Symmetry} \label{suppsec:PH}
A system has PH symmetry ($\mathcal{C}$) if the Hamiltonian transforms as
\begin{equation}
    \mathcal{C}_U^\dagger H^{*}(-\bk)\mathcal{C}_U = - H(\bk)\,,
\end{equation}
where $\mathcal{C}_U$ is a unitary operator such that $\mathcal{C} = \mathcal{C}_U\mathcal{K}$, taken to be independent of $\bk$ (see \ref{suppsec:symmetries}). Consequently, if $|\psi_n(\bk)\rangle$ is an eigenstate of $H(\bk)$ with eigenvalue $E_n$, $\mathcal{C}_U|\psi_n(\bk)\rangle$ is an eigenstate of $H^*(-\bk)$
\begin{equation}
    H^*(-\bk)\mathcal{C}_U|\psi_n(\bk)\rangle = - \mathcal{C}_UH |\psi_n(\bk)\rangle = -E_n \mathcal{C}_U|\psi_n(\bk)\rangle\,,
\end{equation}
 with opposite eigenvalue $E_{-n}:=-E_n$. Thus, $H(-\bk)\left(\mathcal{C}_U|\psi_n(\bk)\rangle\right)^* = E_{-n}\left(\mathcal{C}_U|\psi_n(\bk)\rangle\right)^*$, and up to gauge $|\psi_{-n}(-\bk)\rangle=\left(\mathcal{C}_U|\psi_n(\bk)\rangle\right)^*$. It follows that projectors transform as
\begin{equation}
    \mathcal{C}_U \hat P_n(\bk)\mathcal{C}_U^\dagger = \hat P_{-n}^{*}(-\bk)\,.
\end{equation}
With these properties in hand, we now make the change of variables $\bk \rightarrow - \bk$ in the equation for conductivity:
\begin{align}
   \sigma^{a;bb} &= \frac{e^3}{\hbar}\!\!\int\!\!\!\frac{d^d\bk}{(2\pi)^d}\sum_n \delta(E_n(\bk)-\mu) \Bigg(-\frac{1}{(2\Gamma)^2} \frac{v_n^{a}(\bk)}{m^{bb}_n(\bk)}  \text{tr}[\hat P_n(\bk)]+\frac{1}{2\Gamma}\,v_n^{b}(\bk)\, \Omega_{n}^{ab}(\bk)\nonumber \\ &\hspace{20mm}+2\!\!\sum_{m \neq n} \!\!\frac{v_n^{b}(\bk)\,g_{mn}^{ab}(\bk)- v_n^{a}(\bk)\,g_{mn}^{bb}(\bk)}{\epsilon_{mn}(\bk)}-\frac{1}{2}\partial^a\,g^{bb}_n(\bk)\Bigg) \\
   &= \frac{e^3}{\hbar}\!\!\int\!\!\!\frac{d^d\bk}{(2\pi)^d}\sum_n \delta(E_{-n}(\bk)-(-\mu)) \Bigg(\frac{1}{(2\Gamma)^2} \frac{v_{-n}^{a}(\bk)}{m^{bb}_{-n}(\bk)}  \text{tr}[\hat P_{-n}(\bk)]+\frac{1}{2\Gamma}\,v_{-n}^{b}(\bk)\, \left(\Omega_{-n}^{ab}(\bk)\right)^*\nonumber \\ &\hspace{20mm}+2\!\!\sum_{m \neq n} \!\!\frac{v_{-n}^{b}(\bk)\,\left(g_{(-m)(-n)}^{ab}(\bk)\right)^* - v_{-n}^{a}(\bk)\,\left(g_{(-m)(-n)}^{bb}(\bk)\right)^*}{-\epsilon_{(-m)(-n)}(\bk)}+\frac{1}{2}\partial^a\,\left(g^{bb}_{-n}(\bk)\right)^*\Bigg)\,.
\end{align}
We used $v_{n}^{a}(-\bk)= \partial^a E_n(-\bk) = v_{-n}^{a}(\bk)$ and $m_{n}^{bb}(-\bk)= -m_{-n}^{bb}(\bk)$ (the change of variables also applies to $\partial_{\bk}$). Now, $g_{mn}(\bk)$, $g_{n}(\bk)$ are real and $\Omega_{n}^{ab}(\bk)$ is purely imaginary:
\begin{align}
   \sigma^{a;bb} &= \frac{e^3}{\hbar}\!\!\int\!\!\!\frac{d^d\bk}{(2\pi)^d}\sum_n \delta(E_{-n}(\bk)-(-\mu)) \Bigg(\frac{1}{(2\Gamma)^2} \frac{v_{-n}^{a}(\bk)}{m^{bb}_{-n}(\bk)}  \text{tr}[\hat P_{-n}(\bk)]-\frac{1}{2\Gamma}\,v_{-n}^{b}(\bk)\, \Omega_{-n}^{ab}(\bk)\nonumber \\ &\hspace{20mm}-2\!\!\sum_{m \neq n} \!\!\frac{v_{-n}^{b}(\bk)\,(g_{(-m)(-n)}^{ab}(\bk) - v_{-n}^{a}(\bk)\,g_{(-m)(-n)}^{bb}(\bk)}{\epsilon_{(-m)(-n)}(\bk)}+\frac{1}{2}\partial^a\,g^{bb}_{-n}(\bk)\Bigg)\,.
\end{align}
Finally, we relabel the sums by $-n \rightarrow n$, $-m \rightarrow m$ and extract the overall $-$ sign:
\begin{align}
   \sigma^{a;bb} &= -\frac{e^3}{\hbar}\!\!\int\!\!\!\frac{d^d\bk}{(2\pi)^d}\sum_n \delta(E_{n}(\bk)-(-\mu)) \Bigg(-\frac{1}{(2\Gamma)^2} \frac{v_{n}^{a}(\bk)}{m^{bb}_{n}(\bk)}  \text{tr}[\hat P_{n}(\bk)]+\frac{1}{2\Gamma}\,v_{n}^{b}(\bk)\, \Omega_{n}^{ab}(\bk)\nonumber \\ &\hspace{20mm}+2\!\!\sum_{m \neq n} \!\!\frac{v_{n}^{b}(\bk)\,(g_{mn}^{ab}(\bk) - v_{n}^{a}(\bk)\,g_{mn}^{bb}(\bk)}{\epsilon_{mn}(\bk)}-\frac{1}{2}\partial^a\,g^{bb}_{n}(\bk)\Bigg)\,.
\end{align}
We have thus effectively proven that with PH symmetry, the conductivity verifies
\begin{equation}
    \sigma^{a;bb}(\mu) = - \sigma^{a;bb}(-\mu)\,.
\end{equation}

\subsection{\texorpdfstring{$\mathcal{C}\mathcal{P}\mathcal{T}$}{CPT} Symmetry} \label{suppsec:CPT}
A closely related symmetry is what we refer to as $\mathcal{C}\mathcal{P}\mathcal{T}$ symmetry (PH symmetry combined with $\mathcal{P}\mathcal{T}$), where the Hamiltonian satisfies
\begin{equation}
    \mathcal{S}^\dagger H(-\bk)\mathcal{S} = - H(\bk)\,,
\end{equation}
where $\mathcal{S}$ is a \emph{unitary} operator, as opposed to PH symmetry. The bands can be labeled in the same way, respecting
\begin{equation}
    E_n(\bk) = - E_{-n}(\bk), \quad n\in \{0,\ldots,N\}\,.
\end{equation}
Projectors simply transform as 
\begin{equation}
    \mathcal{S}^\dagger \hat{P}_n(\bk)\mathcal{S} = \hat{P}_{-n}(-\bk)\,.
\end{equation}
The only difference compared to the PH symmetric case is that the BCD contribution now no longer flips sign, i.e. we obtain the relations
\begin{align}
    \sigma^{a;bb}_{\mathrm{NLD}}(\mu) &= - \sigma^{a;bb}_{\mathrm{NLD}}(-\mu)\,, \\ \sigma^{a;bb}_{\mathrm{BCD}}(\mu)&=  \sigma^{a;bb}_{\mathrm{BCD}}(-\mu)\,, \\ \sigma^{a;bb}_{\mathrm{interQMD}}(\mu) &= - \sigma^{a;bb}_{\mathrm{interQMD}}(-\mu)\,, \\\sigma^{a;bb}_{\mathrm{intraQMD}}(\mu)&= - \sigma^{a;bb}_{\mathrm{intraQMD}}(-\mu)\,.
\end{align}

\section{Analytic results for low-energy models}
In this section, we begin by showing how the contributions to the NLHE are calculated analytically in this work. We then present the various low-energy models for which these were obtained, and show plots when those were omitted in the main text.

\subsection{Calculating the NLHE analytically for low-energy models}\label{suppsubsec:HowToCalculateAnalytically}

In this section, we describe how the analytic results for the conductivities were obtained. All low-energy models considered in this work have analytically diagonalizable Hamiltonians, yielding expressions for the energies and band projectors. Throughout this section, we use the 3D Dirac point as an example, the other cases being similar. From analytic expressions of the projectors, one immediately obtains various quantum geometric quantities. For example, the quantum geometric tensor $Q_{12}^{ab}$ ($1,2$ are band indices) is given by
\begin{align}
    Q_{12}^{ab} = \frac{k_y k_z v^4}{2 \left(v^2\left(k_x^2+k_y^2+k_z^2\right)+m^2\right)^2}\,.
\end{align}
From such expressions, all 4 contributions NLD, BCD, interQMD, and intraQMD before integration are obtained. For example, including the Dirac delta function and evaluating the sum over all bands, the intraQMD contribution $\sigma_{\mathrm{intraQMD}}(\bk)$ is 
\begin{align}
    \sigma_{\mathrm{intraQMD}}^{y;xx}(\bk) &= \frac{k_y v^4 \left(v^2 \left(-k_x^2+k_y^2+k_z^2\right)+m^2\right) \delta \left(k_y t-\mu +\sqrt{m^2+\left(k_x^2+k_y^2+k_z^2\right) v^2}\right)}{2 \left(v^2 \left(k_x^2+k_y^2+k_z^2\right)+m^2\right)^3}\\
    &\qquad +\frac{k_y v^4 \left(v^2 \left(-k_x^2+k_y^2+k_z^2\right)+m^2\right) \delta \left(-k_y t+\mu +\sqrt{m^2+\left(k_x^2+k_y^2+k_z^2\right) v^2}\right)}{2 \left(v^2 \left(k_x^2+k_y^2+k_z^2\right)+m^2\right)^3}\,.
\end{align}
We now treat the cases $\mu>0$ and $\mu<0$ seperately, as in each case only one of the Dirac delta functions in nonzero. Assuming for the sake of the argument that $\mu>0$, we resolve the first delta function by integrating over $k_x$, using the change of variable formula
\begin{align}
    \delta(g(x))=\sum_i \frac{\delta\left(x-x_i\right)}{\left|g^{\prime}\left(x_i\right)\right|}\,,
\end{align}
where the sum extends over all roots of $g(x)$. A solution for $k_x$ exists as long as
\begin{equation} \label{eq:supphowtointegrateboundy}
\begin{aligned}
    -\frac{\sqrt{\mu ^2-k_y^2 (v^2-t^2)-2 k_y \mu  t-m^2}}{v}&< k_z < \frac{\sqrt{\mu ^2-k_y^2 (v^2-t^2)-2 k_y \mu  t-m^2}}{v}\,,
    \\
    \frac{-\mu t -\sqrt{m^2 t^2+(\mu ^2-m^2) v^2}}{v^2-t^2} &< k_y < \frac{\mu  (-t)+\sqrt{m^2 t^2+(\mu ^2-m^2) v^2}}{v^2-t^2}\,.
\end{aligned}
\end{equation}
By solution, we mean a $k_x$ that is a root of the argument of the delta function. After this $k_x$ integration, the result is
\begin{align}
    \frac{k_y v^3 \left(k_y^2 \left(t^2-2 v^2\right)-2 k_y \mu  t-2 k_z^2 v^2+\mu ^2-2 m^2\right)}{(k_y t-\mu )^5 \sqrt{k_y^2 (t-v) (t+v)-2 k_y \mu  t-k_z^2 v^2+\mu ^2-m^2}}\,.
\end{align}
We now integrate this over $k_y$, $k_z$ using Eq.~\eqref{eq:supphowtointegrateboundy} as integration bounds. These integrals are performed using \textsc{RUBI}~\cite{Rich2018}, a package for \textsc{Mathematica}~\cite{Mathematica} for symbolic integration. Performing these steps for all contributions yields the results shown in the following sections as well as in the main text.

\subsection{NLHE of a 3D tilted Dirac and Weyl point} \label{suppsec:3DDirac}
We begin by showing how the Dirac Hamiltonian defined in the main text is derived. Defining a set of mutually anticommuiting $\gamma$ matrices
\begin{align}
    \gamma_1 = \sigma_z \otimes \sigma_x,\quad \gamma_2 = \sigma_0 \otimes \sigma_y,\quad \gamma_3 = \sigma_0 \otimes \sigma_z,\quad \gamma_4 = \sigma_x \otimes \sigma_x,\quad \gamma_5 = \sigma_y \otimes \sigma_x\,,
\end{align}
together with the indentity matrix  $\gamma_0 = \sigma_0 \otimes \sigma_0$,
we may represent inversion symmetry by $\mathcal{P} = \gamma_3$ and time-reversal by $\mathcal{T} = (-i\sigma_y \otimes \sigma_0) \mathcal{K}$, where $\mathcal{K}$ is the complex conjugation operator. The set of $\gamma$ matrices above can be expanded by defining
\begin{equation}
    \gamma_{ij} = \frac{1}{2i}\left[\gamma_i, \gamma_j\right]\,.
\end{equation}
Together, the $\gamma_{ij}$ and the $\gamma$ matrices span the space of Hermitian $4 \times 4$ matrices. All $\gamma_i$ are, in fact, $\mathcal{P}\mathcal{T}$ symmetric:
\begin{equation}
    \mathcal{P}\mathcal{T} \gamma_i \left(\mathcal{P}\mathcal{T}\right)^\dagger = \gamma_i\,.
\end{equation}
From this, it follows (by antiunitarity of $\mathcal{T}$) that the $\gamma_{ij}$ flip sign under $\mathcal{P}\mathcal{T}$: $\mathcal{P}\mathcal{T} \gamma_{ij}\left(\mathcal{P}\mathcal{T}\right)^\dagger = - \gamma_{ij}$. 
Up to rotations in $k_i$ or $\gamma_i$, a $\mathcal{P}\mathcal{T}$ symmetric tilted gapped 3D Dirac point must then be given by
\begin{equation}
 \hat H =  t k_y \gamma_0 + k_x \gamma_1 + k_y \gamma_2 + k_z \gamma_3 + m \gamma_5\,,
\end{equation}
which corresponds to the Dirac Hamiltonian shown in the main text.

The model also retains PH symmetry ($\mathcal{C}$, see Sec. \ref{suppsec:PH}) represented by $\sigma_0 \otimes \sigma_x\mathcal{K}$. This enforces that all conductivities are \emph{odd} functions of $\mu$.

For the nonlinear conductivity to be nonzero, the chemical potential must be outside the gap, which corresponds to the condition $v^2 \mu^2 > m^2(v^2-t^2)$. Within this regime, the analytic results for the Hall conductivities $\sigma^{y;xx}=\sigma^{y;zz}$ are given by
\begin{align} \label{eq:3DDiracResultNLD}
    \sigma_{\mathrm{NLD}}^{y;xx} &= \sigma(\mu)\frac{v^2 \left(\frac{2 t \sqrt{m^2 (t^2-v^2) +\mu ^2 v^2} \left(2 m^2 t^2+3 \mu ^2 v^2\right)}{m^2 t^2+\mu ^2 v^2}+\mu  \left(t^2-3 v^2\right) \log \left(\frac{\mu  v^2+t \sqrt{m^2 (t^2-v^2)+\mu ^2 v^2}}{\mu  v^2-t \sqrt{m^2 (t^2-v^2)+\mu ^2 v^2}}\right)\right)}{16 \pi ^2 \Gamma^2 t^4}\,, \\\label{eq:3DDiracResultBCD}
   \sigma_{\mathrm{BCD}}^{y;xx} &= 0\,, \\\label{eq:3DDiracResultinterQMD}
   \sigma_{\mathrm{interQMD}}^{y;xx} &= -\sigma(\mu)\frac{t \left(m^2 (t-v) (t+v)+\mu ^2 v^2\right)^{3/2}}{6 \pi ^2 \left(m^2 t^2+\mu ^2 v^2\right)^2}\,, \\\label{eq:3DDiracResultintraQMD}
   \sigma_{\mathrm{intraQMD}}^{y;xx} &= -\sigma(\mu)\frac{m^2 t v^2 \sqrt{m^2 (t^2-v^2)+\mu ^2 v^2} \left(m^2 \left(t^2+2 v^2\right)+\mu ^2 v^2\right)}{12 \pi ^2 \left(m^2 t^2+\mu ^2 v^2\right)^3}\,. 
\end{align}
$\sigma$ is the sign function here. 

\begin{figure}[htb]
\centerline{\includegraphics[width=0.69\textwidth]{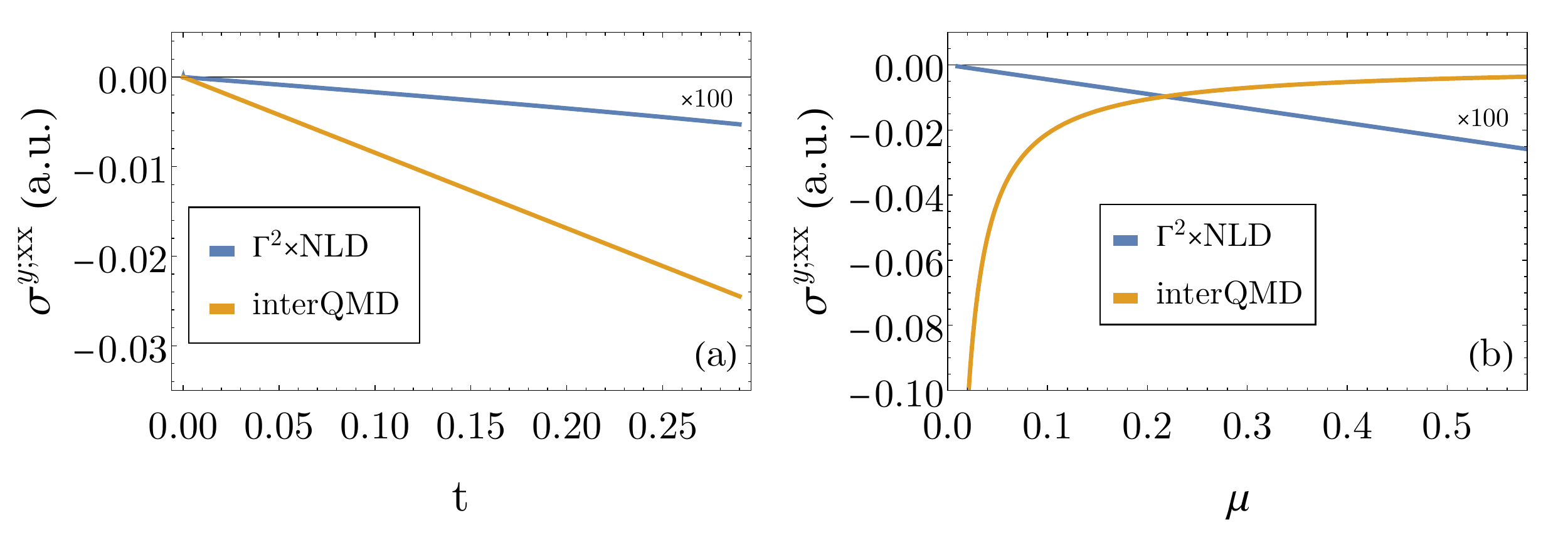}}
\caption{The dependence of contributions to the NLHE for the tilted 3D Weyl model (See Sec.~\ref{suppsec:3DDirac} and Eq.~\eqref{eq:tilted_Weyl}) on (a) $t$ and (b) $\mu$. In both plots, $v=1$. In (a) $\mu=0.1$, while in (b) $t=0.5$. The NLD contribution is proportional to $1/\Gamma^2$, depending on the material-dependent scattering rate $\Gamma$. We therefore multiply the NLD by $\Gamma^2$ to remove this factor, and multiply the result by 100 so that it is visible in the same figure. As we stated in the main text, the intraQMD contribution to the conductivity is zero.}
\label{fig:WeylPlots}
\end{figure}

Now, when the mass $m$ is set to 0, the Hamiltonian can be block diagonalized by expressing it in the basis of eigenvectors of the $\mathcal{C}\mathcal{P}\mathcal{T}$ symmetry. Indeed, in this basis, the Hamiltonian is simply
\begin{equation}
  \hat H =\left(  tk_y \sigma_0 + v(-k_x \sigma_x - k_y \sigma_y + k_z \sigma_z) \right)\oplus  \left(  tk_y \sigma_0 + v(-k_x \sigma_x - k_y \sigma_y - k_z \sigma_z) \right) + m \gamma_{52}\,.
\end{equation}
An additional change of basis $\hat H \rightarrow (\sigma_0 \otimes \sigma_z)^{-1}\hat{H}(\sigma_0 \otimes \sigma_z)$ then yields
\begin{equation} \label{eq:tilted_Weyl}
  \hat H =\left(  tk_y \sigma_0 + v(k_x \sigma_x + k_y \sigma_y + k_z \sigma_z) \right)\oplus  \left(  tk_y \sigma_0 + v(k_x \sigma_x + k_y \sigma_y - k_z \sigma_z) \right) + m \gamma_{52}\,.
\end{equation}
The sign preceding $k_z$ in the second block is inconsequential, as it may be removed by sending $k_z \rightarrow -k_z$. Thus, this Hamiltonian describes two decoupled 3D tilted Weyl points when $m=0$. We conclude that setting $m=0$ in Eqs.~\eqref{eq:3DDiracResultNLD}, \eqref{eq:3DDiracResultBCD}, \eqref{eq:3DDiracResultinterQMD} and \eqref{eq:3DDiracResultintraQMD} yields exactly twice the conductivity of a 3D Weyl point. 
The contributions to this conductivity are given by
\begin{align}\nonumber
    \sigma_{\mathrm{NLD}}^{y;xx} &= \frac{v^2 \mu  \left(3 t v+\left(t^2-3 v^2\right) \tanh ^{-1}\left(\frac{t}{v}\right)\right)}{16 \pi ^2 t^4 \Gamma ^2} \,, \\\nonumber
   \sigma_{\mathrm{BCD}}^{y;xx} &= 0 \,, \\\nonumber
   \sigma_{\mathrm{interQMD}}^{y;xx} &= -\frac{t}{12 \pi ^2 v \mu },\\ \label{eq:WeylResults}
   \sigma_{\mathrm{intraQMD}}^{y;xx} &= 0 \,.
\end{align}
The intraQMD contribution to the 3D Dirac point is proportional to the mass and therefore vanishes in this limit. 
The Weyl point model preserves $\mathcal{C}\mathcal{P}\mathcal{T}$ symmetry, while particle-hole ($\mathcal{C}$) and $\mathcal{P}\mathcal{T}$ symmetries are individually broken. As demonstrated in Sec.~\ref{suppsec:CPT}, this symmetry also enforces $\sigma^{a;bb}(-\mu) = -\sigma^{a;bb}(\mu)$ for all contributions except the BCD (which is zero in this case). Note that the NLD contribution $\sigma_{\mathrm{NLD}}^{y;xx}$ for a nontilted ($t=0$) 3D Weyl point vanishes. The interQMD contribution, $\sigma_{\mathrm{interQMD}}^{y;xx}$ is proportional to $1/\mu$, and diverges at the band crossing point. We plot the obtained conductivities for the 3D Weyl point in Fig.~\ref{fig:WeylPlots}.

\subsection{NLHE of a 3D nodal plane}
The explicit expression of the interQMD contribution (see Tab.~I in the main text) suggests that it is enhanced when two bands are close in energy and near quantum metric hot spots. Both of these features are present in a (gapped) nodal plane, 
which is a two-dimensional band degeneracy~\cite{huber_alpin_CoSi_PRL_22, zhong2016towards, Liang2016, wilde_Nature_21, bzduvsek2017robust, Wu2018,T_rker_2018, zhang2018nodal, yu2019circumventing, fu2019dirac, chang2018topological, ma2021observation, tang2022complete, alpin2023fundamental, frank2024weyl, chen2022large, li2025giant, yamada2025gapping, hu2023hierarchy, xie2021three, Xiao2020chargednodalsurface, scheie2022dirac}.
Indeed, when a perturbative symmetry-breaking term gaps the nodal plane, the two-state quantum metric increases significantly within the gap, while the band gap remains perturbatively small. We give a low-energy model for such a nodal plane, located at $k_z=0$,
\begin{equation} \label{eq:tilted_NP}
   H= \epsilon k_y \sigma_0 + v k_z \sigma_z + a k_z k_x(\sigma_y + \sigma_z) + g (\sigma_x+\sigma_z)\,.
\end{equation}
Here $\epsilon$ represents a tilting of the nodal plane, $v$ is a velocity in the $z$ direction, and $g$ controls the symmetry-breaking gap. The term proportional to $a$ is required for nonvanishing quantum geometric invariants. The model is minimally constructed to yield nonzero contributions to $\sigma^{z;xx}$. In the following, the results are independent of the chemical potential, since it can be absorbed into the tilt term by a momentum shift.

Required by the open Fermi surface, we restrict the integration in $k_z$ to the interval $[-L, L]$ and keep only the leading order behavior for $L\rightarrow \infty$. In a lattice system, such a cutoff arises naturally from the dispersion of the nodal plane. The resulting conductivities are
\begin{alignat}{2}\nonumber
    \sigma_{\mathrm{NLD}}^{z;xx} &= -\frac{ a g L}{16\pi^2 \epsilon \Gamma^2}\,,\quad &\sigma_{\mathrm{interQMD}}^{z;xx} &= \frac{  a \log \left(2+\sqrt{3}\right)}{16\pi^2 \epsilon v}\,, \\
   \sigma_{\mathrm{BCD}}^{z;xx} &= 0\,,&\sigma_{\mathrm{intraQMD}}^{z;xx} &= \frac{  a g}{16\pi^2 \epsilon L v^2 }\,.
\end{alignat}
The BCD is not symmetry-forbidden but vanishes in this minimal model; including higher-order corrections to the Hamiltonian would restore a finite contribution. The nonzero contributions are inversely proportional to the tilt of the nodal plane $\epsilon$, indicating that flatter nodal planes yield stronger contributions, which we attribute to the increased density of states. The NLD and intraQMD contributions are proportional to the gap $g$, whereas the interQMD term is independent (to leading order in $L$). We see that nodal planes are promising candidates for large, QMD-driven NHLE. A clear separation from intra- and interQMD contributions requires a more detailed analysis beyond low-energy models. Note that for degenerate nodal planes ($g=0$), the system recovers $\mathcal{C}_2\mathcal{T}$ symmetry, forcing all contributions to $\sigma^{z;xx}$ to be zero.

Indeed, for $g=0$, the model Eq.~\eqref{eq:tilted_NP} recovers a $\mathcal{C}_2\mathcal{T}$ symmetry, forcing all contributions to $\sigma^{z;xx}$ to be zero (see Tab.~\ref{table:suppSymmetries}). Analytically, however, the interQMD is not integrable in $k_z$ in this limit, because the inverse band gap factor introduces a pole at $k_z=0$. Because of this, the $k_z$ integral and the $g\rightarrow 0$ do not commute. This explains why the expression shown for the interQMD does not vanish in the $g\rightarrow 0$ limit. 

\subsection{NLHE of a 2D Dirac point} \label{suppsec:2DDirac}
\begin{figure}[h] 
\centerline{\includegraphics[width=0.69\textwidth]{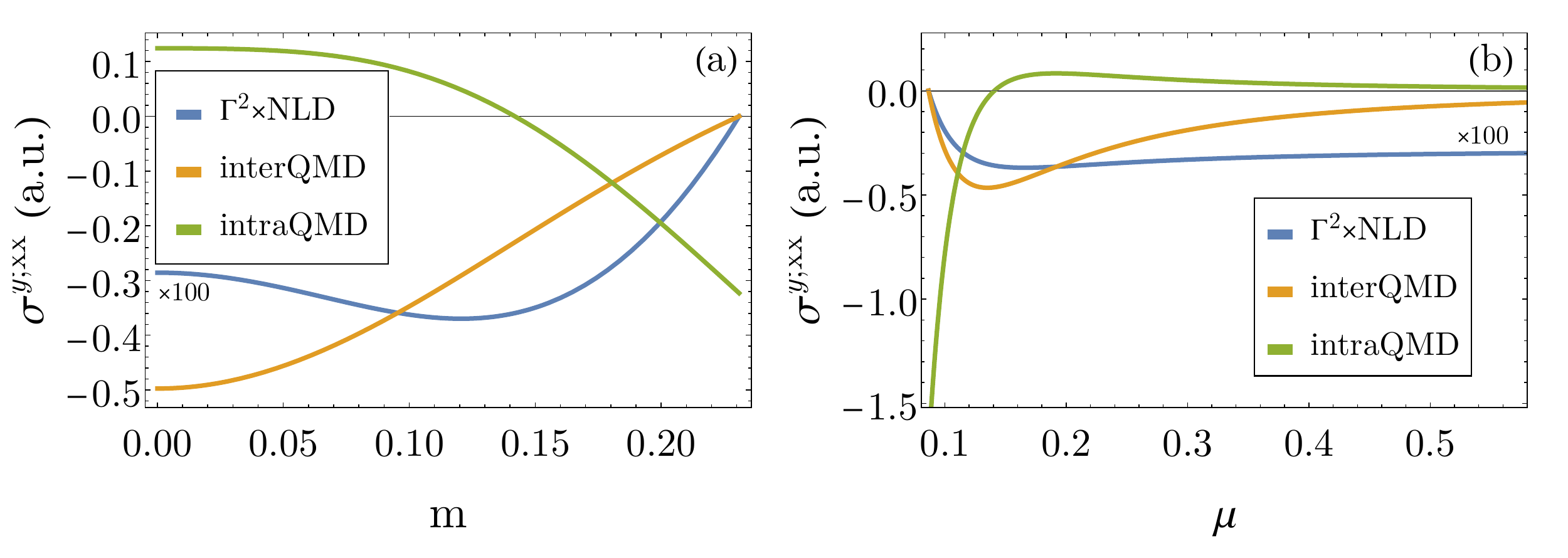}}
\caption{The dependence of contributions to the NLHE for the 2D tilted Dirac point (see Sec.~\ref{suppsec:2DDirac})  on (a) $m$ and (b) $\mu$. In both plots, $v=1$, $t=1/2$. In (a) $m$ is varied while fixing $\mu=0.2$, while in (b) $\mu$ is varied fixing $m=0.1$. The NLD contribution is proportional to $1/\Gamma^2$, depending on the material-dependent scattering rate $\Gamma$. We therefore multiply the NLD by $\Gamma^2$ to remove this factor, and multiply the result by 100 so that it is visible in the same figure. For (b) the chemical potential starts at the edge of the gap, $\mu \geq \frac{m}{v} \sqrt{v^2 -t^2}$. The BCD contribution to $\sigma_{\mathrm{BCD}}^{y;xx}$ is 0 in this case. In both cases, the plot is cut off when $\mu$ enters the mass gap.}
\label{fig:diracPlots}
\end{figure}

We now discuss the 2D case of the tilted Dirac point. 
Up to coordinate system rotations, the Hamiltonian is given by
\begin{align} \label{eq:tilted_Dirac}
    H = t k_y \sigma_0 + v k_x \sigma_y - v k_y \sigma_x + m \sigma_z\,,
\end{align}
where $t$ represents the tilt of the Dirac cone, and $m$ the mass gap. The $\sigma_i$ are Pauli matrices. We assume a type-I Dirac point, where $v>t$. The contribution is only non-zero when the chemical potential $\mu$ is above or below the gap, which is equivalent to $v^2 \mu^2 > m^2 (v^2 -t^2)$. This model was also used by Sodemann and Fu to calculate the Berry curvature dipole~\cite{sodemann_quantum_2015}. We note that the model described by Eq.~\eqref{eq:tilted_Dirac} retains a symmetry combining a mirror reflection and time-reversal, $\mathcal{M}_y \mathcal{T}$, represented in total by $\sigma_x \mathcal{K}$ where $\mathcal{K}$ is complex conjugation.
Focusing on Hall conductivities, we may obtain either $\sigma_{}^{y;xx}$ or $\sigma_{}^{x;yy}$ (one also finds nonzero $\sigma_{}^{y;yy}$). Contributions to the former are found to be
\begin{align}\nonumber
    &\sigma_{\mathrm{NLD}}^{y;xx} = \frac{\frac{\mu  m^2 t^2 \left(t^2-3 v^2\right)+\mu ^3 v^2 \left(t^2-2 v^2\right)}{\left(m^2 t^2+\mu ^2 v^2\right)^{3/2}}+2 \sigma(\mu )
   \sqrt{v^2-t^2}}{16 \pi v^{-3} t^3\Gamma^2}\,, \\\nonumber
   &\sigma_{\mathrm{BCD}}^{y;xx} = 0\,, \\\nonumber
   &\sigma_{\mathrm{interQMD}}^{y;xx} = \frac{  \mu  t v^3 \left(m^2 (v^2-t^2)-\mu ^2 v^2\right)}{8\pi \left(m^2 t^2+\mu ^2 v^2\right)^{5/2}}\,,\\
   &\sigma_{\mathrm{intraQMD}}^{y;xx} = \frac{  \mu  t v^3 \left(m^4 \left(t^4-5 v^4\right)+2 \mu ^2 m^2 t^2 v^2+\mu ^4 v^4\right)}{32\pi \left(m^2 t^2+\mu ^2 v^2\right)^{7/2}}\,,
\end{align}
where $\sigma(x)$ is the sign function, while in the latter case we recover the BCD as the only non-zero contribution. This separation is consistent with the $\mathcal{M}_y \mathcal{T}$ symmetry, see Sec.~\ref{suppsec:symmetries}. The contributions to $\sigma_{\mathrm{NLD}}^{y;xx}$ are in fact non-zero even for $m=0$, as opposed to $\sigma_{}^{x;yy}$~\cite{sodemann_quantum_2015}. These contributions are plotted as a function of $m$ and $\mu$ in Fig.~\ref{fig:diracPlots}. The system also has particle-hole (PH) symmetry. As shown in Sec.~\ref{suppsec:PH} this symmetry enforces $\sigma^{a;bb}(-\mu) = -\sigma^{a;bb}(\mu)$, which is indeed verified here. 

Consistent with the symmetry analysis summarized in Table~\ref{table:suppSymmetries}, depending on the directions measured we either observe exclusively the BCD contribution, or everything except the BCD. Under time-reversal $t\rightarrow -t$ and $m \rightarrow -m$, and equations above make apparent that $\sigma^{y;xx}$ only contributes in time-reversal symmetry broken systems. 

\subsection{NLHE of a nodal line}
\begin{figure}[H] 
\centerline{\includegraphics[width=0.69\textwidth]{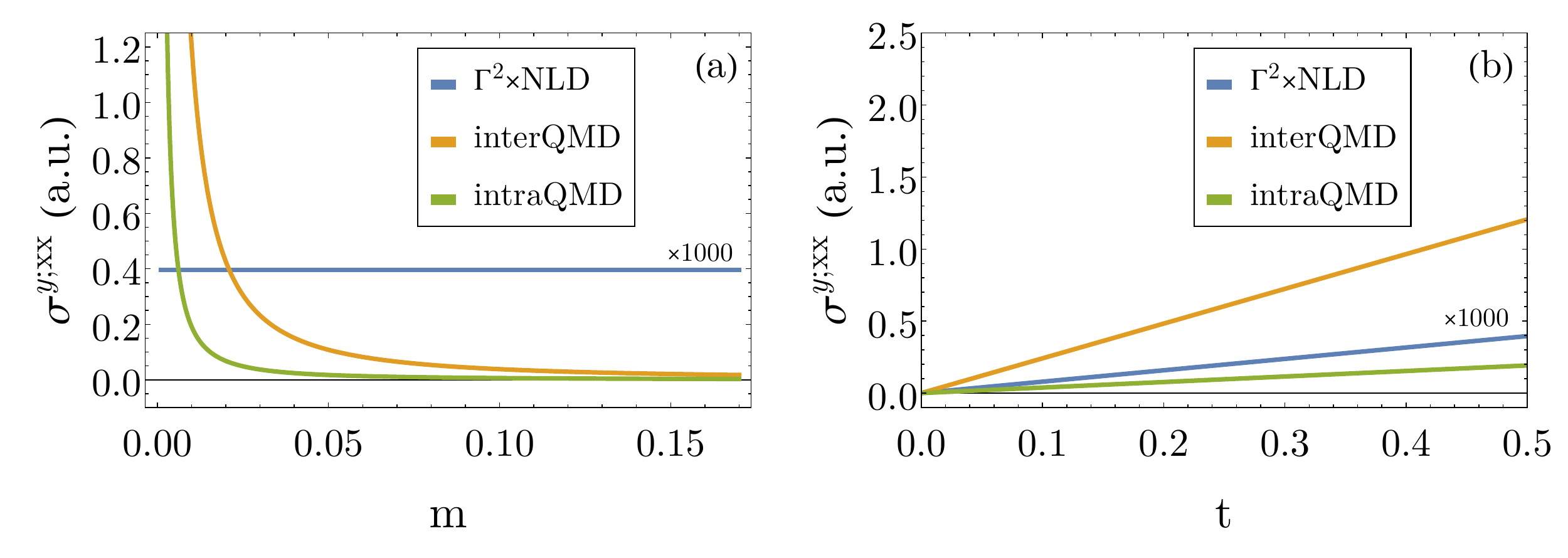}}
\label{fig:NLplots}
\caption{The dependence of contributions to the NLHE for the tilted nodal line~\eqref{eq:nodal_line} on (a) $m$ and (b) $t$. In both plots, $a=1$. For (a) $v=1$, $t=0.5$, while for (b) $v=1/2$, $m=0.01$. The NLD contribution is proportional to $1/\Gamma^2$, depending on the material-dependent scattering rate $\Gamma$. We therefore multiply the NLD by $\Gamma^2$ to remove this factor, and multiply the result by 1000 so that it is visible in the same figure. We also set $L=0.5$.}
\end{figure}

A model closely related to the 2D tilted Dirac point is that of an open nodal line. Indeed, adding a dispersion depending on $k_z$ to the Hamiltonian of the 2D (tilted) Dirac point Eq.~\eqref{eq:tilted_Dirac} leads to a nodal line on the $k_z$ axis:
\begin{align} \label{eq:nodal_line}
    H = t k_y \sigma_0 + v k_x \sigma_y - v k_y \sigma_x + m \sigma_z + \epsilon(k_z)\sigma_0\,.
\end{align}
As the simplest possible model, we choose $\epsilon(k_z) = a k_z^2$ with $a>0$, as using only a linear term results in no contribution to the NL conductivity, because the 2D Dirac point conductivities obtained are odd in $\mu$. Conveniently, this new $k_z$-dependent term can be treated as an effective chemical potential for the 2D Dirac point defined in Eq.~\eqref{eq:tilted_Dirac}, ${\mu_{\textrm{eff}}=\mu - a k_z^2}$~\cite{PhysRevB.111.L081115}. In this way, the integral over $k_z$ can be transformed into an integral over $\mu_\textrm{eff}$, with integration bounds such that the chemical potential is outside the gap. For simplicity, we only consider $\mu = 0$. Furthermore, to obtain finite results for the NLD term, the integration in $k_z$ must be limited to a range above and below the nodal line, $k_z\in [-L, L]$. A realistic choice of $L$ is system-dependent, so for clarity, we keep $L$ arbitrary and consider only the leading-order behavior as $L\rightarrow \infty$. Explicit expressions for the conductivities may then be obtained using elliptic integrals. For simplicity, we only show the leading order contributions in tilt $t$ and gap $m$ around 0 for each term. The resulting conductivities are
\begin{alignat}{2}\nonumber
    \sigma_{\mathrm{NLD}}^{y;xx} &= \frac{L t}{64 \pi ^2\Gamma^2}\,,\quad  &\sigma_{\mathrm{interQMD}}^{y;xx} &= \frac{t}{42 \pi ^2 a^{1/2}m^{3/2}}\,, \\
   \sigma_{\mathrm{BCD}}^{y;xx} &= 0\,,&\sigma_{\mathrm{intraQMD}}^{y;xx} &= \frac{t}{264 \pi ^2 a^{1/2}m^{3/2}}\,.
\end{alignat}
Note that contrary to~\cite{PhysRevB.111.L081115}, which applies a formula from~\cite{kaplan_unification_2024}, we do  not obtain any logarithmic divergence in the gap for $\sigma_{\mathrm{interQMD}}^{y;xx}$, but rather a polynomial one. This is dependent on the dispersion $\epsilon(k_z)$ chosen, for which Ref.~\cite{PhysRevB.111.L081115} used $\epsilon(k_z)= k_z+k_z^2$. The contributions are plotted in Fig.~\ref{fig:NLplots}. As for the 2D tilted Dirac point, if we had calculated $\sigma^{x;yy}$ instead, only a BCD contribution would have appeared.

As $a\rightarrow 0$ the two QMD terms diverge in an inverse square root law. That the nonlinear Drude term seems unaffected is an artifact of the expansion in $t,m$, and it equally diverges in an inverse square root law when the full expressions are taken into account. Since at $a=0$ we recover a 2D (in $k_x,k_y$) Dirac point for all $k_z$, it is natural we obtain a divergent conductivity.

\section{Efficient numerical evaluation of NLHE}
\label{SM_num_eval}
To avoid the numerical evaluation of derivatives of projectors via finite differences, which is computationally costly and potentially unstable, it is useful to have an expression for the NLHE in which derivatives act only on the Hamiltonian (not on the wave functions). These derivatives may be evaluated analytically beforehand or using automatic differentiation methods. We obtain such expressions by expanding various traces that contain only projectors (without derivatives) and derivatives of the Hamiltonian, thereby generalizing the Hellmann-Feynman theorem. To begin, the derivative of the energy can simply be calculated as
\begin{align}
    v_n^a = E_n^a = \text{tr}[\hat{P}_n (\partial^a \hat{H})]\,.
\end{align}
For $m\neq n$, the QGT can be calculated from derivatives of the Hamiltonian as 
\begin{align}
    Q_{nm}^{ab} = -\frac{1}{\epsilon_{nm}^2}\text{tr}[\hat{P}_m (\partial^{a}\hat{H})\hat{P}_n (\partial^{b}\hat{H})]\,, \quad (m \neq n)\,.
\end{align}
Both of these identities can easily be verified using Eq.~\eqref{eq:dHformulaProj}. From the latter, we immediately obtain the two-state quantum metric and the Berry curvature:
\begin{equation}
	g^{ab}_{mn} = -\frac{1}{\epsilon_{nm}^2}\,\text{Re}\,\text{tr}[\hat{P}_n (\partial^{a}\hat{H})\hat{P}_m (\partial^{b}\hat{H})]\,,
\end{equation}
and
\begin{equation}
	\Omega^{ab}_{n} = 2\, \text{Im} \sum_{m\neq n} Q_{mn}^{ab} = 2\, \text{Im} \sum_{m\neq n}-\frac{1}{\epsilon_{mn}^2}\text{tr}[\hat{P}_n (\partial^{a}\hat{H})\hat{P}_m (\partial^{b}\hat{H})]= 2\, \text{Im} \sum_{m\neq n}\frac{1}{\epsilon_{mn}^2}\text{tr}[\hat{P}_m (\partial^{a}\hat{H})\hat{P}_n (\partial^{b}\hat{H})]\,.
\end{equation}
These can be used to calculate the BCD and interQMD terms.
A useful identity can be extracted from Eq.~\eqref{eq:doubleDerHamiltonianTraceGeometric} with $n=m$:
\begin{align}
\text{tr}[\hat{P}_n(\partial^{ a }\hat{\hat{H}})\hat{P}_n (\partial^{ b }\partial^{ c }\hat{\hat{H}})]&=   E^{ a }_n E_{n}^{ b  c }\text{tr}[\hat{P}_n]- E^{ a }_n\sum_{l\neq n}\epsilon_{ln} (Q_{ln}^{ b  c }+Q_{ln}^{ c  b }) \\
&=E^{ a }_n E_{n}^{ b  c }\text{tr}[\hat{P}_n]- 2\text{tr}[\hat{P}_n (\partial^a \hat{H})]\sum_{l\neq n}\epsilon_{ln} \text{Re}\,Q_{ln}^{ b  c }\,.
\end{align}
Substituting one of the energy derivatives and the QGT by the traces above,
\begin{align}
\text{tr}[\hat{P}_n(\partial^{ a }\hat{\hat{H}})\hat{P}_n (\partial^{ b }\partial^{ c }\hat{\hat{H}})]=&   E^{ a }_n E_{n}^{ b  c }\text{tr}[\hat{P}_n]+ 2\text{tr}[\hat{P}_n (\partial^a \hat{H})]\sum_{l\neq n}\frac{1}{\epsilon_{ln}}  \text{Re}\,\text{tr}[\hat{P}_l (\partial^{b}\hat{H})\hat{P}_n (\partial^{c}\hat{H})]\,.
\end{align}
We used that the real part of this trace is symmetric in $l,n$. Isolating the remaining energy derivatives, we obtain
\begin{equation}
    E^{ a }_n E_{n}^{ b  c }\text{tr}[\hat{P}_n]=-\text{tr}[\hat{P}_n(\partial^{ a }\hat{\hat{H}})\hat{P}_n (\partial^{ b }\partial^{ c }\hat{\hat{H}})]   + 2\text{tr}[\hat{P}_n (\partial^a \hat{H})]\sum_{l\neq n}\frac{1}{\epsilon_{ln}}  \text{Re}\,\text{tr}[\hat{P}_l (\partial^{b}\hat{H})\hat{P}_n (\partial^{c}\hat{H})]\,.
\end{equation}
This is the NLD term. Note that a sum over all bands is required in this calculation, which increases the numerical effort. However, it must be compared to performing a full diagonalization at multiple points to obtain a finite-difference approximation of this derivative, which is less accurate and possibly more resource-intensive. 

Finally, we present a formula for the intraQMD contribution, proportional to the skewness tensor Eq.~\eqref{eq:QGdefSkewness}
\begin{align}
    \text{Re}\,\partial^{ a }Q_{n}^{ b  b} = 2\,\text{Re}\,Q^{b;ab}\,.
\end{align}
We focused on the most relevant case of $b=c$ as in the main text, and used Eq.~\eqref{eq:QGTdipoleAsSkewness}. We thus need to express 
\begin{align}
    Q^{b;ab} = \text{tr}[\hat{P}_n(\partial^ b  \hat{P}_n)(\partial^ a  \partial^ b  \hat{P}_n) ]
\end{align}
using derivatives of the Hamiltonian only. This case is the most complicated, but the method used to find a formula can be generalized to more complex cases.
To begin, we insert an identity matrix as a sum over projectors:
\begin{align}
    Q_{n}^{b;ab} &= \text{tr}[\hat{P}_n(\partial^b \hat{P}_n)(\partial^a \partial^b \hat{P}_n) ] \\
    &= \sum_m \text{tr}[\hat{P}_n(\partial^b \hat{P}_n)\hat{P}_m(\partial^a \partial^b \hat{P}_n) ]\\
    &= \sum_{m\neq n} \text{tr}[\hat{P}_n(\partial^b \hat{P}_n)\hat{P}_m(\partial^a \partial^b \hat{P}_n) ]\,.
\end{align}
Now a lengthy but straightforward calculation leads to
\begin{equation}
  \begin{aligned}
    \hat{P}_m (\partial^{ a}\partial^{b }\hat{H}) \hat{P}_n
    =& \epsilon_{nm}^{ b } \hat{P}_m (\partial^{ a }\hat{P}_n) \hat{P}_n +\epsilon_{nm}^{ a } \hat{P}_m (\partial^{ b }\hat{P}_n) \hat{P}_n - \epsilon_{mn} \hat{P}_m (\partial^{ a }\partial^{ b }\hat{P}_n) \hat{P}_n\\
&+ \sum_{k\neq m,n}
\epsilon_{mk} \hat{P}_m (\partial^{ a }\hat{P}_k)\hat{P}_k(\partial^{ b }\hat{P}_n) \hat{P}_n + \sum_{k\neq m,n}
\epsilon_{mk} \hat{P}_m (\partial^{ b }\hat{P}_k)\hat{P}_k(\partial^{ a }\hat{P}_n) \hat{P}_n\,.
\end{aligned}  
\end{equation}
Now for $m \neq n$ we may use Eq.~\eqref{eq:dHformulaProj} to insert derivatives of the Hamiltonian, which yields
\begin{equation}
    \begin{aligned}
    \hat{P}_m (\partial^a\partial^b \hat{H}) \hat{P}_n =& -\frac{\epsilon_{nm}^{ b }}{\epsilon_{mn}} \hat{P}_m (\partial^a \hat{H}) \hat{P}_n -\frac{\epsilon_{nm}^{ a }}{\epsilon_{mn}} \hat{P}_m (\partial^b \hat{H}) \hat{P}_n - \epsilon_{mn} \hat{P}_m (\partial^{ a }\partial^{ b }\hat{P}_n) \hat{P}_n\\
&+ \sum_{k\neq m,n}
\frac{1}{\epsilon_{mn}\epsilon_{kn}} \hat{P}_m (\partial^a \hat{H}) \hat{P}_k (\partial^b \hat{H}) \hat{P}_n + \sum_{k\neq m,n}
\frac{1}{\epsilon_{mn}\epsilon_{kn}} \hat{P}_m (\partial^b \hat{H}) \hat{P}_k (\partial^a \hat{H}) \hat{P}_n\,.
\end{aligned}
\end{equation}
We may isolate the second derivative of a projector in the above:
\begin{equation} \label{eq:suppdoublederProjExpressedwithH}
    \begin{aligned}
     \hat{P}_m (\partial^{ a }\partial^{ b }\hat{P}_n) \hat{P}_n =& -\frac{1}{\epsilon_{mn}}\hat{P}_m (\partial^a\partial^b \hat{H}) \hat{P}_n -\frac{\epsilon_{nm}^{ b }}{\epsilon_{mn}^2} \hat{P}_m (\partial^a \hat{H}) \hat{P}_n -\frac{\epsilon_{nm}^{ a }}{\epsilon_{mn}^2} \hat{P}_m (\partial^b \hat{H}) \hat{P}_n \\
&+ \sum_{k\neq m,n}
\frac{1}{\epsilon_{mn}\epsilon_{kn}} \hat{P}_m (\partial^a \hat{H}) \hat{P}_k (\partial^b \hat{H}) \hat{P}_n + \sum_{k\neq m,n}
\frac{1}{\epsilon_{mn}\epsilon_{kn}} \hat{P}_m (\partial^b \hat{H}) \hat{P}_k (\partial^a \hat{H}) \hat{P}_n\,.
\end{aligned}
\end{equation}
We now insert this back into $Q_n^{b;ab}$, also using Eq.~\eqref{eq:dHformulaProj} again:
\begin{align}
    Q_{n}^{b;ab} &= \sum_{m\neq n} \text{tr}[\hat{P}_n(\partial^b \hat{P}_n)\hat{P}_m(\partial^a \partial^b \hat{P}_n) ] \\
    &= \sum_{m\neq n}\frac{1}{\epsilon_{mn}}\text{tr}[\hat{P}_n(\partial^b \hat{H})\hat{P}_m(\partial^a \partial^b \hat{P}_n) ] \\
    &= -\sum_{m\neq n}\frac{1}{\epsilon_{mn}^2}\text{tr}[\hat{P}_n(\partial^b \hat{H})\hat{P}_m (\partial^a\partial^b \hat{H})] -\sum_{m\neq n}\frac{\epsilon_{nm}^{ b }}{\epsilon_{mn}^3}\text{tr}[\hat{P}_n(\partial^b \hat{H})\hat{P}_m (\partial^a \hat{H}) ]-\sum_{m\neq n}\frac{\epsilon_{nm}^{ a }}{\epsilon_{mn}^3}\text{tr}[\hat{P}_n(\partial^b \hat{H})\hat{P}_m (\partial^b \hat{H}) ] \\\nonumber
    &\quad + \sum_{m\neq n}\sum_{k\neq m,n}\frac{1}{\epsilon_{mn}^2\epsilon_{kn}}\text{tr}[\hat{P}_n(\partial^b \hat{H})\hat{P}_m (\partial^a \hat{H}) \hat{P}_k (\partial^b \hat{H}) ]+ \sum_{m\neq n}\sum_{k\neq m,n}\frac{1}{\epsilon_{mn}^2\epsilon_{kn}}\text{tr}[\hat{P}_n(\partial^b \hat{H})\hat{P}_m (\partial^b \hat{H}) \hat{P}_k (\partial^a \hat{H}) ]\,.
\end{align}
For Hamiltonians with many bands, the (multiple) sums over band indices may be costly numerically. However, due to the factors of $1/\epsilon_{mn}$ inside the sums, these may be truncated to only include bands which are close by in energy. For general more complicated quantum geometric quantities, one only needs to use formulas such as Eq.~\eqref{eq:suppdoublederProjExpressedwithH} for the required order of derivatives, and compose them to form the tensor as we did here.

\end{document}